\documentclass[11pt,a4paper]{article}

\usepackage{ifthen} 
\newboolean{pdflatex}
\setboolean{pdflatex}{true} 

\newboolean{articletitles}
\setboolean{articletitles}{true} 

\newboolean{uprightparticles}
\setboolean{uprightparticles}{false} 

\newboolean{inbibliography}
\setboolean{inbibliography}{false} 

\usepackage{microtype}
\usepackage{lineno}  
\usepackage{xspace} 
\usepackage{caption} 

\usepackage{graphicx}  
\usepackage{color}
\usepackage{colortbl}
\graphicspath{{./figs/}} 
\usepackage{pdflscape}

\usepackage{siunitx} 
\usepackage{amsmath} 
\usepackage{amssymb}
\usepackage{amsfonts}
\usepackage{upgreek} 

\usepackage{jinstpub}

\usepackage{xspace} 
\usepackage{upgreek}

\def\MagUp {\mbox{\em Mag\kern -0.05em Up}\xspace}

\ifthenelse{\boolean{uprightparticles}}%
{

 \def\PDelta      {\ensuremath{\Delta}\xspace}                 
 \def\PXi      {\ensuremath{\Xi}\xspace}                 
 \def\PLambda      {\ensuremath{\Lambda}\xspace}                 
 \def\PSigma      {\ensuremath{\Sigma}\xspace}                 
 \def\POmega      {\ensuremath{\Omega}\xspace}                 
 \def\PUpsilon      {\ensuremath{\Upsilon}\xspace}                 
 
 \def\PB      {\ensuremath{\mathrm{B}}\xspace}                 
                  
 \def\PD      {\ensuremath{\mathrm{D}}\xspace}

 \def\PK      {\ensuremath{\mathrm{K}}\xspace}

 \def\Pi      {\ensuremath{\mathrm{i}}\xspace}

}
{

 \mathchardef\PDelta="7101
 \mathchardef\PXi="7104
 \mathchardef\PLambda="7103
 \mathchardef\PSigma="7106
 \mathchardef\POmega="710A
 \mathchardef\PUpsilon="7107
                  
 \def\PB      {\ensuremath{B}\xspace}                 
                  
 \def\PD      {\ensuremath{D}\xspace}

 \def\PK      {\ensuremath{K}\xspace}

 \def\Pi      {\ensuremath{i}\xspace}

}

\makeatletter
\ifcase \@ptsize \relax
  \newcommand{\miniscule}{\@setfontsize\miniscule{4}{5}}
\or
  \newcommand{\miniscule}{\@setfontsize\miniscule{5}{6}}
\or
  \newcommand{\miniscule}{\@setfontsize\miniscule{5}{6}}
\fi
\makeatother

\DeclareRobustCommand{\optbar}[1]{\shortstack{{\miniscule (\rule[.5ex]{1.25em}{.18mm})}
  \\ [-.7ex] $#1$}}

  \def\Kbar    {{\kern 0.2em\overline{\kern -0.2em \PK}{}}\xspace}

\def\KorKbar    {\kern 0.18em\optbar{\kern -0.18em K}{}\xspace}

  \def\Dbar    {{\kern 0.2em\overline{\kern -0.2em \PD}{}}\xspace}

\def\DorDbar    {\kern 0.18em\optbar{\kern -0.18em D}{}\xspace}

\def\Bbar    {{\ensuremath{\kern 0.18em\overline{\kern -0.18em \PB}{}}}\xspace}

\def\BorBbar    {\kern 0.18em\optbar{\kern -0.18em B}{}\xspace}

  \def\Y#1S{\ensuremath{\PUpsilon{(#1S)}}\xspace}

\def\Lbar        {{\ensuremath{\kern 0.1em\overline{\kern -0.1em\PLambda}}}\xspace}
\def\LorLbar    {\kern 0.18em\optbar{\kern -0.18em \PLambda}{}\xspace}

\def\AT#1     {\ensuremath{A_{\mathrm{T}}^{#1}}\xspace}           

\def\C#1      {\ensuremath{\mathcal{C}_{#1}}\xspace}                       
\def\Cp#1     {\ensuremath{\mathcal{C}_{#1}^{'}}\xspace}                    
\def\Ceff#1   {\ensuremath{\mathcal{C}_{#1}^{\mathrm{(eff)}}}\xspace}        
\def\Cpeff#1  {\ensuremath{\mathcal{C}_{#1}^{'\mathrm{(eff)}}}\xspace}       
\def\Ope#1    {\ensuremath{\mathcal{O}_{#1}}\xspace}                       
\def\Opep#1   {\ensuremath{\mathcal{O}_{#1}^{'}}\xspace}                    

\newcommand{\unit}[1]{\ensuremath{\mathrm{ \,#1}}\xspace}          

\newcommand{\tev}{\ifthenelse{\boolean{inbibliography}}{\ensuremath{~T\kern -0.05em eV}\xspace}{\ensuremath{\mathrm{\,Te\kern -0.1em V}}}\xspace}
\newcommand{\gev}{\ensuremath{\mathrm{\,Ge\kern -0.1em V}}\xspace}
\newcommand{\mev}{\ensuremath{\mathrm{\,Me\kern -0.1em V}}\xspace}
\newcommand{\kev}{\ensuremath{\mathrm{\,ke\kern -0.1em V}}\xspace}
\newcommand{\ev}{\ensuremath{\mathrm{\,e\kern -0.1em V}}\xspace}
\newcommand{\gevc}{\ensuremath{{\mathrm{\,Ge\kern -0.1em V\!/}c}}\xspace}
\newcommand{\mevc}{\ensuremath{{\mathrm{\,Me\kern -0.1em V\!/}c}}\xspace}
\newcommand{\gevcc}{\ensuremath{{\mathrm{\,Ge\kern -0.1em V\!/}c^2}}\xspace}
\newcommand{\gevgevcccc}{\ensuremath{{\mathrm{\,Ge\kern -0.1em V^2\!/}c^4}}\xspace}
\newcommand{\mevcc}{\ensuremath{{\mathrm{\,Me\kern -0.1em V\!/}c^2}}\xspace}

\def\cm   {\ensuremath{\mathrm{ \,cm}}\xspace}

\def\mum  {\ensuremath{{\,\upmu\mathrm{m}}}\xspace}

\def\invfb   {\ensuremath{\mbox{\,fb}^{-1}}\xspace}

\def\sec  {\ensuremath{\mathrm{{\,s}}}\xspace}

\def\neutroneq {\ensuremath{\mathrm{ \,n_{eq}}}\xspace}

\def\gsim{{~\raise.15em\hbox{$>$}\kern-.85em
          \lower.35em\hbox{$\sim$}~}\xspace}
\def\lsim{{~\raise.15em\hbox{$<$}\kern-.85em
          \lower.35em\hbox{$\sim$}~}\xspace}

\def\tell1  {TELL1\xspace}
\def\ukl1   {UKL1\xspace}

\usepackage{caption}
\usepackage{subcaption}

\def\dg{\ensuremath{^\circ}\xspace}

\def\np {{n-on-p}\xspace}
\def\nn {{n-on-n}\xspace}

\def\maxfluence {\ensuremath{8 \times 10^{15} 1 \mev \neutroneq {\mathrm{ \,cm}}^{-2}}\xspace} 
\def\fluence {\ensuremath{{10^{15}  1 \mev \neutroneq {\mathrm{ \,cm}}^{-2}}}\xspace}

\newcommand{\mevneq}{\text{1~MeV~n$_\text{eq}$ cm$^{-2}$}} 
\usepackage[noabbrev, capitalise]{cleveref}
\usepackage{siunitx}
\usepackage{etoolbox}
\usepackage{appendix}
\usepackage{todonotes}
\usepackage{subcaption}

\title{Spatial resolution and efficiency of prototype sensors for the LHCb VELO Upgrade}

\author[a,c,1]{E.~Buchanan\note{Corresponding author},}
\author[b]{ K.~Akiba,}
\author[b]{ M.~van Beuzekom,}
\author[c]{ P.~Collins,}
\author[b,d]{ E.~Dall'Occo,}
\author[c,e]{ T.~Evans,}
\author[f]{ V.~Franco Lima,}
\author[b]{R.~Geertsema,}
\author[g]{P.~Kopciewicz,}
\author[a]{E.~Price,}
\author[g]{B.~Rachwal,}
\author[a]{S.~Richards,}
\author[a]{D.~Saunders,}
\author[c]{H.~Schindler,}
\author[g]{T.~Szumlak,}
\author[b,h]{P.~Tsopelas,}
\author[a]{J.~Velthuis,}
\author[i]{and M.R.J.~Williams}

\affiliation[a]{University of Bristol, Beacon House, Queens Road, BS8 1QU, Bristol, United Kingdom}
\affiliation[b]{Nikhef, Science Park 105, 1098 XG Amsterdam, the Netherlands}
\affiliation[c]{CERN, 1211 Geneve, Switzerland}
\affiliation[d]{TU Dortmund, Otto-Hahn-Straße 4, 44227 Dortmund, Germany}
\affiliation[e]{University of Oxford, Particle Physics Department,\\Denys Wilkinson Bldg., Keble Road, Oxford OX1 3RH, United Kingdom}
\affiliation[f]{Oliver Lodge Laboratory, University of Liverpool, \\Liverpool, L69 7ZE, United Kingdom}
\affiliation[g]{AGH University of Science and Technology, Faculty of Physics and \\Applied Computer Science, Krak\'ow, Poland}
\affiliation[h]{Spectricon, Science and Technology Park of Crete, Heraklion, Greece}
\affiliation[i]{School of Physics and Astronomy, University of Edinburgh, \\Edinburgh, United Kingdom }

\emailAdd{emma.buchanan@cern.ch}

\abstract{
A comprehensive study of the spatial resolution and detection efficiency of sensor prototypes developed for the LHCb VELO upgrade is presented. 
Data samples were collected at the CERN SPS H8 beam line using a hadron mixture of protons and pions with momenta of approximately \SI{180}{GeV/c}.
The sensor performance was characterised using both irradiated and non-irradiated sensors. Irradiated samples were subjected to a maximum fluence of \maxfluence, of both protons and neutrons.
The spatial resolution is measured comparing the detected hits to the 
position as predicted by tracks reconstructed by the Timepix3 telescope. 
The resolution is presented for different applied bias voltages and track angles,  sensor thickness and implant size.}


\begin{document}
\maketitle




\section{Introduction}

The LHCb experiment is upgrading its VErtex LOcator (VELO) detector 
during Long Shutdown 2 of the LHC to allow the experiment to operate at 
an instantaneous luminosity of $\mathcal{L} = 2 \times10^{33}~\cm^{-2}\sec^{-1}$, 
five times higher than previous runs~\cite{LHCb-TDR-013}. 
The VELO requires very precise tracking and fast pattern recognition
in order to reconstruct collisions and decay vertices in real
time as the first step of the LHCb trigger decision.
The VELO upgrade will replace the original detector's
silicon strips with hybrid pixel detectors, 
which consist of planar silicon sensors bump-bonded to
VeloPix~\cite{VELOPIX} readout ASICs (Application Specific Integrated Circuits). 

 The  region of the detector closest to the collision point will be exposed to a total integrated fluence of $\phi =$ \maxfluence \cite{LHCb-TDR-013}. 
To maintain an excellent tracking and vertexing performance through its lifetime,  the detector must remain fully efficient ($> 99\%$) up to the maximum fluence levels.

This paper presents the spatial resolution and the hit efficiency 
of several different prototype devices designed for the VELO upgrade, 
 before and after being irradiated up to \maxfluence with both uniform and non-uniform fluence profiles. 

The sensor prototypes are planar silicon pixel sensors, with square pixels of   
55$\mum$ pitch, 
bonded to  Timepix3 ASICs~\cite{Timepix3}, which can  measure 
simultaneously the particles Time-Of-Arrival (TOA) and Time-Over-Threshold (TOT).
The TOA is measured with 1.56~ns TDC bin width and the TOT is 
proportional to the charge collected by each pixel, and calibrated as described in  Refs.\cite{Akiba_2019,VicenteBarretoPinto:2134709}. The threshold is set to 1000\unit{e^-} to ensure negligible noise.
The sensor performance has been evaluated using a charged
hadron beam. 
Tracks are reconstructed using the Timepix3 Telescope~\cite{Akiba_2019}.

This paper is organised as follows.
Firstly, the experimental setup is described in \Cref{sec:setup}. The main results are then presented in 
\Cref{sec:bias_clus_sizes,sec:efficiency,sec:resolution}, describing the 
cluster distributions, efficiency and spatial resolution, respectively.
Finally, the applicability 
of the prototype sensors regarding the running conditions in the LHCb experiment is discussed in \Cref{sec:conclusion}.


\section{Prototype sensors and experimental setup}
\label{sec:setup}

\subsection{Prototype devices}
\label{section:PD}
The prototype sensors were produced by two different manufacturers, Hamamatsu~Photonics~K.~K. (HPK)\footnote{Hamamatsu Photonics K. K., 325-6, Sunayama-cho, Naka-ku, Hamamatsu City, Shizuoka, 430-8587, Japan} and Micron Semiconductor Ltd\footnote{Micron Semiconductor Ltd, 1 Royal Buildings, Marlborough Road, Lancing BN158UN, United Kingdom}.
The prototypes have different design features in terms of pixel implant size, sensor thickness, bulk type and the pixel-to-edge (PTE) distance (see \Cref{tab:assemblies}).
The HPK sensors have been produced with n$^+$-type implants separated by p$^+$-stop implants on a $200\pm 20 $\mum\ thick float-zone p-doped silicon substrate   
(resistivity of 3-8 k$\Omega$ cm). The back of the sensor consists of a thin p$^+$-doped layer and is fully metallised. 
Two different guard ring designs with PTE of 450 and 600\mum. 
The pixel implants are either 35 or 39\mum squares with rounded corners. The Micron prototypes have been produced with 36\mum\ wide n$^+$-type implants with rounded corners and p$^+$-spray isolation. Two different types of substrates have been used: 200\mum\ p-type ($>$ 5 k$\Omega$ cm) and 150\mum\ n-type ($>$ 1.5 k$\Omega$ cm). The latter is double sided processed with guard rings on the backside of the sensor partially implanted underneath the edge pixels. The back of these sensors consist of a thin p$^+$-doped layer and the back for the p-type substrate is fully metallised, while for the n-type substrate the back is metallised in a grid structure.
For these sensors there are two PTE variants with corresponding distances of 250\mum and 450\mum. The n$^+$-p-p$^+$ (n$^+$-n-p$^+$) sensors are from now on referred to as \np (\nn) sensors. 
The details of individual assemblies and their identification numbers, which are used in the following sections, can be found in \Cref{sec:appendix1}.

\begin{table}[!h]
\caption{Prototype assemblies.}
\label{tab:assemblies}
\centering
\begin{tabular}{llccc}
 \hline
Vendor & Type & Thickness & PTE & Implant width \\
\hline
\hline
HPK    & \np & 200\mum & 450, 600\mum & 35, 39\mum \\
Micron & \np & 200\mum & 450\mum & 36\mum   \\
Micron & \nn & 150\mum & 250, 450\mum & 36\mum   \\
\hline
\end{tabular}
\end{table}

The depletion voltages were measured to be around $\sim$120~V for HPK devices and $\sim$40~V for both types of Micron devices~\cite{dallocco2021temporal,Geertsema_2021}.
A subset of devices were uniformly irradiated with neutrons at JSI~\cite{JSI}, up to \maxfluence, referred to as full fluence. Another group of devices were non-uniformly irradiated at the IRRAD facility at CERN~\cite{Gkotse_2237333}, which provides a 24~\gev proton beam that has approximately a two-dimensional Gaussian distribution.
This is used to emulate the variation in the fluence anticipated across the inner VELO sensors at the end of their lifetime.
The reconstructed fluence profile~\cite{DallOccoThesis} is shown in \Cref{sec:appendix3}.


\subsection{The Timepix3 Telescope and track reconstruction}
\label{section:TPX3}

An extensive test beam programme has been carried out at the SPS H8 beamline at CERN~\cite{sps} to characterise the sensors. 
The beam is composed of a mixture of charged hadrons
($\sim$67\% protons, $\sim$30\% pions) at 180~\gevc. 
The Timepix3 telescope~\cite{Akiba_2019} is a high rate, data-driven beam telescope, composed of two arms of four planes each.
Each plane is equipped with a 300\mum p-on-n silicon sensor 
that is bump bonded to a Timepix3 ASIC, which is operated with a 1000\unit{e^-} threshold. 
The centre of the telescope is reserved for the Device Under Test (DUT). 
The DUT area is equipped with remotely controlled motion stages able 
to translate in $x$ and $y$ directions, with $z$ as the beam axis and rotate about the $y$ axis. 
A vacuum box can also be installed on the central stage  to facilitate 
testing of irradiated devices at high voltage. The cooling block connected to the ASIC can be cooled down to temperatures of  $ -35^{\circ}$C, which keeps the sensors below $ -20^{\circ}$C. The temperature of the cold box is measured with a Pt100, which is converted to the corresponding sensor temperature using a calibration curve. This calibration curve is determined before the testbeam in a dedicated lab setup. In this setup a Pt100 was glued to the sensor itself and another Pt100 was glued to the cold box in order to measure the sensor temperature for different cold box temperatures.
The pointing resolution at the DUT position is about $1.6\mum$, 
enabling intrapixel studies of the DUT.
The typical temporal resolution on a track using only the Timepix3 timestamps 
is about $350$~ps~\cite{Heijhoff_2020}.


Tracks are only reconstructed if hits are found
in all eight telescope planes within 20~ns. 
The maximum cluster size 
on each plane is set to 10 and the track
fit $\chi^2$ is required to be below 10.
A DUT cluster is associated to a track if its position is within a 100\mum and \SI{35}{ns} from the track intercept reconstructed by the telescope.

\subsection{Spatial resolution measurement}
\label{sec:res_description}
The orthogonal, $x$ (or $y$), position of a traversing particle through 
the sensor is reconstructed using the  Centre of Gravity (CoG) method as: 
\begin{equation}
x = \frac{ \sum_{cluster} S_{i} \cdot x_{i} }{ \sum_{cluster} S_{i}},
\label{eq:COG}
\end{equation}
where $x_{i}$ is the coordinate of the $i^{th}$ pixel in the cluster and $S_{i}$ is the signal measured in the corresponding pixel. 
The resolution of this interpolation depends on the signal-to-noise ratio
of the sensor and on the digitisation precision of the front-end, as well as on the threshold value set. For the purposes of this paper the resolution on the charge measurement performed by the Timepix3 front-end gives a negligible contribution to the CoG uncertainty.

The figure of  merit adopted  to compare the spatial performance of the prototypes is the resolution given by the width of the residual distribution, 
where residual is defined as the difference between the track intercept estimated from the trajectory fit, provided by the
telescope~(\Cref{section:TPX3}) and the cluster position reconstructed using the sensor.  The  width of the residual distribution, $\sigma_{residual}$, incorporates not only the intrinsic resolution of the sensor $\sigma_{sensor}$
but also the telescope pointing resolution $\sigma_{telescope}$:
\begin{equation}
\sigma_{residual}^{2} = \sigma_{sensor}^{2} + \sigma_{telescope}^{2}.
\label{eq:res_total}
\end{equation} 
The intrinsic spatial resolution of the sensor is therefore obtained by subtracting in quadrature from the measured width of the residual distribution the telescope resolution. 

\begin{figure}[ht]
   \begin{subfigure}{0.48\textwidth}
     \includegraphics[width=0.99\textwidth]{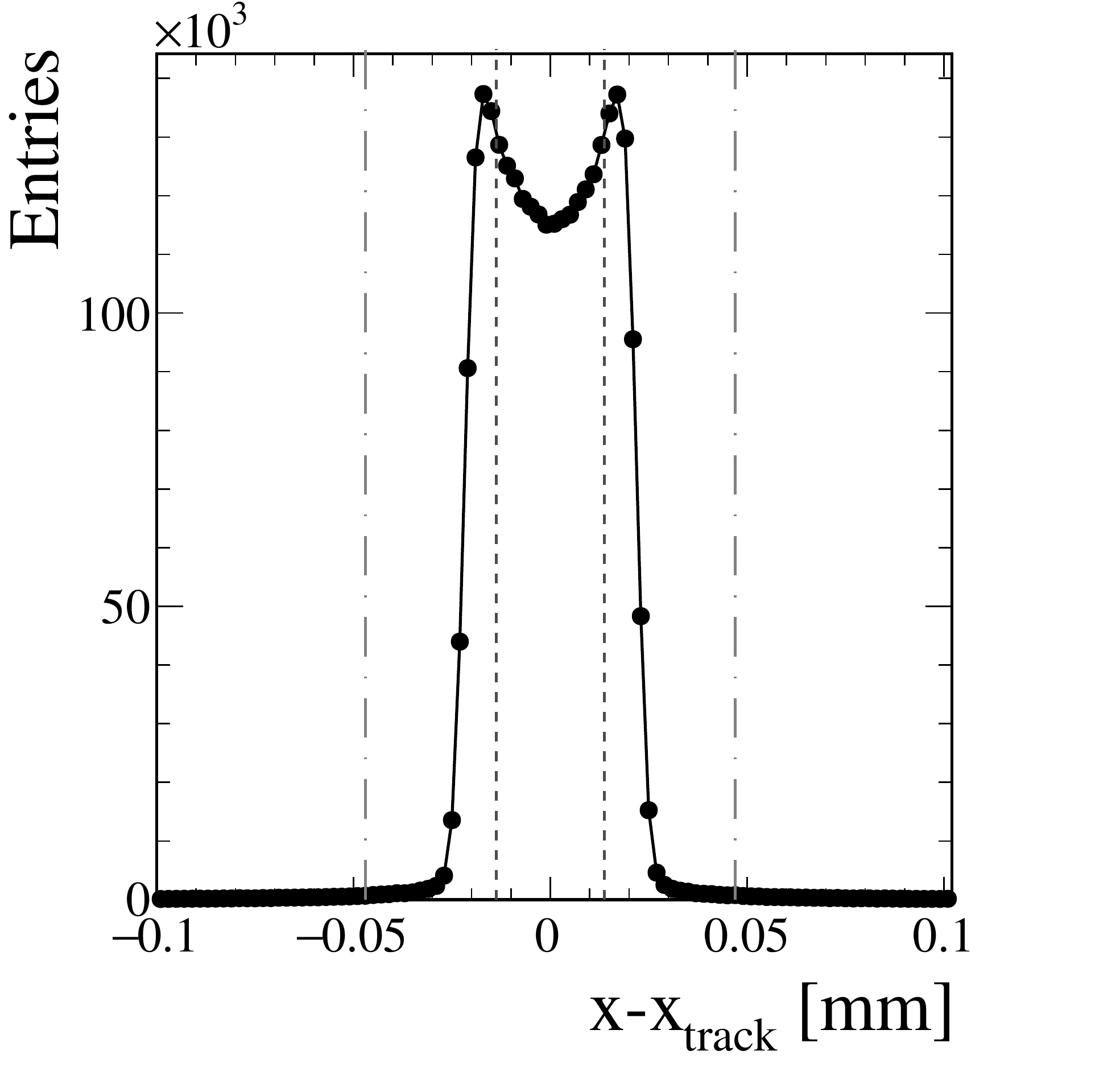}
     \caption{} 
     \end{subfigure}
   \begin{subfigure}{0.48\textwidth}
     \includegraphics[width=0.99\textwidth]{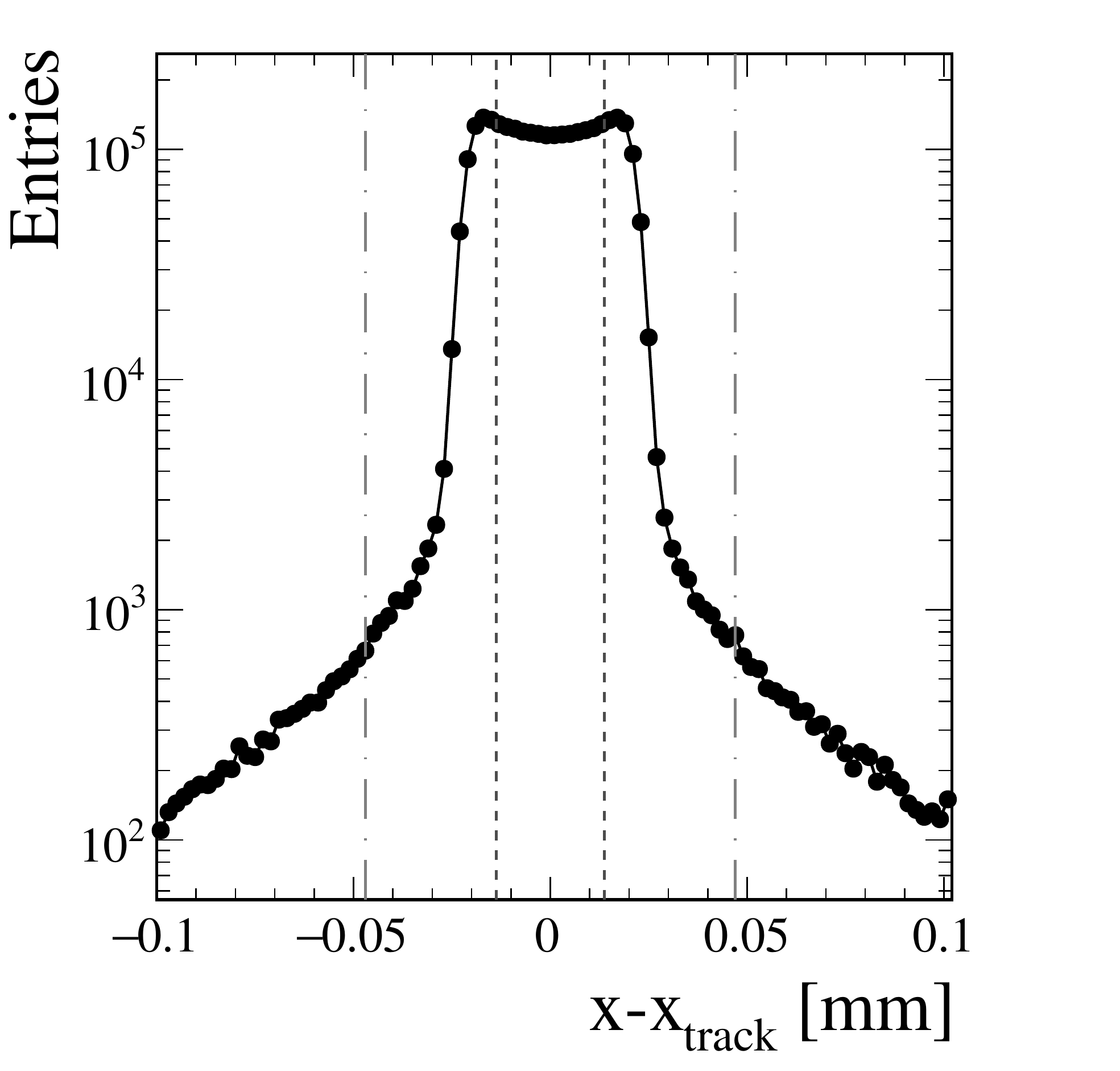}
     \caption{}
     \end{subfigure}
    
    \caption[]{Example residual distribution for a HPK n-on-p sensor (S6), placed perpendicular to the beam and operated near the full depletion voltage (a), logarithmic scale (b). The outer dashed lines indicate $\pm3\times$RMS and the inner dashed lines indicate the {\it truncated RMS}.}
  \label{fig:res_double_peak}
\end{figure}

The residual distribution for a Micron n-on-p sensor is shown in \Cref{fig:res_double_peak}.
The distribution has a distinctive double peak feature due to fact that the fraction of charge shared between two pixels is not linearly related to the distance between the intercept and the mid point of the pixels~\cite{Buchanan}. 
In this analysis the non-linear effect is not corrected.
The spatial resolution is quantified as the RMS of the residual distribution considering only a range of $\pm 3$ times the raw Root Mean Square (RMS) of the entire distribution.
This quantity is referred to as the {\it truncated RMS}. 


For charged particles traversing at an angle, the path of the particle
through the sensor bulk is greater than that of a perpendicular track. This results in an increase in the liberated charge and an increased probability
that particles will cross multiple pixels. Neglecting the threshold and diffusion effects and considering only two pixels, the angle that leads to the optimal spatial resolution
($\theta _{\rm{opt}}$)
is given in terms of the
pixel pitch and detector thickness by:
\begin{equation}
\label{eq:bestres}
\theta_{opt} = \tan^{-1}\left( \frac{\text{pitch}}{\text{thickness}}\right).
\end{equation}
From this, $\theta_{opt}$ is 15$^{\circ}$ for 200$\mum$ thick sensors and 20$^{\circ}$ for 150$\mum$ thick sensors.

\subsection{Cluster finding efficiency}
\label{sec:sec_eff}
The efficiency is defined as the ratio of the number of clusters matched to a track and the total number of reconstructed tracks. The
error on the efficiency is calculated as the error from the
binomial distribution. The telescope allows the efficiency to be measured as a function of intrapixel position. 



\section{Analysis of cluster sizes and charge sharing}
\label{sec:bias_clus_sizes}
\begin{figure}[h]
 \centering
 \begin{subfigure}{0.49\textwidth}
    \includegraphics[width=0.99\textwidth]{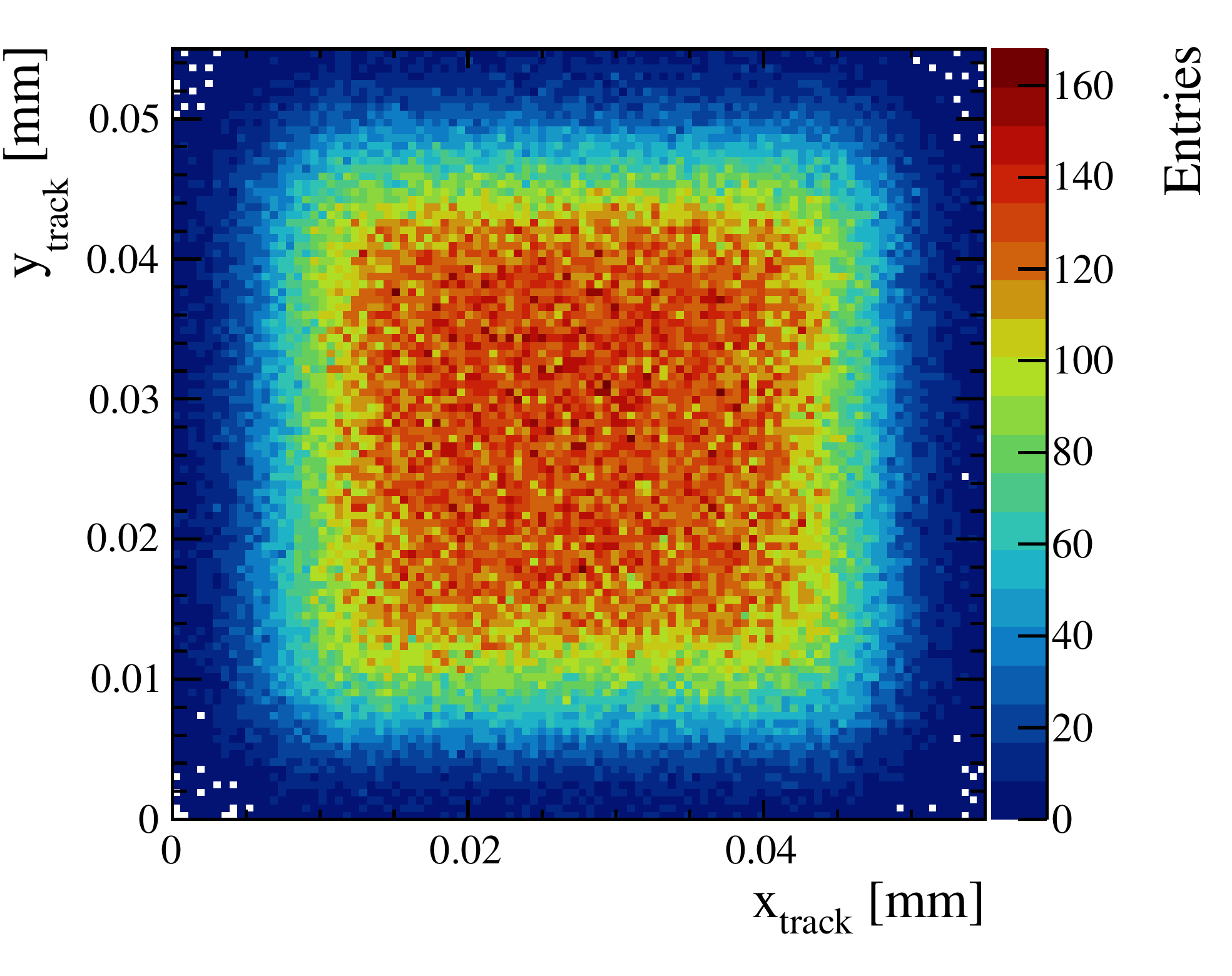}
      \caption{}
 \end{subfigure}
 \begin{subfigure}{0.49\textwidth}
    \includegraphics[width=0.99\textwidth]{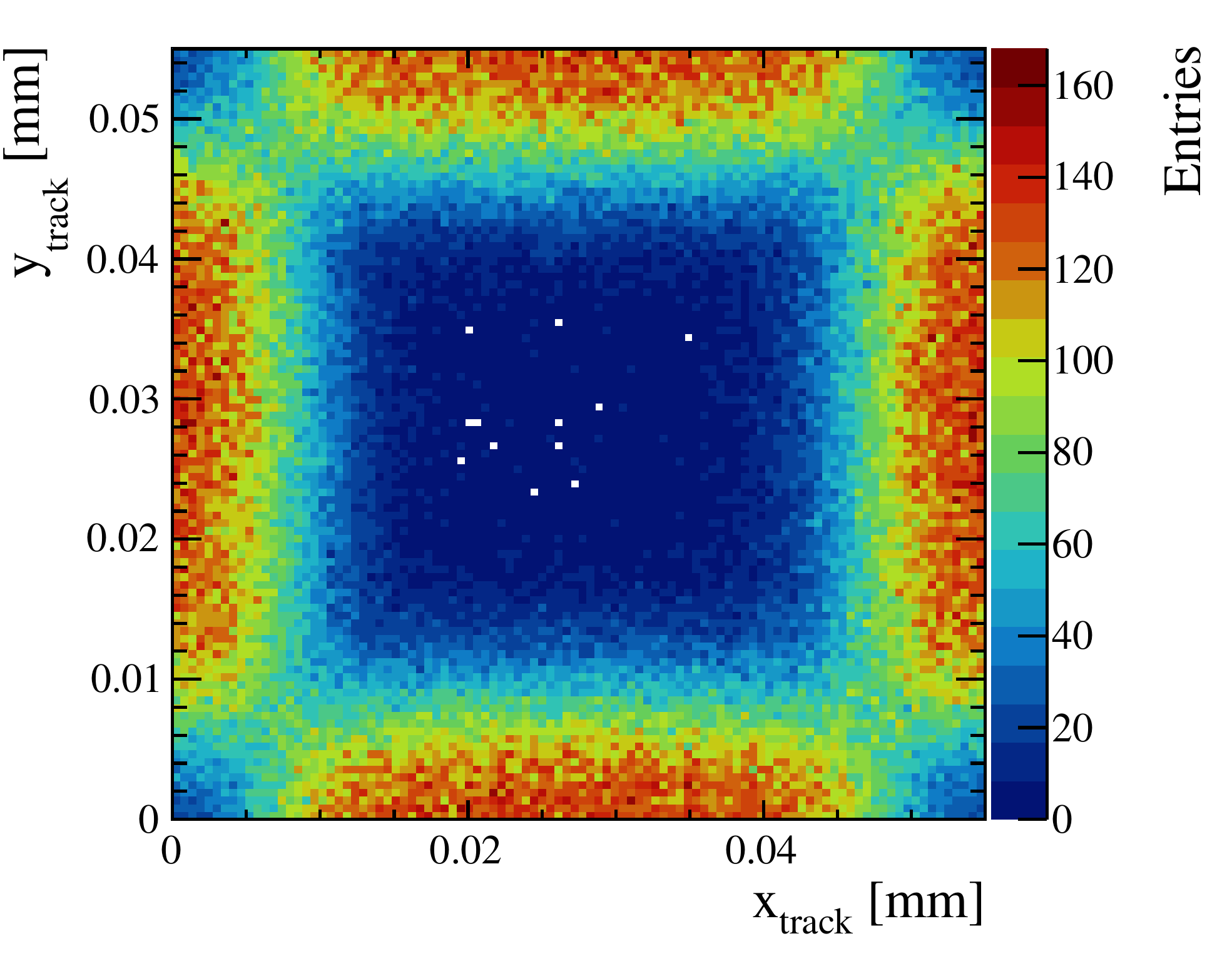}
     \caption{}
 \end{subfigure}
 \begin{subfigure}{0.49\textwidth}
    \includegraphics[width=0.99\textwidth]{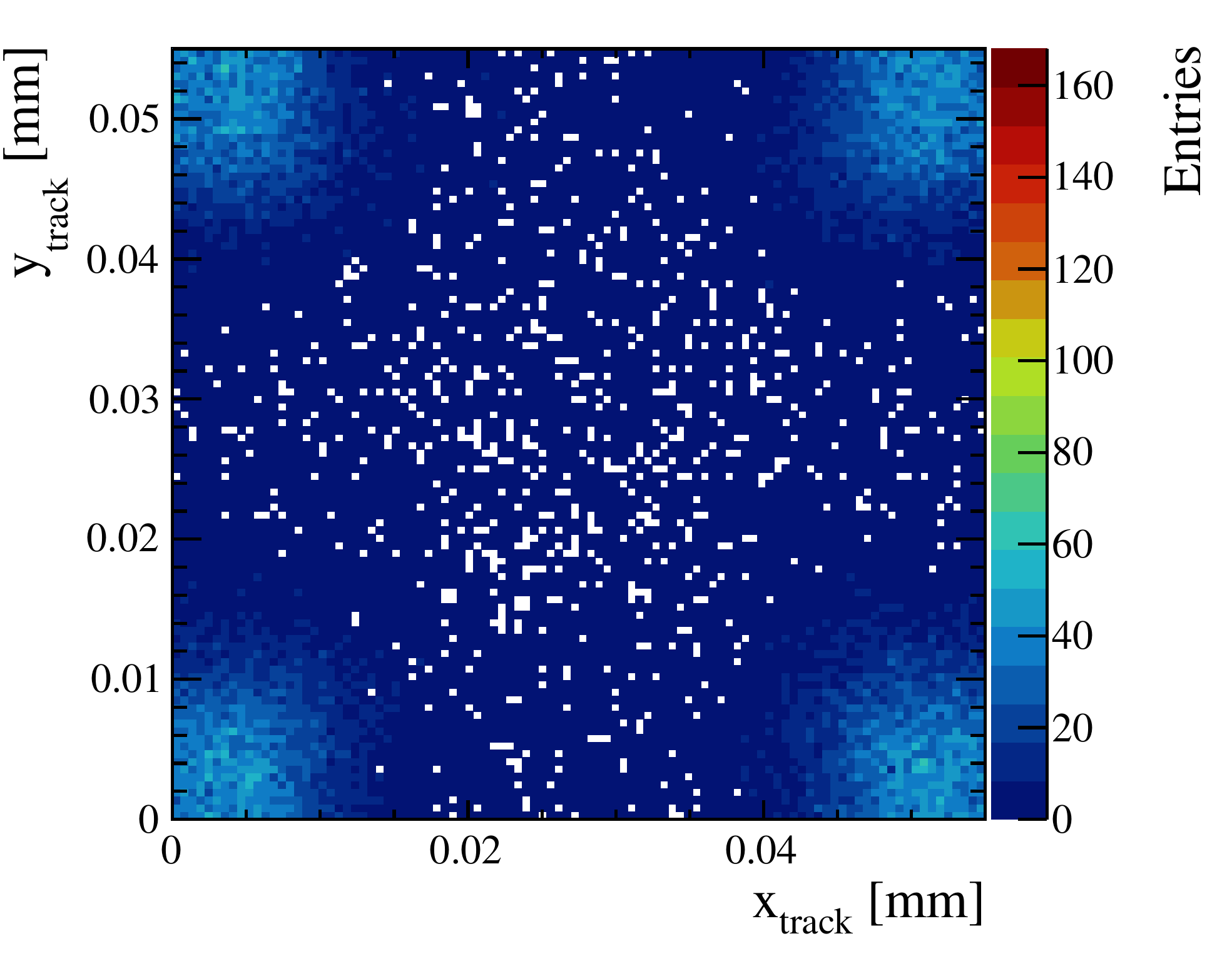}
     \caption{}
 \end{subfigure}
\begin{subfigure}{0.49\textwidth}
    \includegraphics[width=0.99\textwidth]{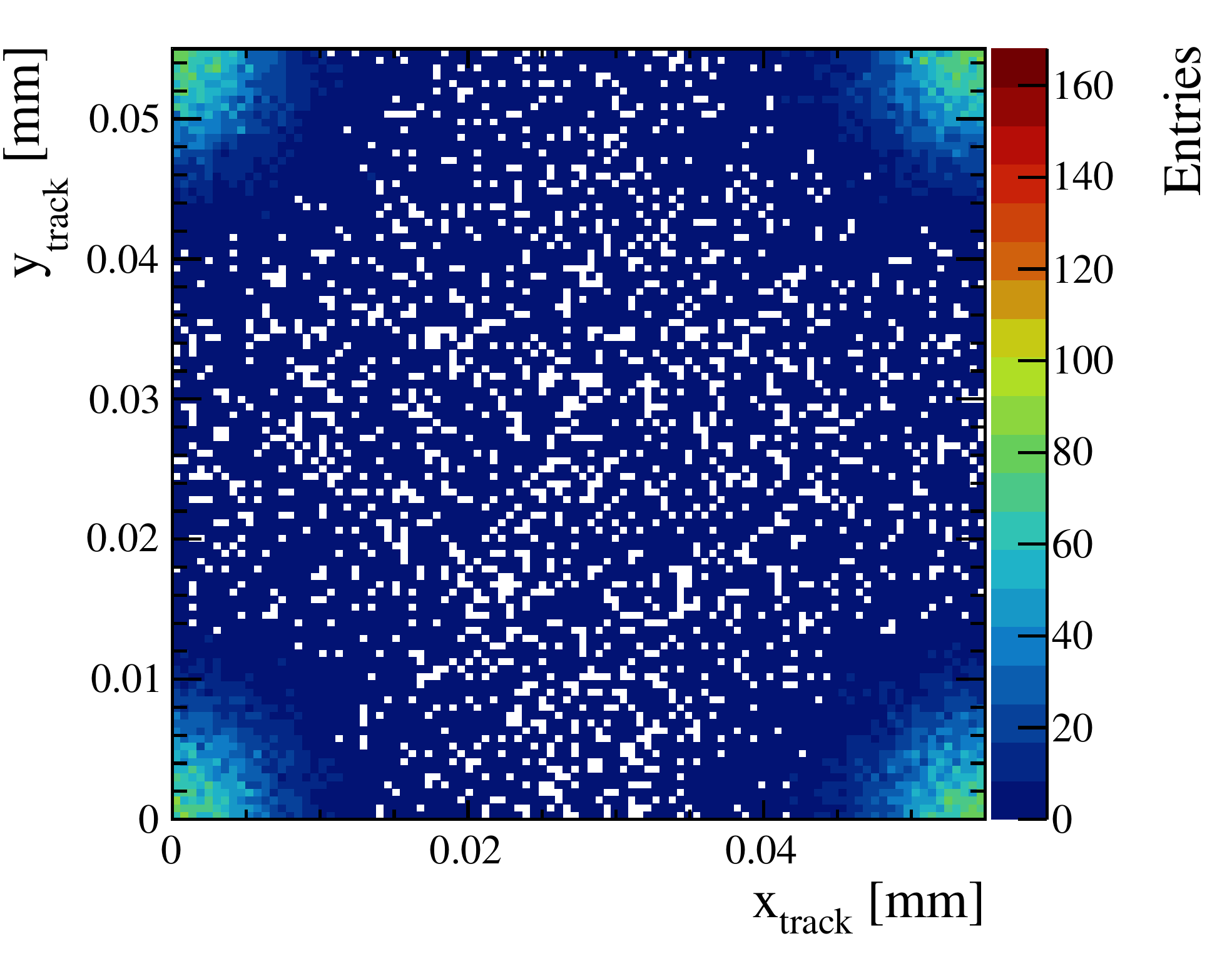}
     \caption{}
 \end{subfigure}
  \caption[]{The intrapixel track positions for cluster sizes: size~1~(a), size~2~(b), size~3~(c) and size~4~(d) for an Micron n-on-p sensor (S25).}
  \label{fig:hits}
 \end{figure}

The position with which a track traverses within a pixel influences 
charge sharing between pixels. \Cref{fig:hits} shows the track intercept locations within the pixel, known as intrapixel track position, for
size~1,~2,~3~$\&$~4 for a Micron n-on-p sensor placed perpendicular to the beam and operated 
near the full depletion voltage.
Size 1 clusters are mainly from tracks passing through the centre of the pixel.  Size two clusters originate from tracks hitting the pixel edges, while tracks at the corners result in cluster sizes of 3 and 4 pixels. In general cluster size~3~and~4 only account for $\sim$5\% of all clusters for perpendicular tracks. 


\subsection{Bias voltage}
\label{sec:bias_clus_section}
\begin{figure}[ht]
 \centering
\begin{subfigure}{0.49\textwidth}
 \includegraphics[width=0.99\textwidth]{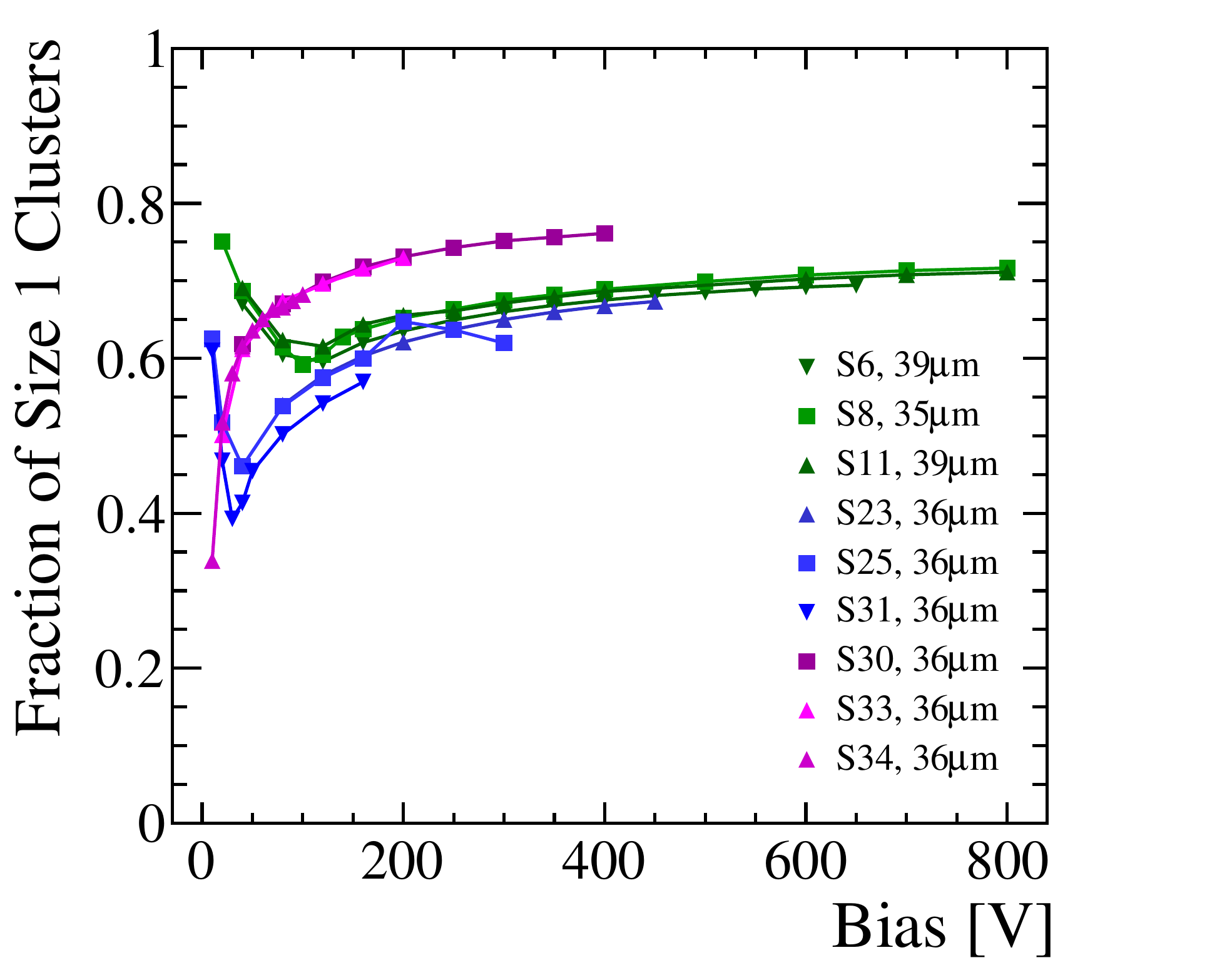}
    \caption{}
 \end{subfigure}
\begin{subfigure}{0.49\textwidth}
    \includegraphics[width=0.99\textwidth]{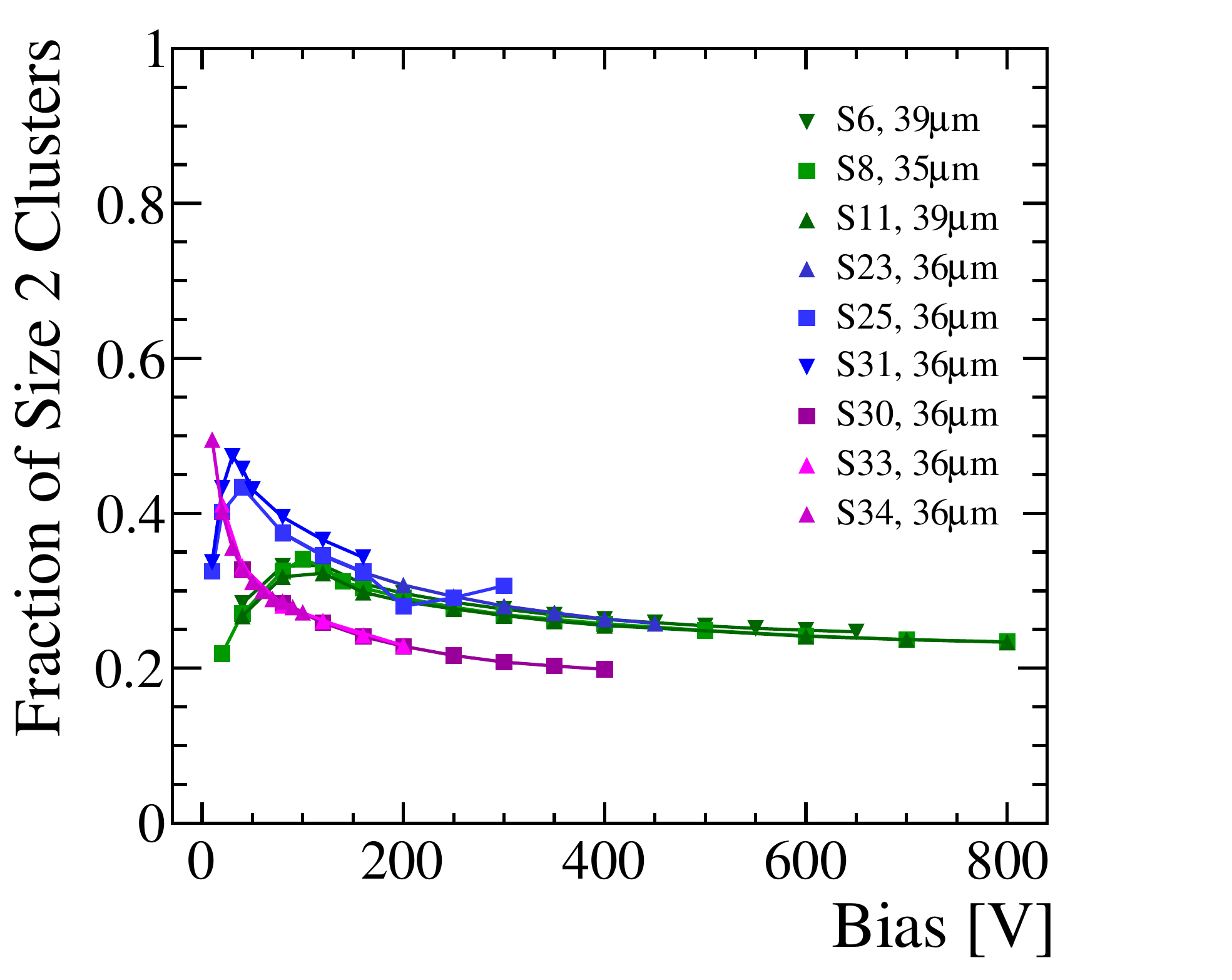}
    \caption{}
\end{subfigure}
 \caption[]{The fraction of size~1 (a) and size~2 (b) clusters as a function of the applied bias voltage. Green for HPK n-on-p, blue for Micron n-on-p and purple for Micron n-on-n.  }
 \label{fig:bias_frac_clus}
\end{figure}

Higher electric fields increase the drift velocity and subsequently reduce the lateral diffusion due to a shorter charge collection time~\cite{Spieler:1010490}. 
This effect is studied by measuring the cluster size distribution as a function of the applied bias voltage for sensors placed perpendicular to the beam, shown in Figures~\ref{fig:bias_frac_clus}~(a) and~(b). 
For the 200\mum n-on-p devices and voltages below full depletion (40~V for Micron sensors and 120~V for HPK), the diffusion between pixels is largest but the majority of clusters remain of size~1. 
As the sensor is not fully depleted the collected charge is small, and hence the charge shared with neighbouring pixels does not reach the threshold. 
Increasing the bias up to the full depletion voltage improves the charge collection efficiency~\cite{Geertsema_2021},  
which in turn increases the probability that the shared charge will cross the threshold,
leading to higher fractions of size~2 clusters, and consequently decreasing the fraction of size~1. The minimum in the fraction of cluster size~1 and hence the maximum in the fraction of size~2, corresponds to the full depletion voltage. 
The peak of size~2 clusters is overall higher and appears at a lower voltage for the Micron n-on-p sensors because the depletion voltage is lower compared to the HPK n-on-p sensors and hence the electric fields are weaker, leading to more diffusion.
Above full depletion voltage, the electric field strength continues to increase, 
but now reduces the charge sharing due to shorter drift times and therefore lower lateral diffusion. 
 
For the Micron n-on-n sensors the trend differs
because the junction develops from the backplane of the sensor instead of the pixel implants. 
Below full depletion (40~V), the field is weak in the vicinity of the implants and thus the charge diffuses further, which enhances charge sharing and hence the fraction of size~2 clusters. 
The weak field region is always narrow enough, about 20\mum~\cite{dallocco2021temporal}, that the charge will diffuse towards the implants and thus be collected.
As the applied bias voltage increases, the field strengthens, the diffusion and charge sharing is suppressed and hence the fraction of size~2 clusters monotonically decreases. 

For the irradiated prototypes used in this study the bias voltage 
limit was set at 1000~V, due to unstable operation at higher values and
to prevent thermal runaway.
For irradiated sensors the measured charge is lower due to the changes in the electric field and charge trapping, the likelihood of shared charge reaching threshold is greatly reduced,
affecting cluster sizes and eventually also the efficiency.

\begin{figure}[ht]
  \centering
     \includegraphics[width=0.48\textwidth]{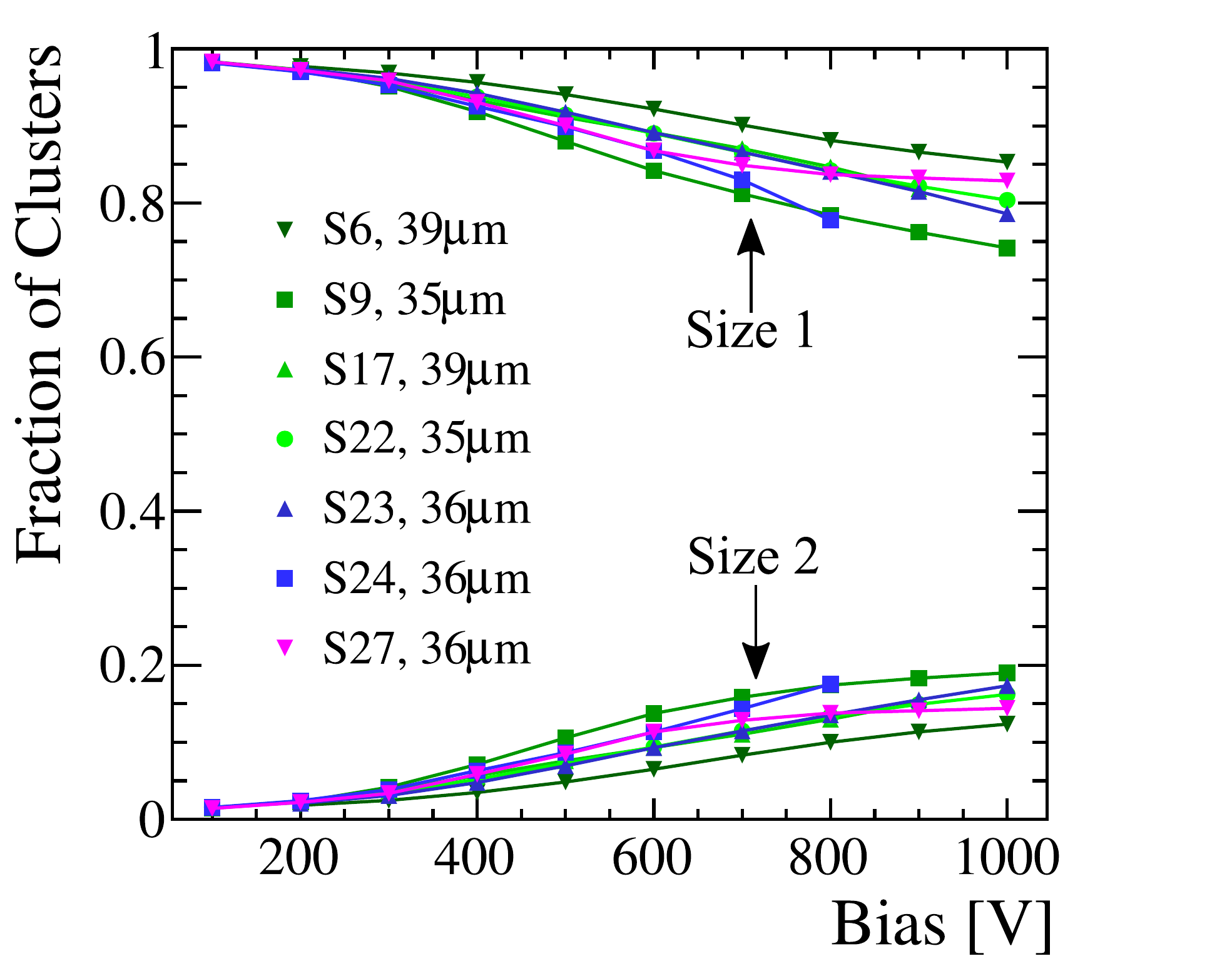}
    \caption[]{The fraction of size~1 and size~2 clusters for sensors uniformly irradiated to \maxfluence, placed perpendicular to the beam. Green for HPK n-on-p, blue for Micron n-on-p and purple for Micron n-on-n.}
  \label{fig:post_bias_clus}
\end{figure}

In \Cref{fig:post_bias_clus}, the fraction of size~1 and size~2 clusters as a function of applied bias voltage are compared for several prototype sensors placed perpendicular to the beam, uniformly irradiated to the full fluence. At bias voltages below 200~V, the charge is predominantly measured in one pixel. The effect of the irradiation is mitigated by increasing the bias voltage, partially regaining the fraction of size~2 clusters. These effects can be observed in the intrapixel track positions (see \Cref{sec:appendix2}). A minimum in the fraction of size~1 clusters is not observed due to the sensors not being fully depleted at 1000~V~\cite{Geertsema_2021, dallocco2021temporal}. 

\subsection{Track angle}
\begin{figure}[ht]
 
\centering
\begin{subfigure}{0.49\textwidth}
     \includegraphics[width=0.98\textwidth]{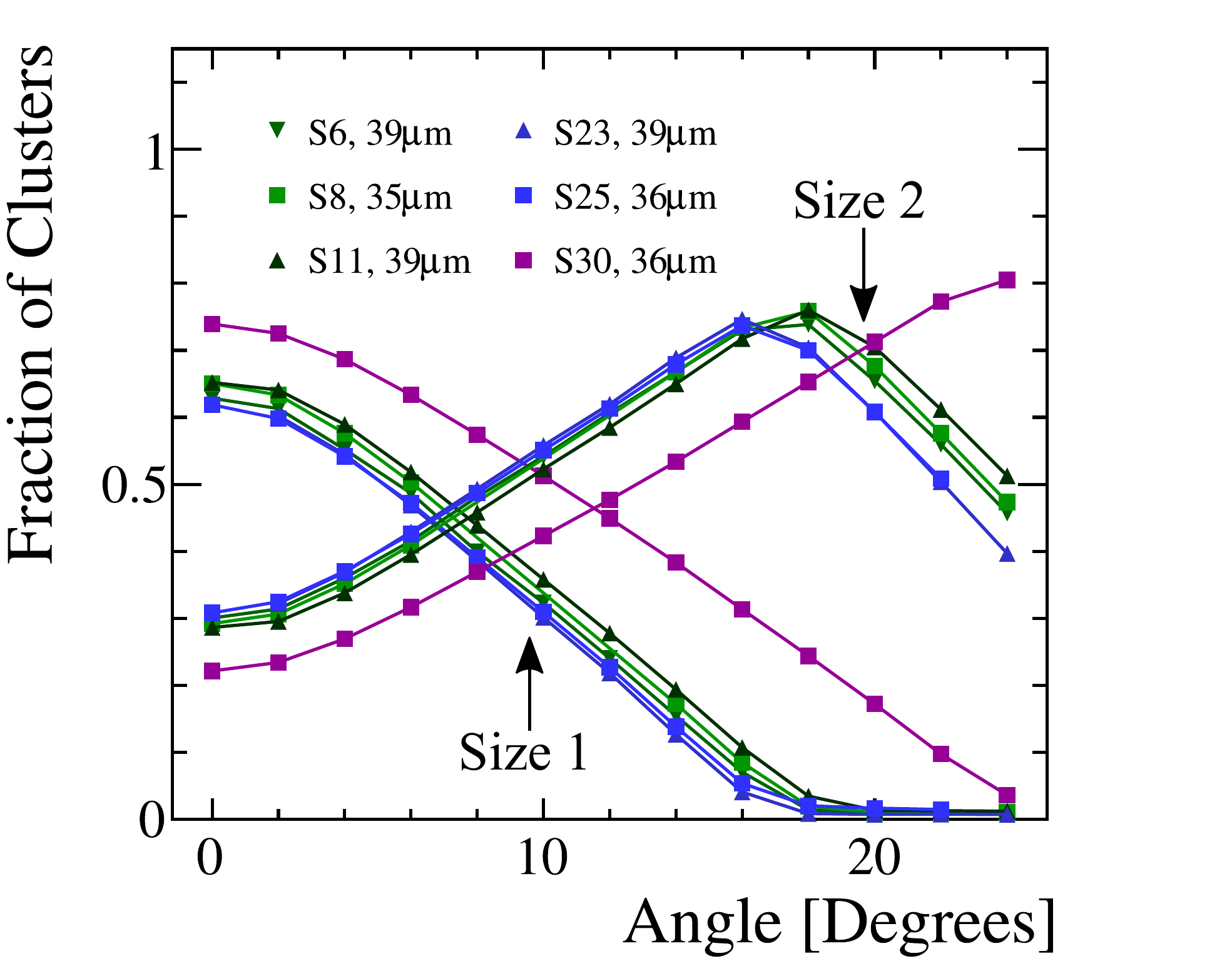}
     \caption{}
     \end{subfigure}
\begin{subfigure}{0.49\textwidth}
     \includegraphics[width=0.98\textwidth]{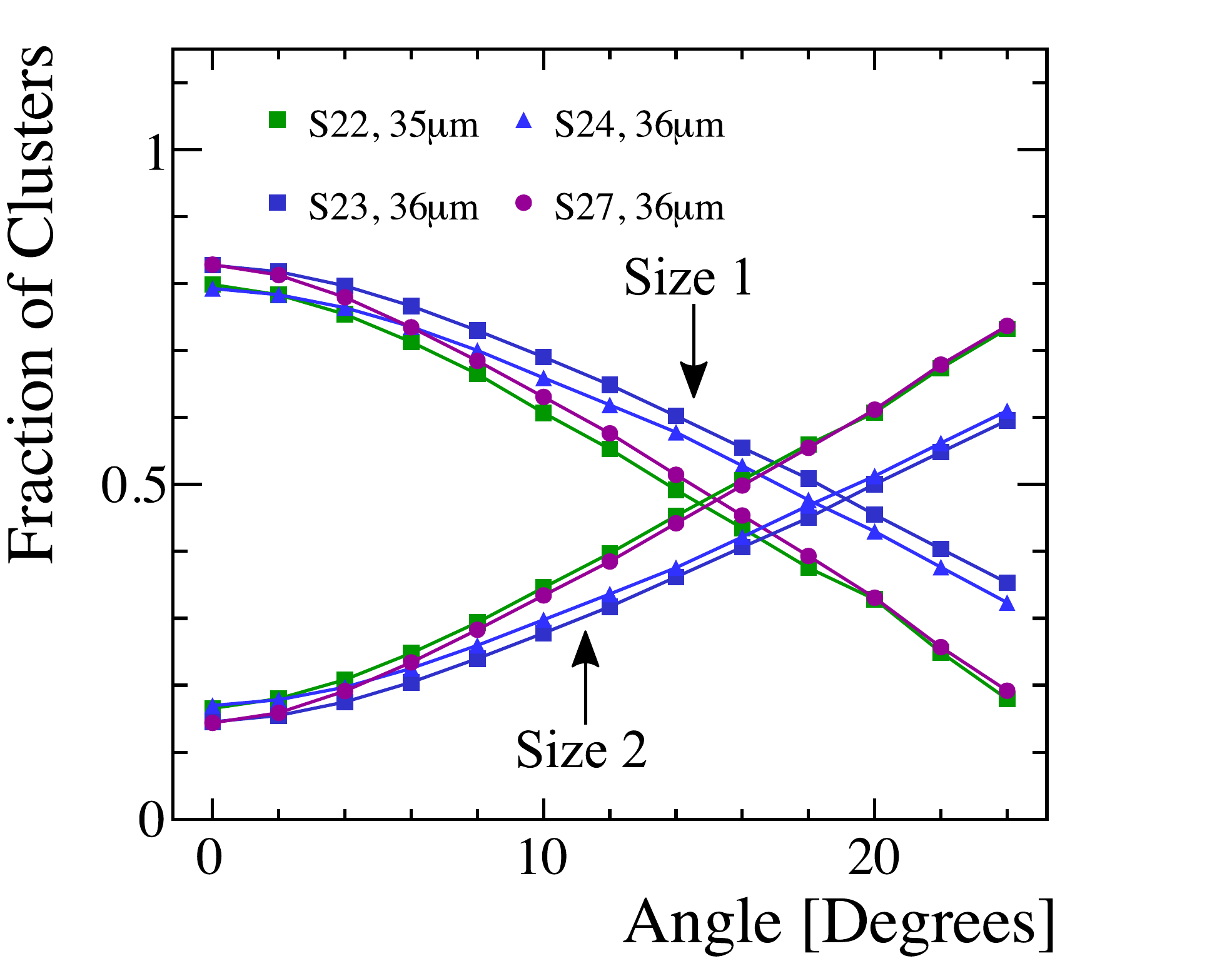}
     \caption{}
     \end{subfigure}
     \caption[]{The fraction of size~1 and size~2 clusters as a function of angle, (a) for non-irradiated and (b) for sensors irradiated uniformly to \maxfluence. Green for HPK n-on-p, blue for Micron n-on-p and purple for Micron n-on-n.}
  \label{fig:angle_clus_pre_clus}
\end{figure}

To study the sensor performance for angled tracks, angle scans were performed focusing on the range between approximately 0$^{\circ}$ and 24$^{\circ}$, in order to cover the LHCb acceptance~\cite{LHCb-TDR-013}. 
The fraction of size~1 and size~2 clusters as a function of angle 
is shown in \Cref{fig:angle_clus_pre_clus}.
For non-irradiated sensors (a), the largest fraction of 
size~2 clusters occurs at an angle of around 16$^{\circ}$
for the Micron n-on-p devices and around 18$^{\circ}$ 
for the HPK n-on-p devices, which are roughly compatible with the naive calculation of 15$^{\circ}$ (\Cref{sec:res_description}). 
The difference between the vendors is consistent with the slight variations in the thickness of the sensors~\cite{DallOccoThesis}. For the Micron n-on-n devices, 150\mum thick, 
the largest fraction is expected to be around 20\dg. However the fraction does not seem to decrease below angles of 24\dg which is the limit of the scan.
After irradiation (\Cref{fig:angle_clus_pre_clus} (b)), the curves show a
behaviour similar to that of a thinner sensor, because of the reduced effective depletion depth. The intrapixel track positions depending on cluster size for three different track angles are shown in \Cref{sec:appendix2}

\section{Efficiency results}
\label{sec:efficiency}
\begin{figure}[ht]
  \centering
    \begin{subfigure}{0.49\textwidth} 
    \includegraphics[width=0.90\textwidth]{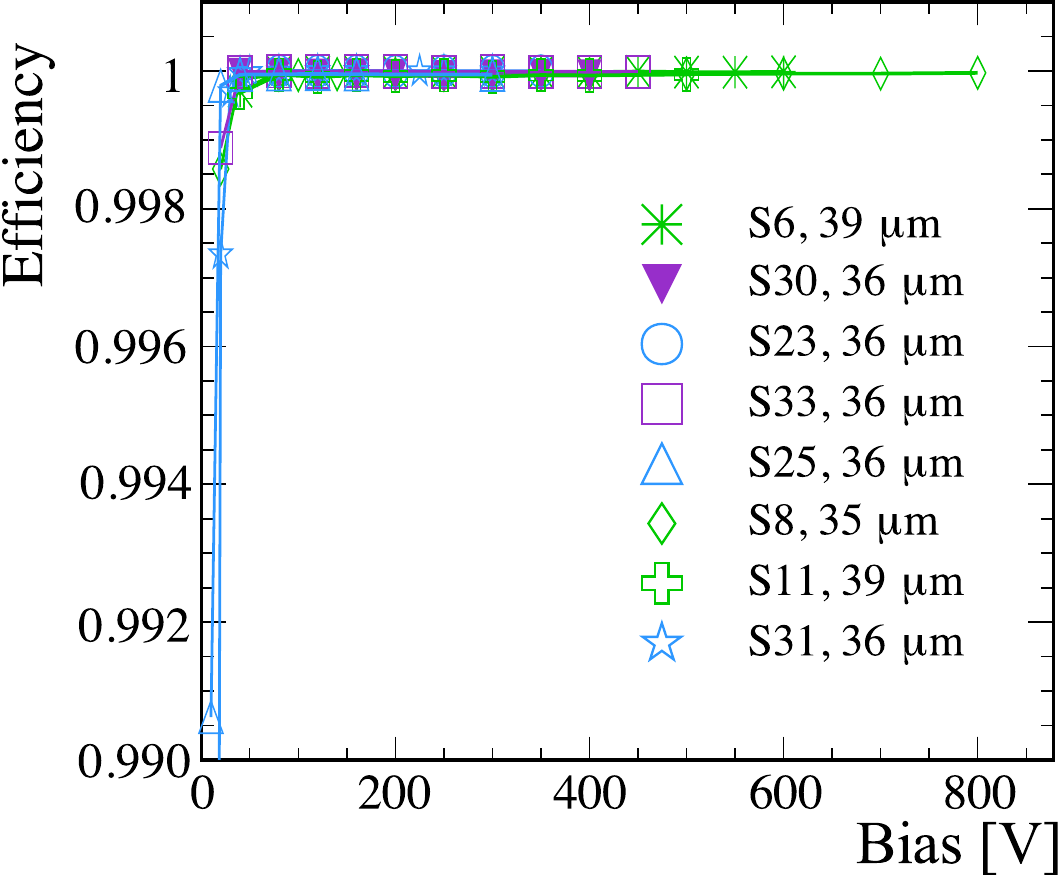}
    \caption{}
    \hfill
    \end{subfigure}
    \begin{subfigure}{0.49\textwidth} 
    \includegraphics[width=0.90\textwidth]{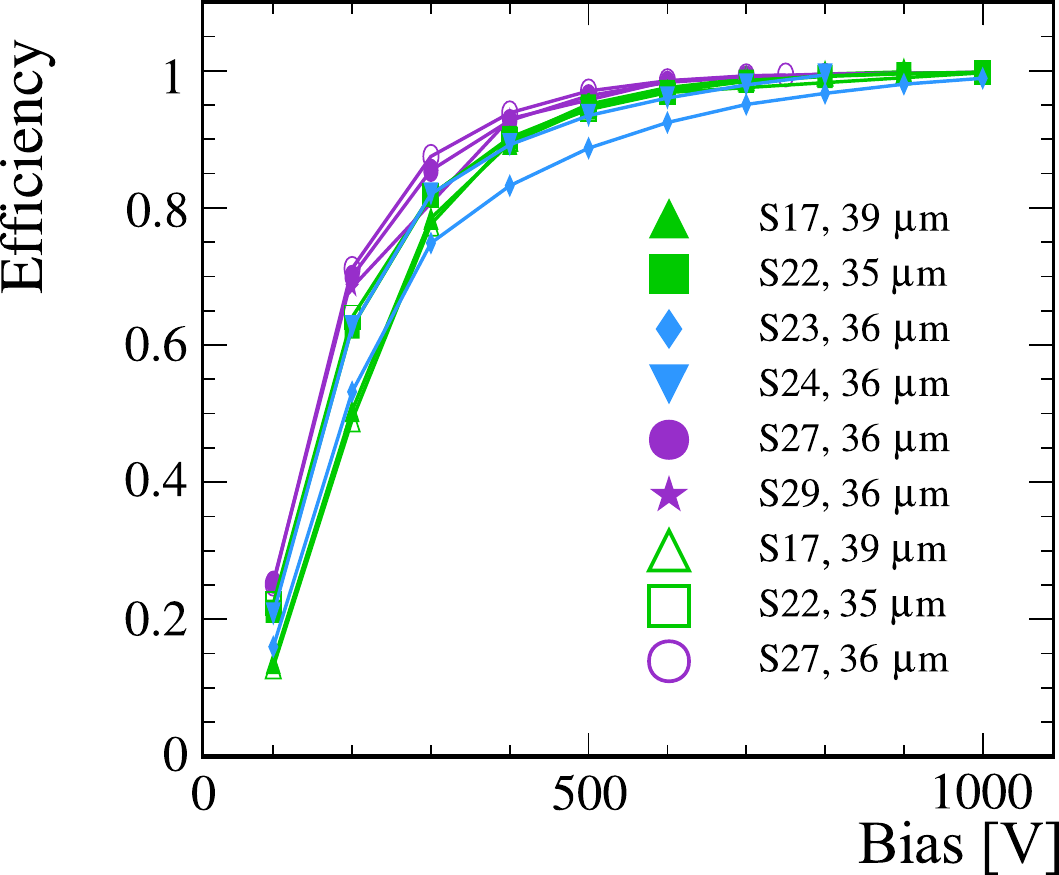}
    \caption{}
    \end{subfigure}
    \caption[]{Efficiency as a function of bias voltage, for non-irradiated sensors (a) and irradiated sensors (b). The colour of the markers indicates the vendor and sensor type, where green is for HPK n-on-p, blue is for Micron n-on-p and purple for Micron n-on-n.  }
  \label{fig:eff2}
\end{figure}

\begin{figure}[ht]
  \centering
      \begin{subfigure}{0.49\textwidth} 
    \includegraphics[width=0.99\textwidth]{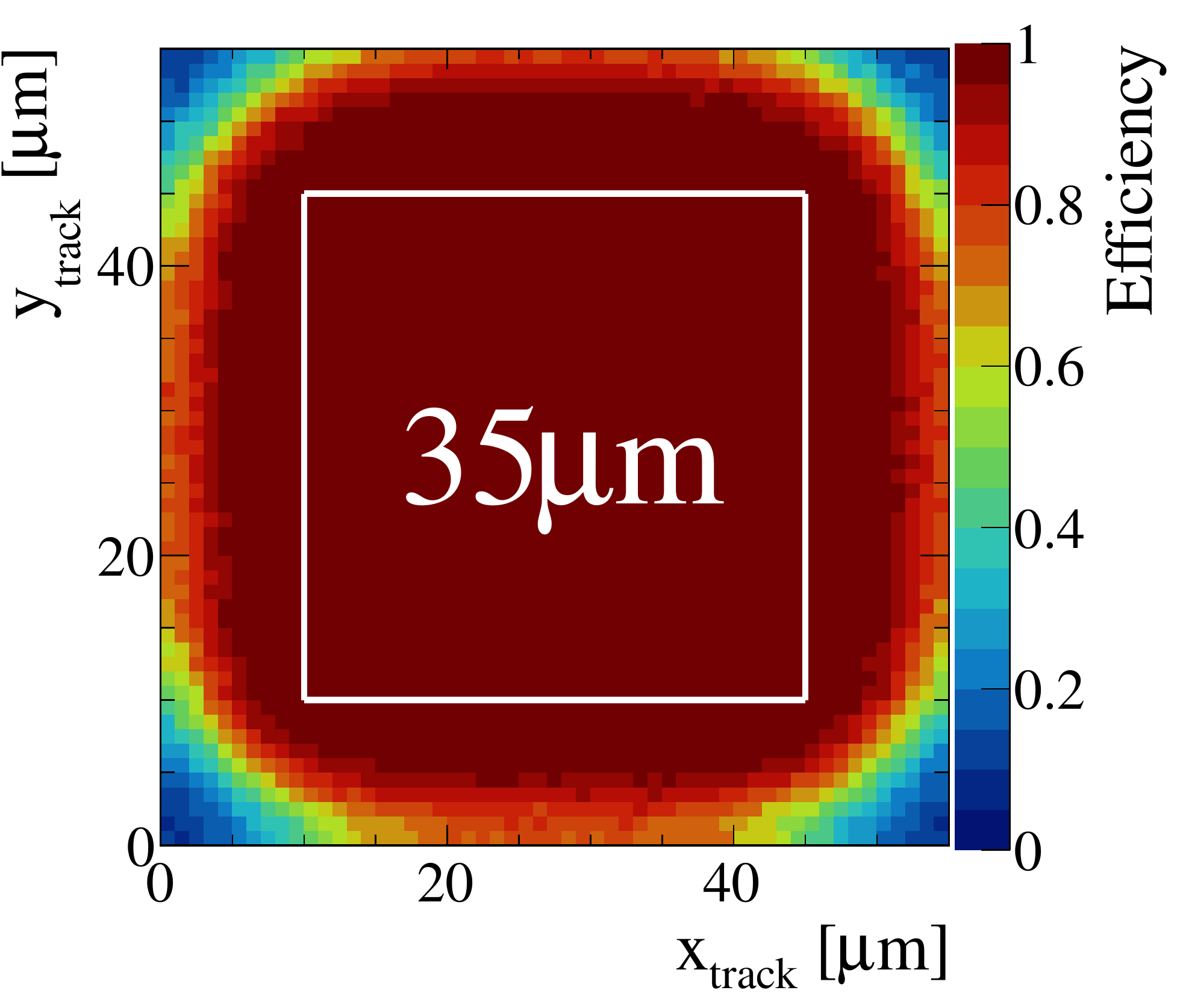}
    \caption{}
    \hfill
    \end{subfigure}
    \begin{subfigure}{0.49\textwidth} 
    \includegraphics[width=0.99\textwidth]{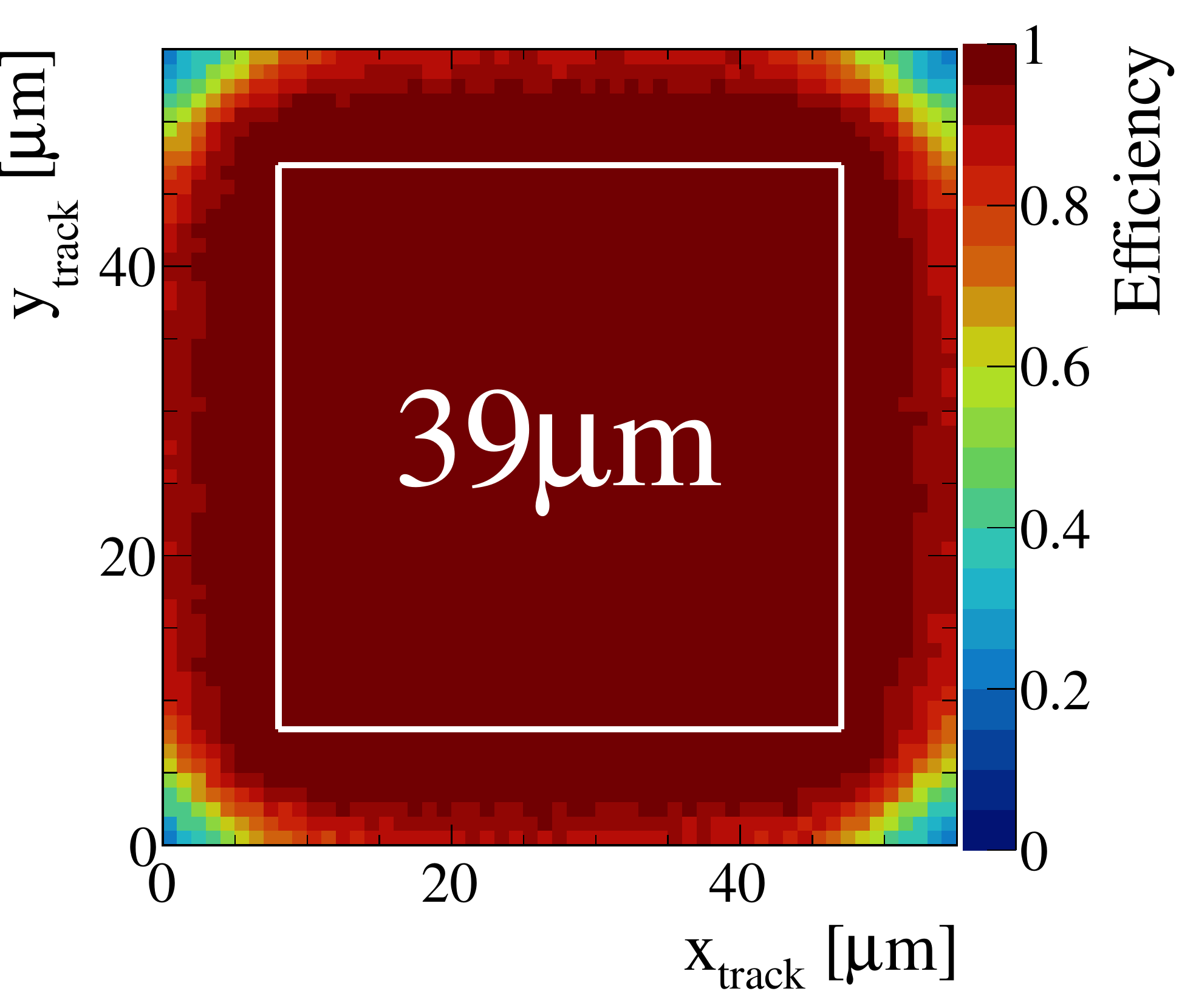}
    \caption{}
    \end{subfigure}
    \caption[]{Cluster finding efficiency as a function of the intrapixel position for two 200\mum thick HPK sensors operated at 300~V uniformly irradiated to the full fluence of \maxfluence. The two sensors differ in implant size: S22 has an implant size of 35\mum (a) and S17 has an implant size of 39\mum (b).}
  \label{fig:eff1}
\end{figure}

Non-irradiated sensors are found to be $\geq$99$\%$ efficient when the
applied bias voltage is above $\sim$20~V as shown in \Cref{fig:eff2}~(a)~\cite{Richards}. However the efficiency
degrades due to radiation damage. The efficiency as a function of applied bias voltage is  shown in 
\Cref{fig:eff2}~(b), for several sensors uniformly irradiated to a
fluence of $\maxfluence$. 
The efficiency is reduced to as low as 20\% at 100~V, but is recovered up to 99\% at bias voltages higher than 900~V due to increased charge collection
\cite{Geertsema_2021}. The efficiency loss is not uniform over the pixel cell, 
but higher at the corners, as can be seen in \Cref{fig:eff1}. After irradiation the charge collection is much lower. For size~3 and size~4 clusters where the charge is divided it becomes less likely that the signal will be sufficient to cross the threshold.  
\Cref{fig:eff1} compares the intrapixel efficiencies of two uniformly irradiated 200\mum thick HPK sensors with different implant sizes, operated at 300~V.  
The difference in implant sizes shows that the efficiency degradation in the corner is probably due to the lower electric field. Overall, the sensor with an implant size of 35\mum is 6\% less efficient than a sensor with an implant size of 39\mum when operated at 300~V. 

%






\section{Spatial resolution results} 
\label{sec:resolution}

The spatial resolution depends on operational parameters,
such as bias voltage and threshold, as well as the signal-to-noise ratio, the angle of incidence of the particle and the accumulated radiation damage. In this section the spatial resolution is presented as a function of the bias voltage, track angle and fluence.   

\subsection{Bias voltage}
\begin{figure}[ht]
  \centering
        \begin{subfigure}{0.49\textwidth} 
  \includegraphics[width=0.99\textwidth]{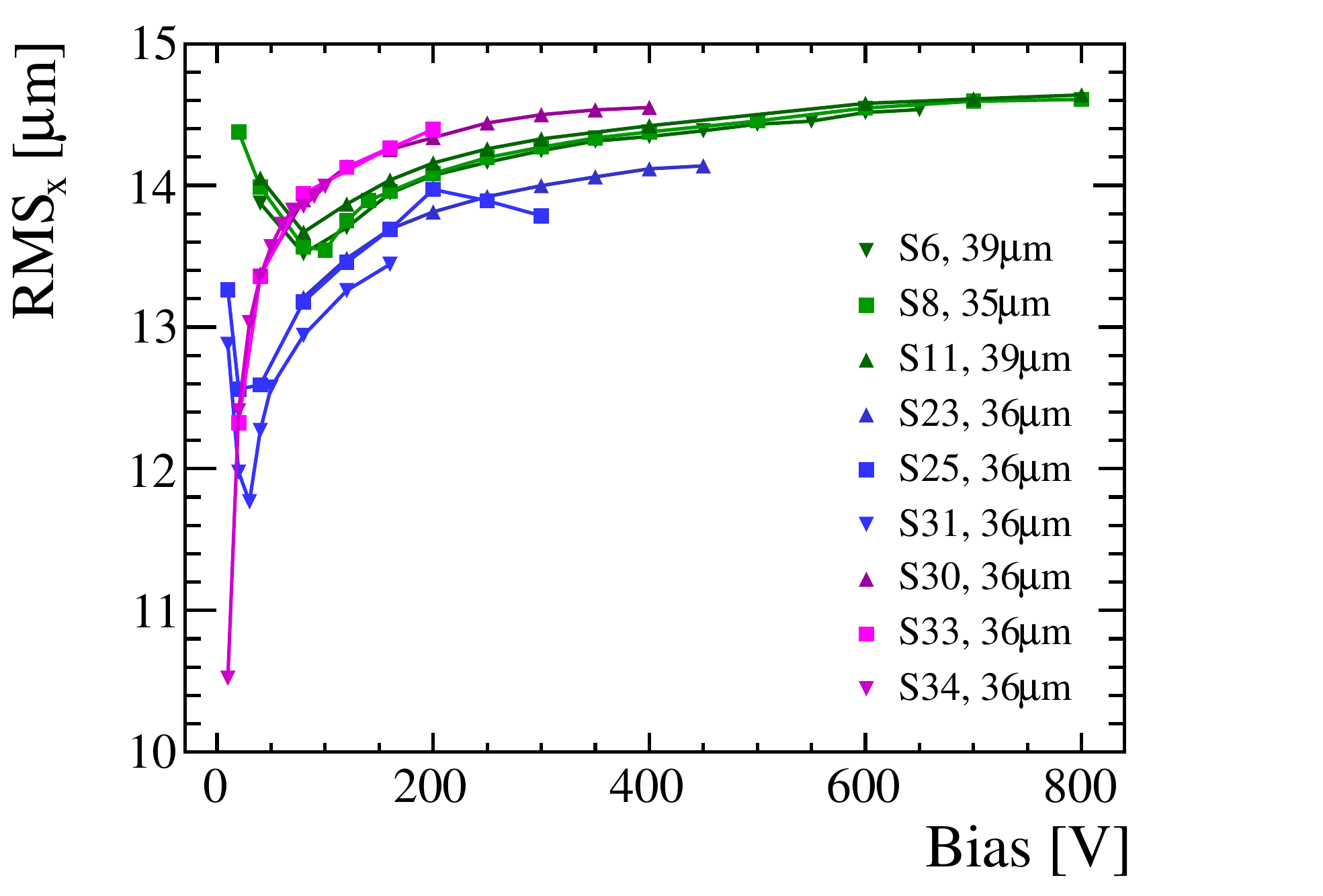}
    \caption{}
    \hfill
    \end{subfigure}
    \begin{subfigure}{0.49\textwidth} 
\includegraphics[width=0.99\textwidth]{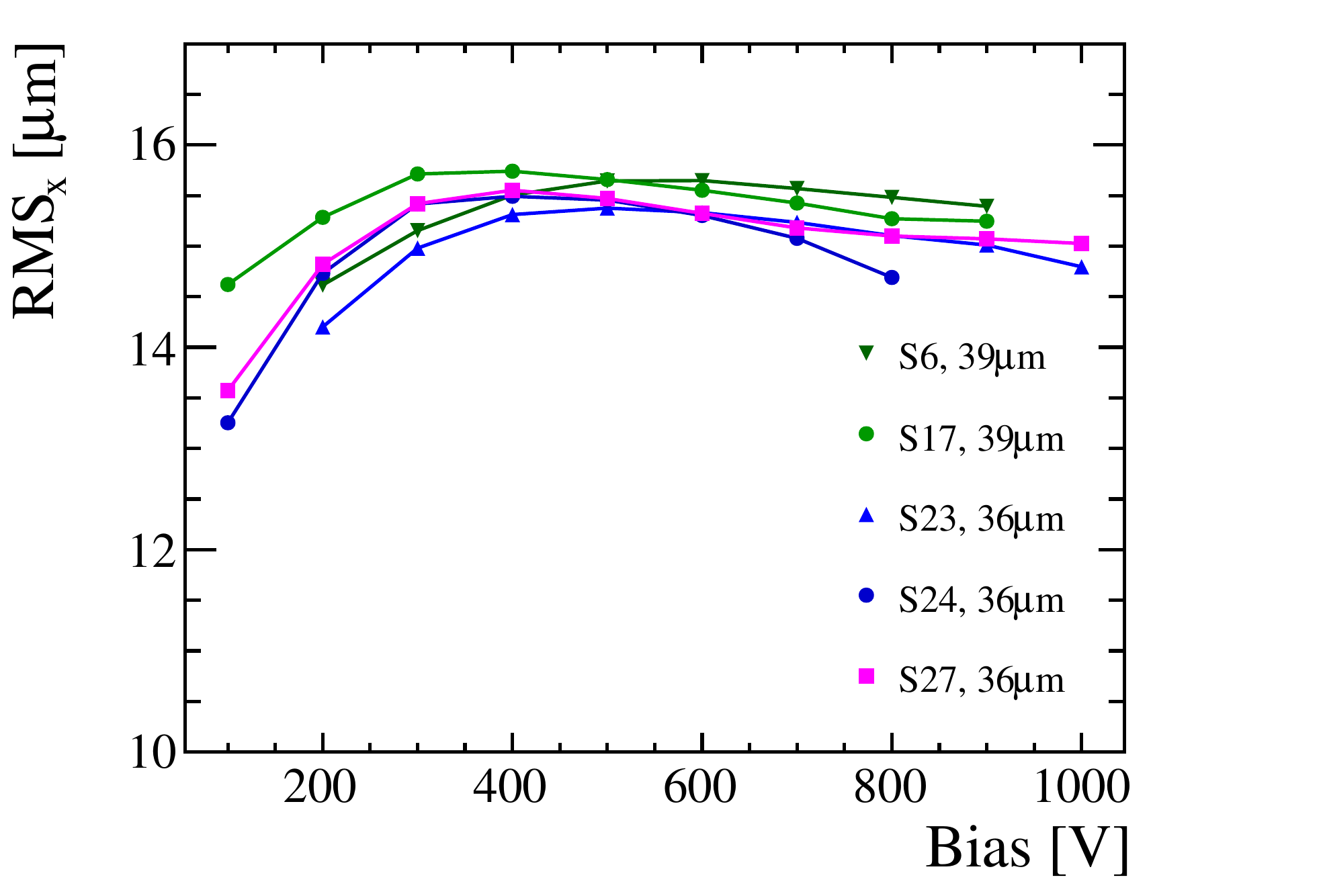}
    \caption{}
    \end{subfigure}

  \caption[]{The spatial resolution as a function of bias voltage for non-irradiated sensors (a) and for uniformly irradiated sensors to the \maxfluence (b). Green for HPK n-on-p, blue for Micron n-on-p and purple for Micron n-on-n.  }
  \label{fig:bias_res_all} 
\end{figure}

The spatial resolutions
are measured for all the prototype variants as
a function of different applied bias voltages for both
non-irradiated and irradiated sensors. 
For the non-irradiated sensors, the resolution in $x$ as a function of the bias voltage is shown in \Cref{fig:bias_res_all}~(a).
The spatial resolution is directly correlated to 
the cluster size distribution and therefore it follows similar
trends as seen in \Cref{fig:bias_frac_clus}. 
For the irradiated sensors, shown in \Cref{fig:bias_res_all}~(b), the worst resolution is observed between $\sim$300~V and 500~V depending on the sensor.  
At lower bias voltages, this is due to reduced efficiency especially at the corners, effectively decreasing the pixel area (\Cref{sec:efficiency}). 
Only tracks that pass through the centre of pixel are likely to lead to a measured signal, yielding a smaller residual difference for size~1 clusters thus artificially improving the spatial resolution. At higher bias voltages, the collected charge becomes larger, thereby recovering the efficiency in the corners but also increasing the cluster size. As described in \Cref{sec:bias_clus_sizes}, this improves the resolution. 




\subsection{Track angle}
\begin{figure}[ht]
  \centering
\begin{subfigure}{0.49\textwidth} 
 \includegraphics[width=0.99\textwidth]{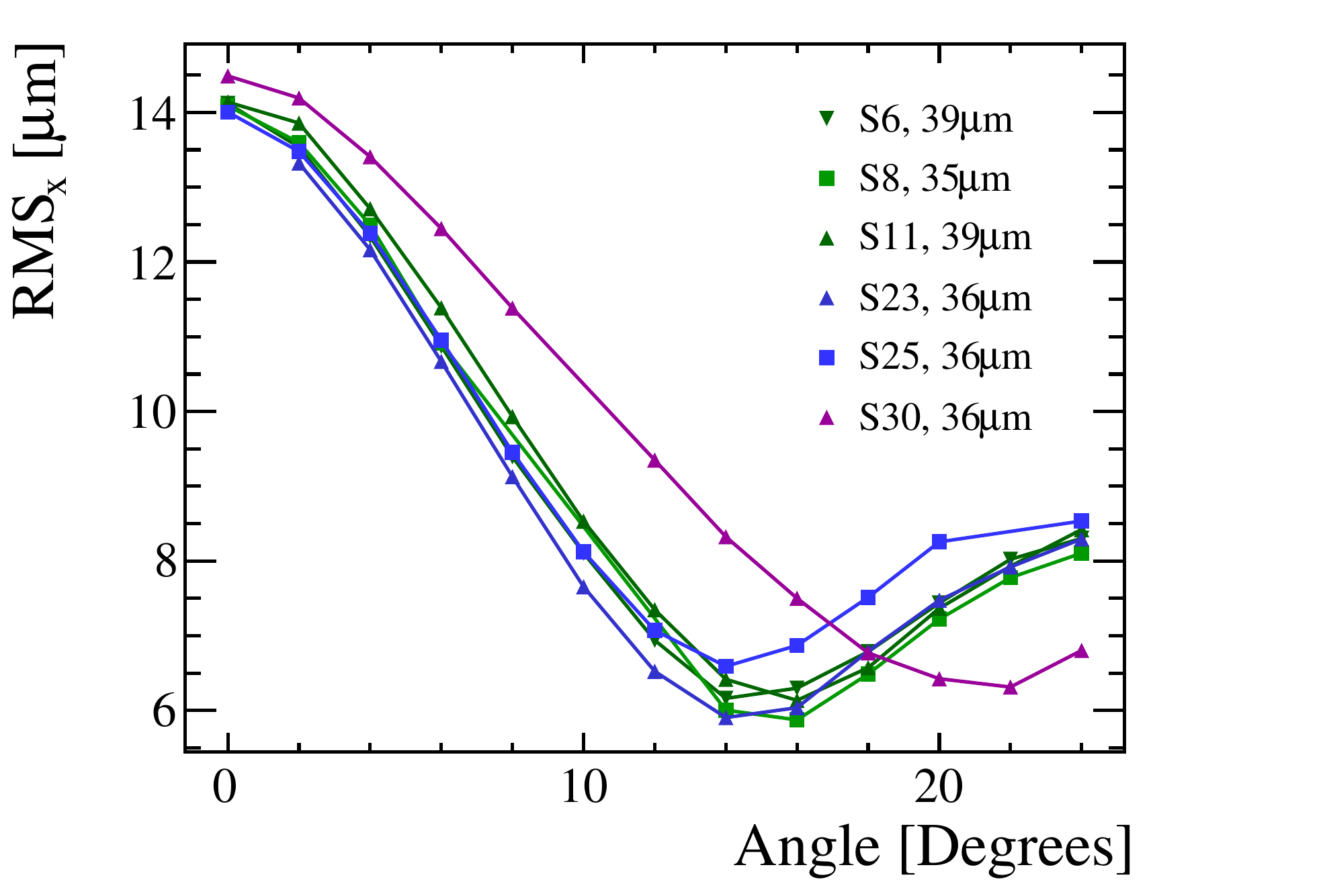}
    \caption{}
    \hfill
    \end{subfigure}
    \begin{subfigure}{0.49\textwidth} 
\includegraphics[width=0.99\textwidth] {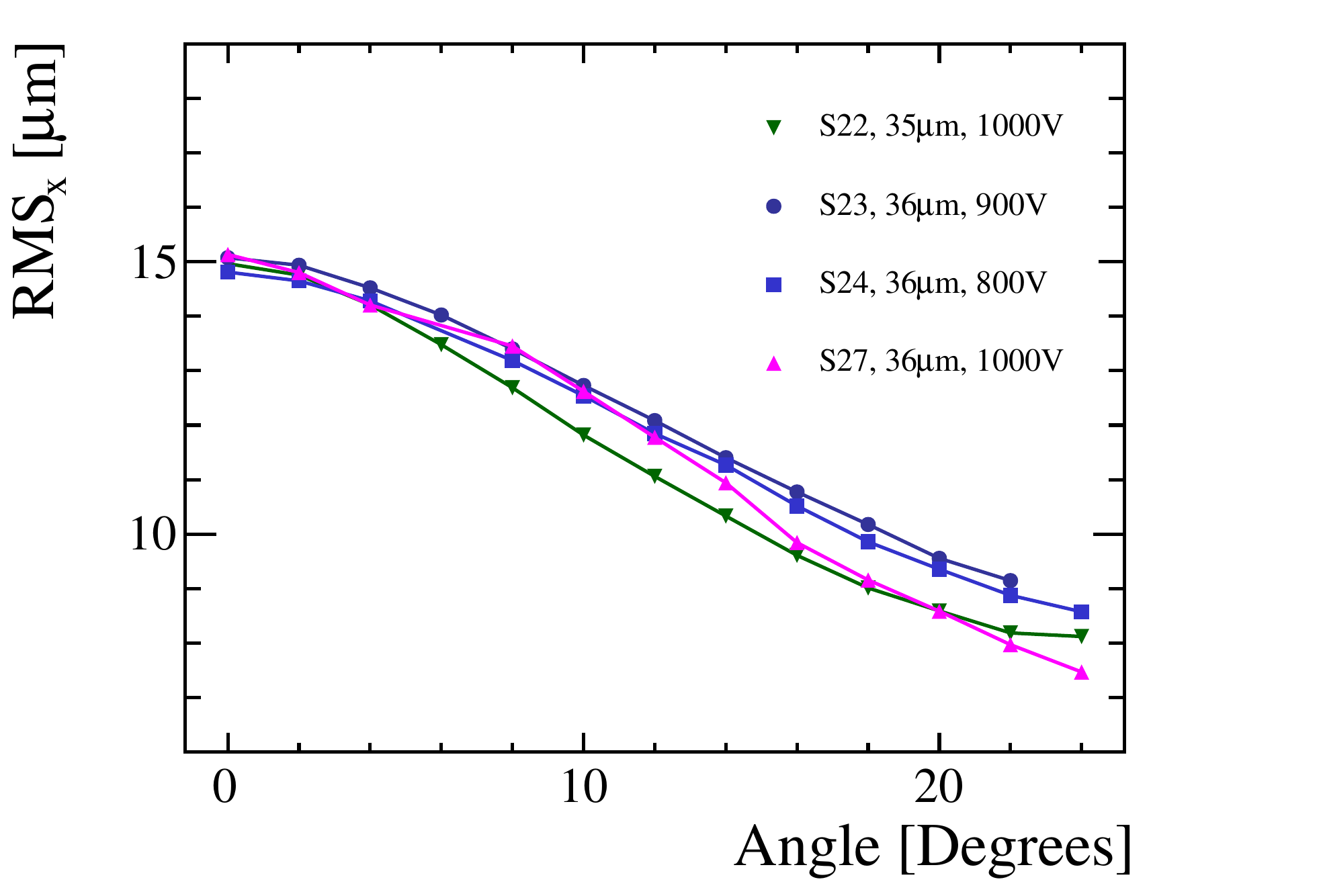}
    \caption{}
    \end{subfigure}

  \caption[]{The spatial resolution as a function of angle for non-irradiated sensors operated at 200~V (a) and uniformly irradiated sensors to \maxfluence (b). Green for HPK n-on-p, blue for Micron n-on-p and purple for Micron n-on-n.  }
  \label{fig:angle_res_all}
\end{figure}

The spatial resolution is also measured as a function of the 
incident track angle. The results for non-irradiated samples are shown in \Cref{fig:angle_res_all}~(a). All sensors show a similar trend which depends on the thickness, as predicted by \Cref{eq:bestres}. The optimum resolution is 6.5$\pm$0.5\mum at about an angle of 15\dg and 21\dg. The quoted uncertainty is from the variation in the measured resolution of the different prototype devices tested of the same type. 
As described in \Cref{sec:setup}, the Micron n-on-n prototypes
are thinner, and hence the best resolution occurs at a larger angle.
After irradiation the active sensor volume is shallower, due to the change in the effective doping concentration~\cite{dallocco2021temporal}. This leads to a less pronounced dependency as a function of 
the incident angle, even at the highest operational bias voltage, as presented in \Cref{fig:angle_res_all}~(b).  
In this plot it can also be seen that
the n-on-n prototype behaves quite similar to the thicker sensors, which
supports the conclusion that at these fluence levels the sensor thickness is less relevant to the total collected charge. 

\subsection{Combined bias and track angle}
\begin{figure}[ht]
  \centering
  \begin{subfigure}{0.49\textwidth} 
   \includegraphics[width=0.99\textwidth]{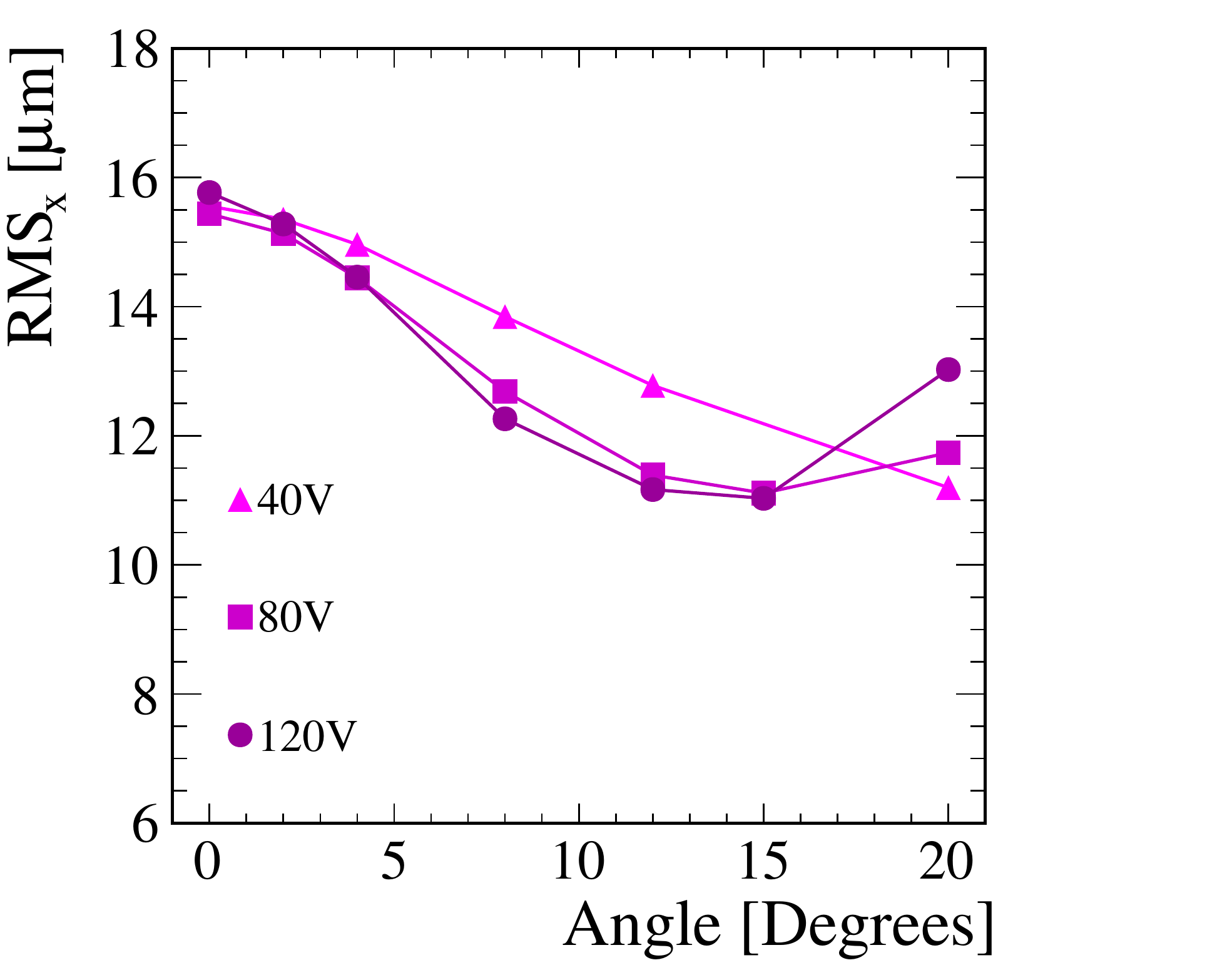}
    \caption{}
    \hfill
    \end{subfigure}
    \begin{subfigure}{0.49\textwidth} 
    \includegraphics[width=0.99\textwidth]{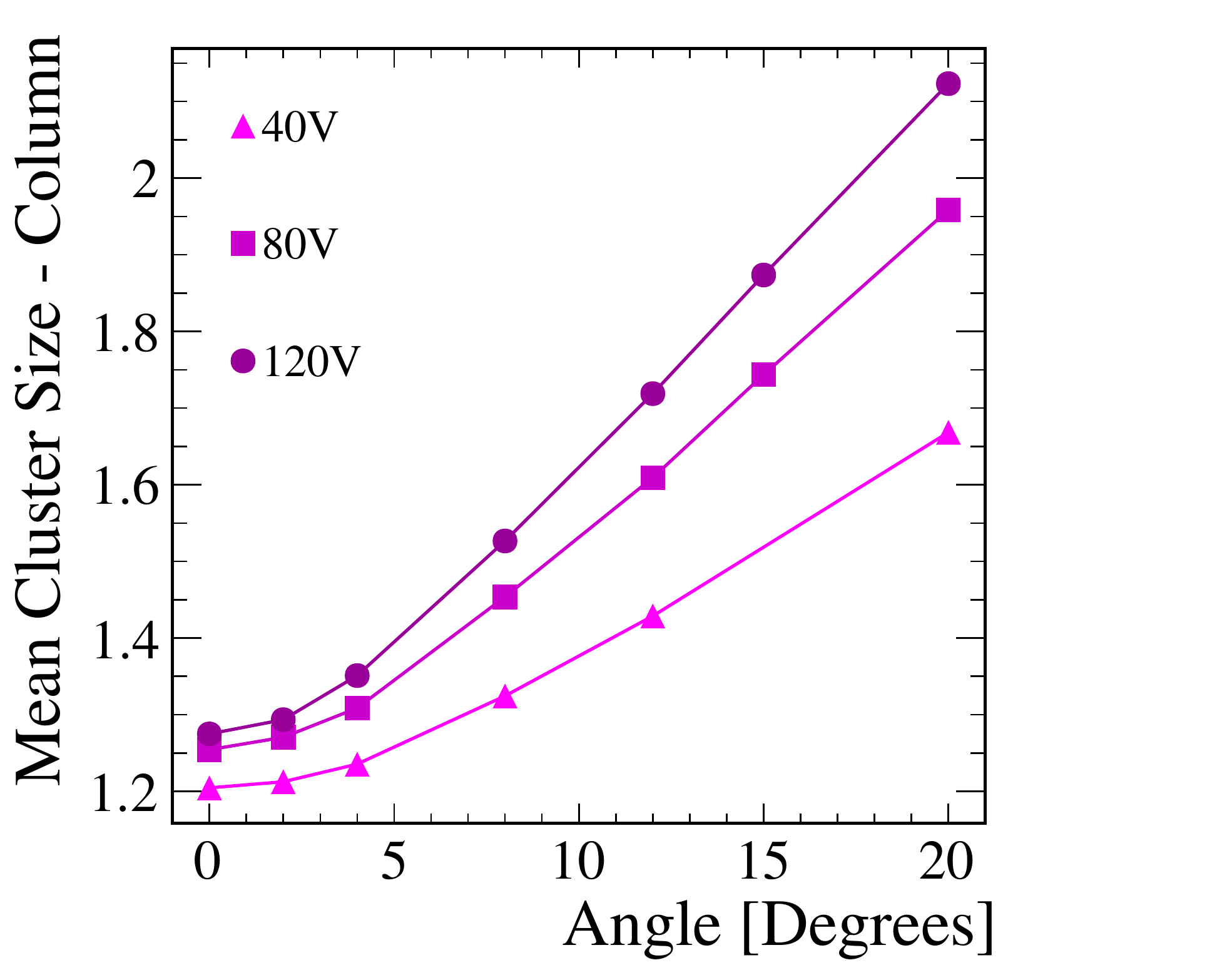}
    \caption{}
    \end{subfigure}
  
  \caption[]{Spatial resolution (a) and the mean cluster size (b) as a function angle, for three different bias voltages for an non-irradiated HPK sensor (S6).}
  \label{fig:bias_angle_res_all}
\end{figure}

In this section the cross dependence of bias voltage and angle of incidence is discussed. At bias voltages lower than the full depletion, the effective active region of the sensor is thinner. This leads to the effect shown in \Cref{fig:bias_angle_res_all}, where the spatial resolution is plotted as a function of angle for three different bias voltages for an non-irradiated sensor.  


\begin{figure}[ht]
  \centering
   \begin{subfigure}{0.49\textwidth} 
 \includegraphics[width=0.99\textwidth]{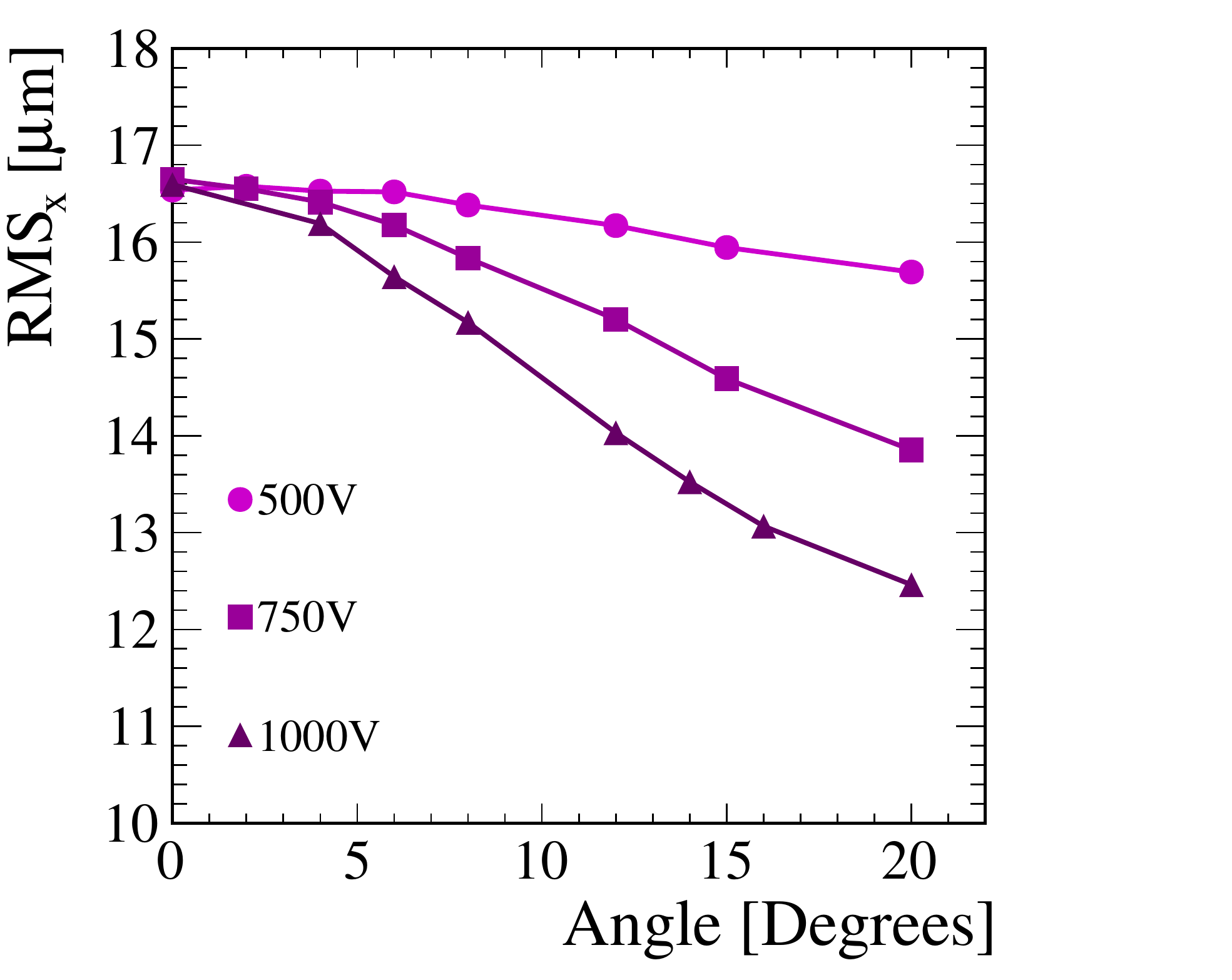}
    \caption{}
    \hfill
    \end{subfigure}
    \begin{subfigure}{0.49\textwidth} 
    \includegraphics[width=0.99\textwidth]{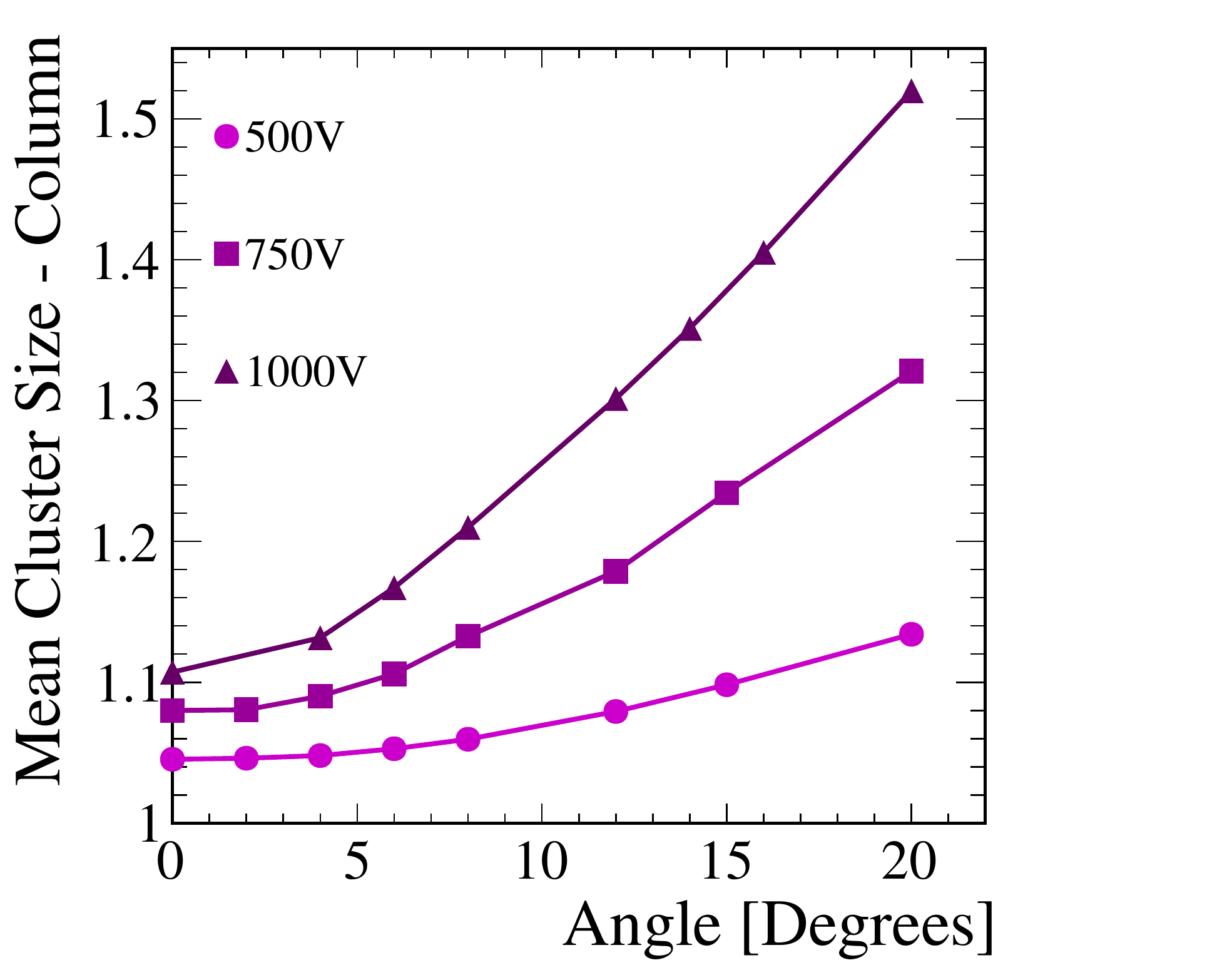}
    \caption{}
    \end{subfigure}

  \caption[]{Spatial resolution (a) and the mean cluster size (b) as a function angle, for three different bias voltages for a HPK sensor (S6) uniformly irradiated to the full fluence of \SI{8e15}{\mevneq.}. }
  \label{fig:post_bias_angle_res_all}
\end{figure}

For the irradiated sensors in~\Cref{fig:post_bias_angle_res_all} a similar trend is seen, 
with the caveat that even at the highest voltages the n-on-p
sensors are not fully depleted, and only a fraction of 
the charge is collected. 
At bias voltages smaller than the maximum operational value 
the cluster size is smaller since more often a fraction of the shared charge is smaller than the threshold, and 
the resolution versus angle continues to improve as the applied voltage increases.

\subsection{Non-uniform irradiation }
\begin{figure}[ht]
  \centering
     \begin{subfigure}{0.49\textwidth} 
    \includegraphics[width=0.99\textwidth]{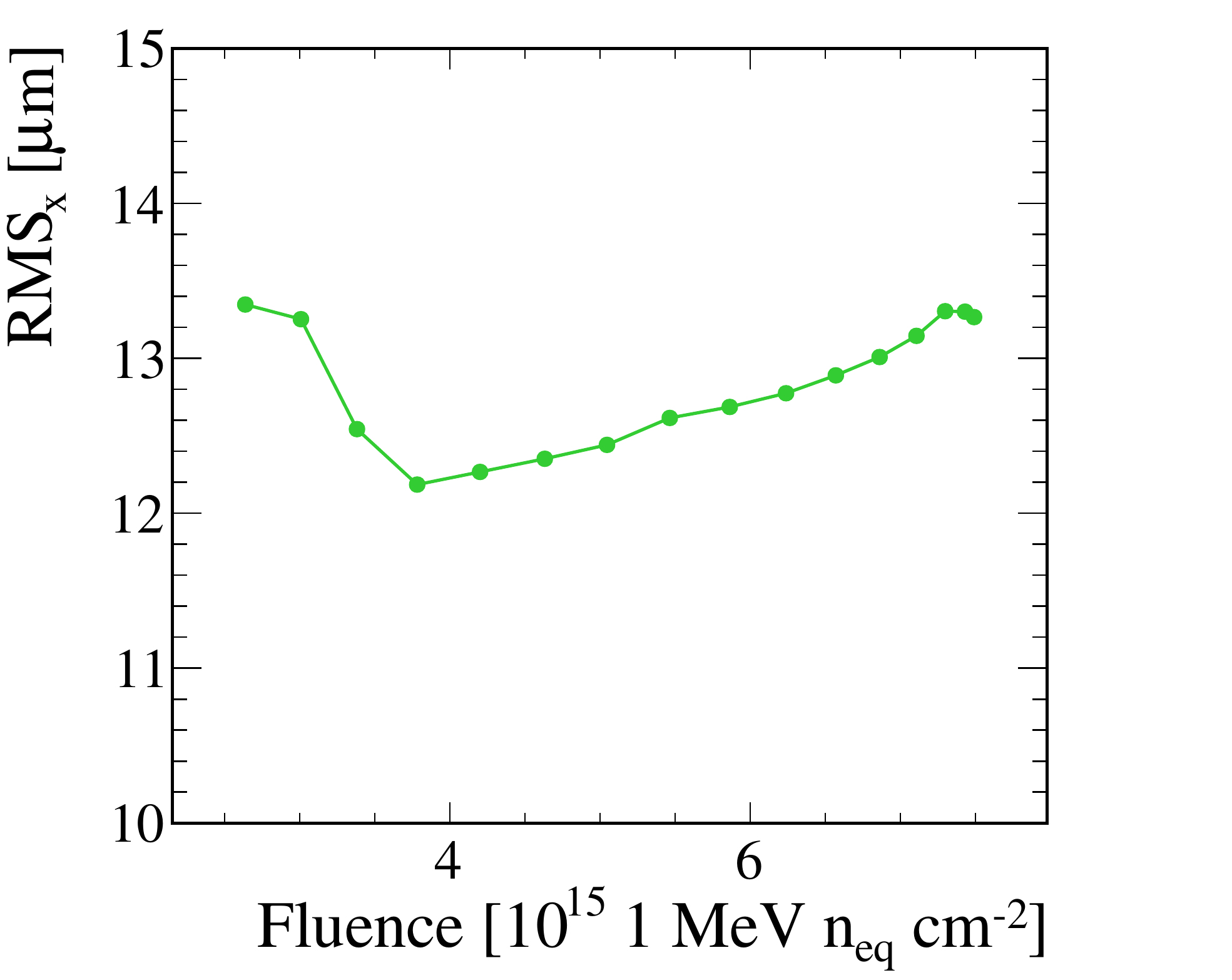}
    \caption{}
    \hfill
    \end{subfigure}
    \begin{subfigure}{0.49\textwidth} 
      \includegraphics[width=0.99\textwidth]{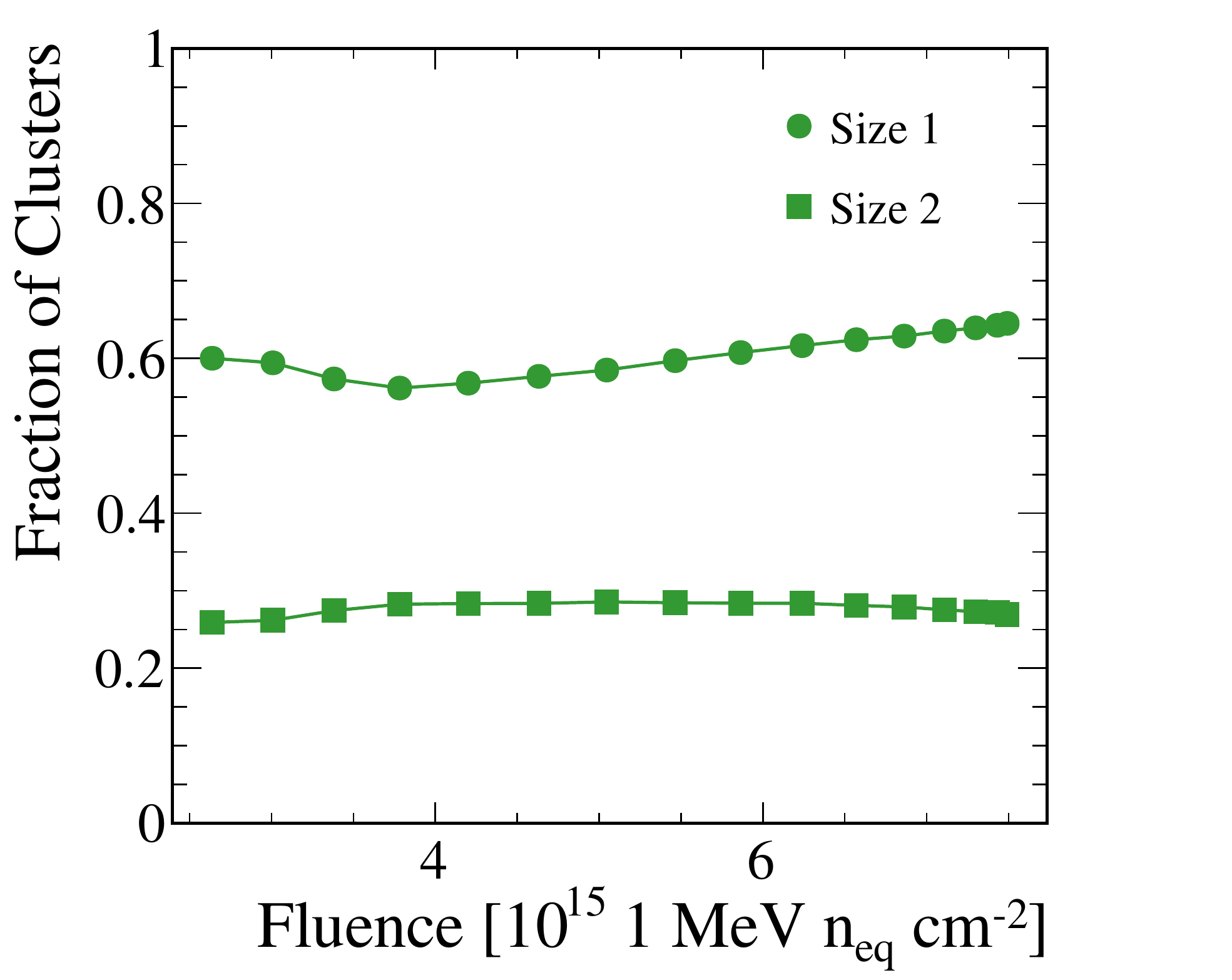}
    \caption{}
    \end{subfigure}
  
  \caption[]{The spatial resolution (a) and the fraction of clusters (b) as a function of fluence, for an HPK sensor operated at 1000~V placed perpendicular to the beam.}
  \label{fig:non_uniform}
\end{figure}

The spatial resolution and the mean cluster size as a function of the fluence are shown in \Cref{fig:non_uniform}~(a) and~(b), respectively, for an HPK sensor, placed perpendicular to the beam, operated at 1000~V and non-uniformly irradiated as described in~\Cref{sec:appendix3}.
For fluences below \SI{4e15}{\mevneq}, the spatial resolution degrades inversely proportional to the fluence. 
This is because the whole sensor is biased with a single voltage, which means that there are over-depleted regions, precisely where this degradation occurs.
At the over-depleted regions the charge sharing decreases as described in \Cref{sec:bias_clus_sizes}, leading to a decrease in cluster size.
With increasing fluence, the sensor becomes less over depleted,
leading to regions with lower electric and therefore increasing diffusion and consequently more charge sharing. 
Above \SI{4e15}{\mevneq}, the spatial resolution starts to degrade with increasing fluence and the cluster size decreases due to the irradiation effects, such as charge trapping.

\section{Conclusions and outlook}
\label{sec:conclusion}
In this paper the spatial performance of pixel sensor
prototypes bump-bonded to Timepix3 ASICs is presented
for sensors before and after irradiation. Irradiated sensors were exposed to either a uniform neutron flux or non-uniform proton irradiation. The maximum fluence was \maxfluence in each case. Sensors were provided by two vendors and the impact of the design differences on the
efficiency, cluster size and spatial resolution is assessed.

The non-irradiated sensor measurements indicate a very 
high efficiency even below full depletion voltage for all 
prototype variants.
The best spatial resolution obtained is $6.5\pm0.5$\mum at about an angle of 15\dg and 21\dg for the 200\mum (n-on-p) and 150\mum (n-on-n)
thick prototypes, respectively. 

For n-on-p sensors, the best resolution for perpendicular tracks 
is achieved at the depletion voltage.
For the n-on-n prototypes, the best resolution is below the depletion since the sensor depletes from the opposite side to the pixel implants, as discussed in \Cref{sec:bias_clus_section}.

Measurements of the irradiated sensors show a clear
decline in efficiency and resolution, which can be mitigated by operating at higher bias voltages. At 
the fluence of \maxfluence, which is the expected top
value after a total delivered luminosity of \SI{50}{\invfb} at the innermost region of the LHCb VELO detector, all sensor variants remain operational 
and exceed the minimum efficiency requirement for the LHCb Upgrade. As shown in \Cref{fig:eff2}, higher efficiencies are obtained for sensors with a larger implant width (39\mum HPK) and at a lower bias voltage values 
when compared directly to sensors with smaller implant widths (36\mum Micron or 35\mum HPK).

The results presented in this paper validate the sensor candidates as
suitable for the application in the LHCb VELO upgrade. Amongst other performance advantages~\cite{dallocco2021temporal, Geertsema_2021}, the HPK sensor design with a 39\mum implant width was chosen for the final detector due to its higher efficiency at lower operational bias voltages. 

\section*{Acknowledgments}

We would like to express our gratitude to our  colleagues in the CERN accelerator departments for the excellent performance of the beam in the SPS North Area.
We would like to acknowledge Jan Buytaert, Wiktor Byczynski and Raphael Dumps for their extensive and continuous support to keep the telescope operational. We would also like to thank all people that took part in the  data taking effort throughout the years of 2014 to 2016.
We gratefully acknowledge the financial support from CERN and from the national agencies: CAPES, CNPq, FAPERJ (Brazil); the Netherlands Organisation for Scientific Research (NWO); The Royal Society and the Science and Technology Facilities Council (U.K.). This project has received funding from the European Union’s Horizon 2020 Research and Innovation programme under Grant Agreement no. 654168.


\clearpage
\begin{appendices}
\crefalias{section}{appendix}
\section{List of assemblies}
\label{sec:appendix1}
The details of the sensors that are tested during this analysis are listed here. Two different types of substrates (n-on-p and n-on-n) are used from two different vendors. Further differences between individual assemblies are indicated in the table below.

\begin{table}[!h]
\centering
    \begin{tabular}{  c c c c c c c c }
      \hline
       Sensor & Vendor & Type &Thickness   & Implant  & Guard Ring   & Irradiation &  Fluence  \\ 
                   &              &          &	[$\muup$m] &  [$\muup$m] & [$\muup$m] &   Profile              &   $\mathrm{1MeVn_{eq}cm^{-2}}$ \\  
      	\hline
      	\hline
 		
 		S6 & HPK & n-on-p & 200 & 39 & 450 & uniform n & $8\times10^{15}$ \\ 	
 	    S8 & HPK & n-on-p & 200 & 35 & 450 & non-uniform p & $8\times10^{15}$ \\
 	    S9 & HPK & n-on-p & 200 & 35 & 600 & uniform n & $8\times10^{15}$ \\
	    S11 & HPK & n-on-p & 200 & 39 & 450 & non-uniform p & $8\times10^{15}$ \\
	    S17 & HPK & n-on-p & 200 & 39 & 450 & uniform n & $8\times10^{15}$ \\ 	
	    S22 & HPK & n-on-p & 200 & 35 & 450 & uniform n & $8\times10^{15}$ \\ 		
	  \hline	
	  	S23 & Micron & n-on-p & 200 & 36 & 450 & uniform n & $8\times10^{15}$ \\ 
	  	S24 & Micron & n-on-p & 200 & 36 & 450 & uniform n & $8\times10^{15}$ \\ 
        S25 & Micron & n-on-p & 200 & 36 & 450 & non-uniform p & $8\times10^{15}$  \\ 
        S31 & Micron & n-on-p & 200 &35 &250 & non-irradiated & - \\
       
        \hline
        S27	& Micron & n-on-n & 150 & 36 & 450 & uniform n & $8\times10^{15}$  \\ 
        S29 & Micron & n-on-n & 150 & 36 & 450 & uniform n & -  \\ 
		S30 & Micron & n-on-n & 150 & 36 &450 & non-uniform p & $8\times10^{15}$ \\
		S33 & Micron & n-on-n & 150 & 36 &250 & non-irradiated & - \\
		S34 & Micron & n-on-n & 150 & 36 &250 & non-irradiated  & -\\
 
 \hline
   \end{tabular}
   
    \caption{Table of the prototype sensors tested. For the irradiated profiles the n denotes neutron irradiated and p denotes proton irradiated.  }
    \label{tab:prot_tab} 
\end{table}

\section{Non-uniform irradiation profiles}
\label{sec:appendix3}
\begin{figure}[ht]
 \centering
     \includegraphics[width=0.49\textwidth]{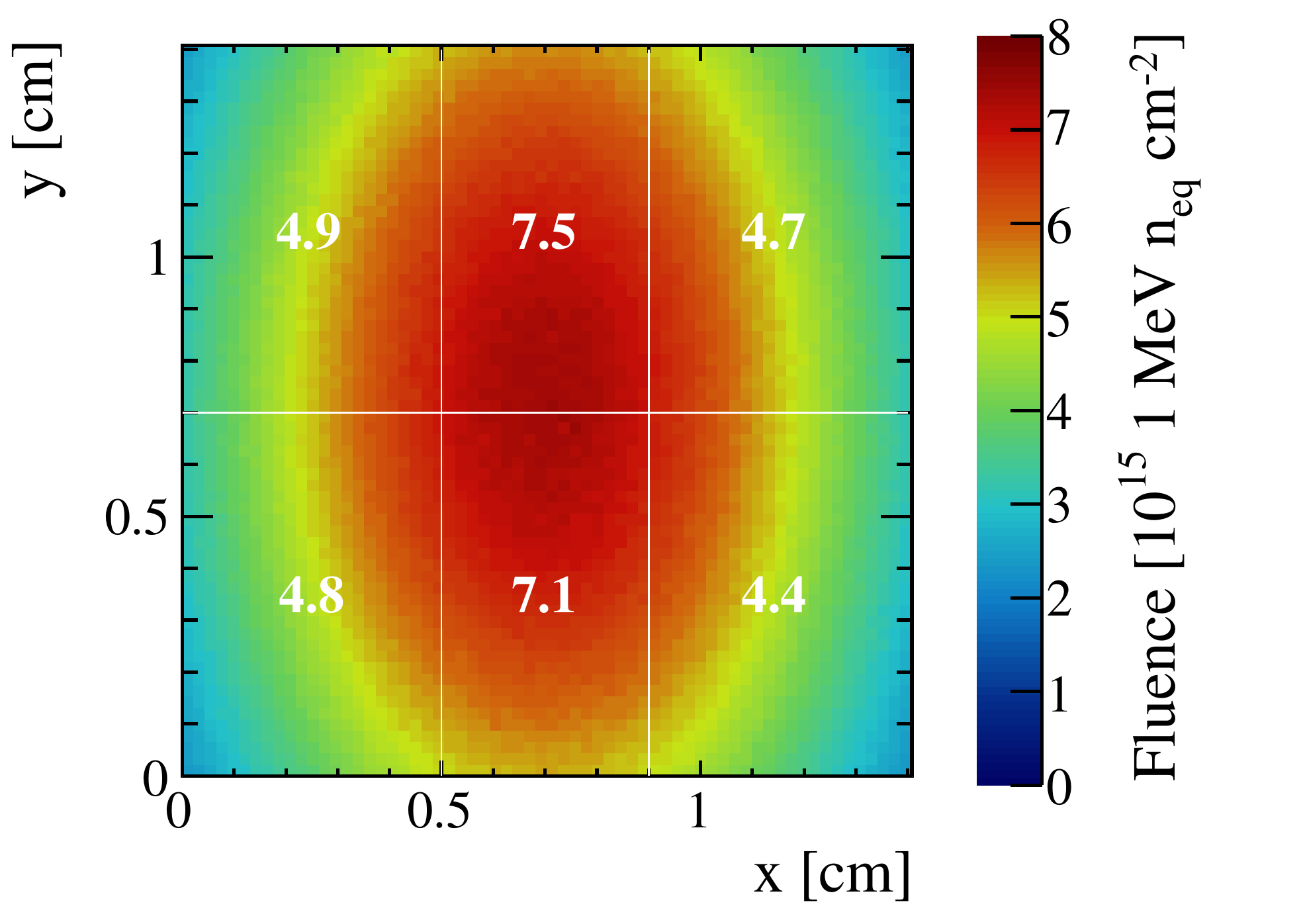}
    
    \caption[]{The reconstructed fluence profile from irradiation at IRRAD. The dosimetry results, converted to neutron equivalent fluence, for the six regions are indicated by the white numbers ($\times$\fluence).}
  \label{fig:IRRAD_Fluence}
\end{figure}

The IRRAD facility at CERN~\cite{Gkotse_2237333} provides a 24~\gev proton beam that is approximately distributed as a two-dimensional Gaussian, hence yielding a nonuniform fluence profile. After irradiation, the residual activation of the assemblies is obtained by measuring the distribution of hits in the sensor when not exposed to the beam.
The hits are caused by the radioactivity induced in the assembly and thus their rate is proportional to the fluence. 
The radiation profile can then be  determined from the activation map, modelled using a two-dimensional Gaussian distribution, further described in Ref.~\cite{DallOccoThesis}. The reconstructed fluence profile of a sensor is shown in \Cref{fig:IRRAD_Fluence}. 

\section{Intrapixel studies}
\label{sec:appendix2}
\begin{figure}[ht]
\centering
  \includegraphics[width=0.32\textwidth]{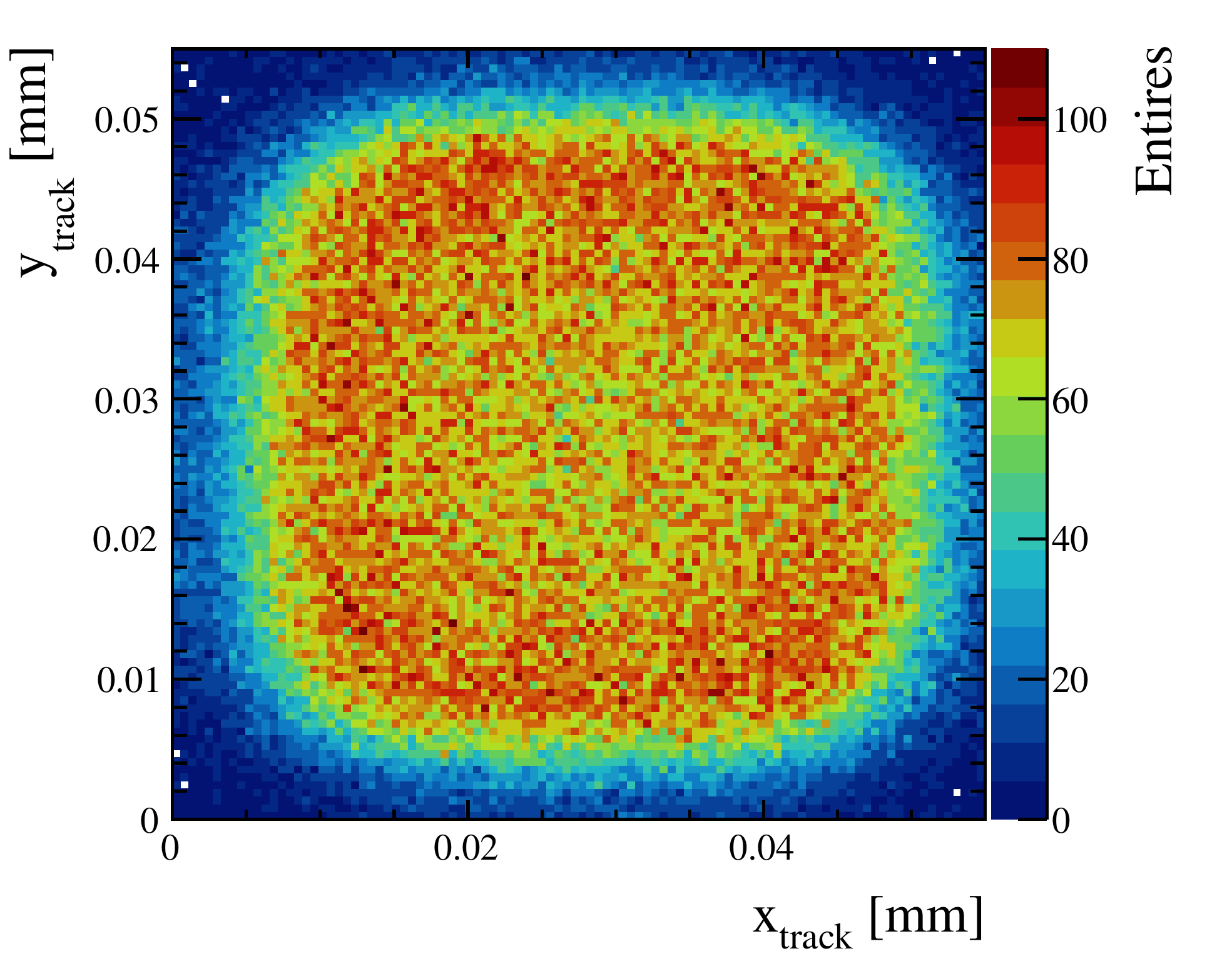}
  \includegraphics[width=0.32\textwidth]{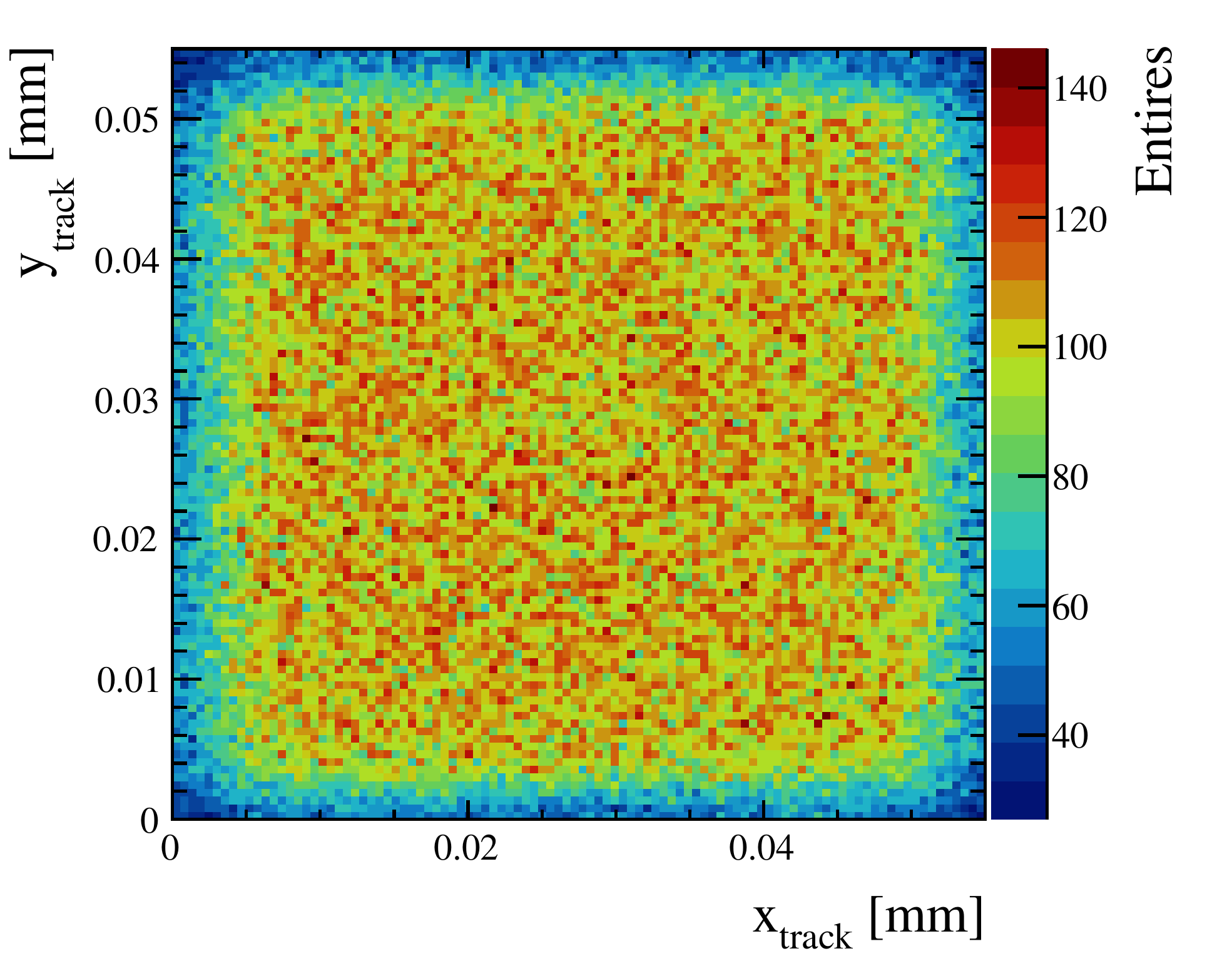}
  \includegraphics[width=0.32\textwidth]{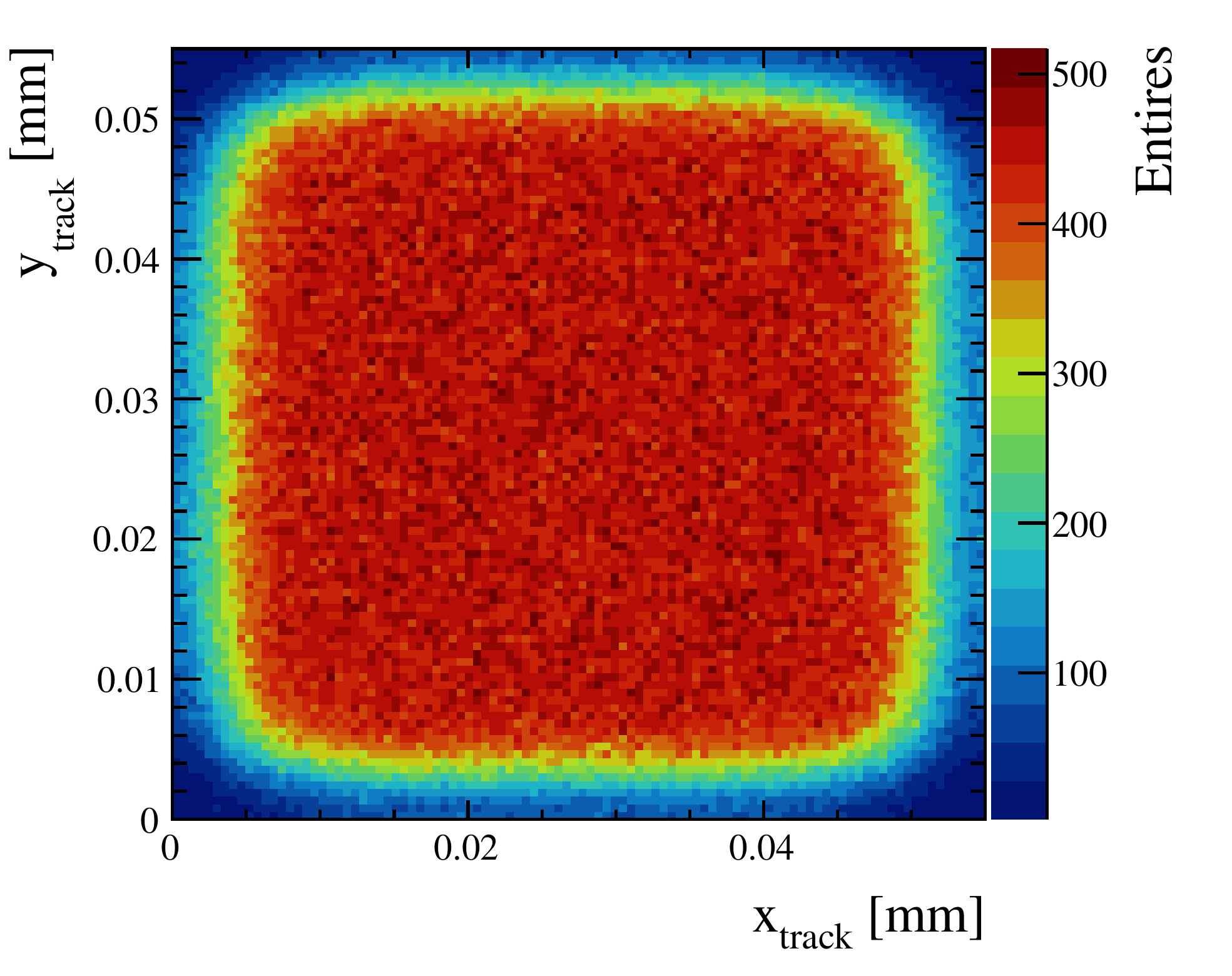}
  \includegraphics[width=0.32\textwidth]{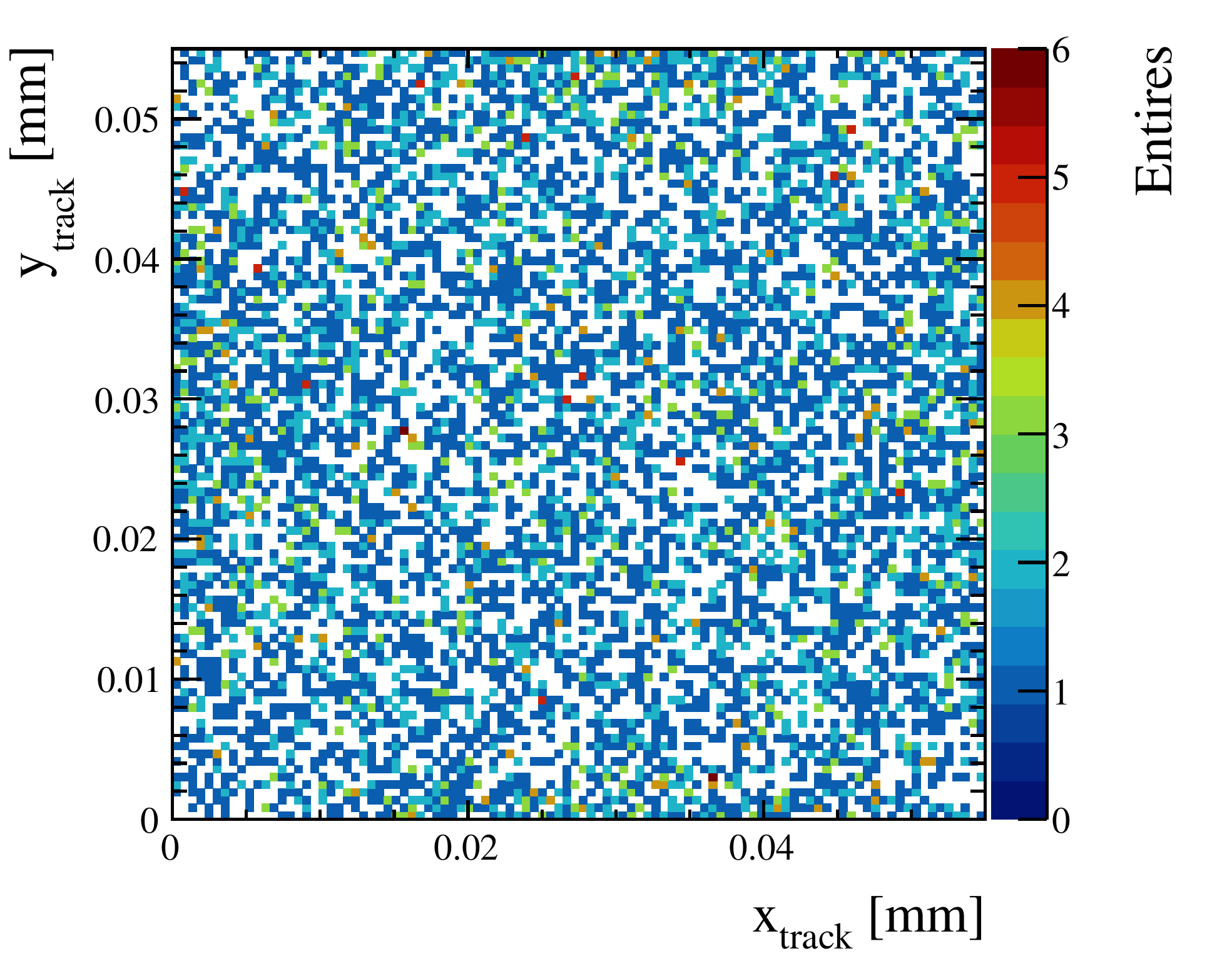}
  \includegraphics[width=0.32\textwidth]{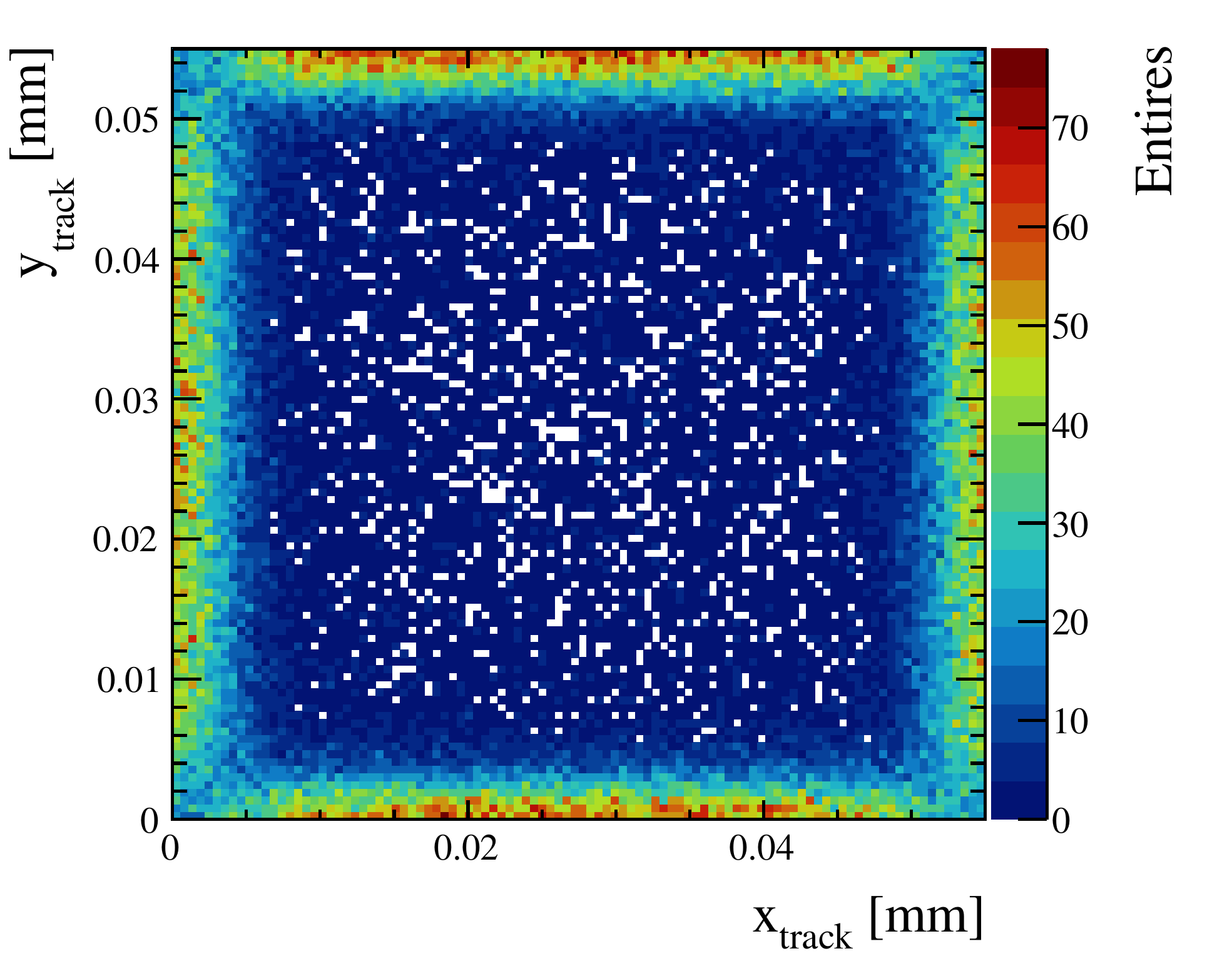}
  \includegraphics[width=0.32\textwidth]{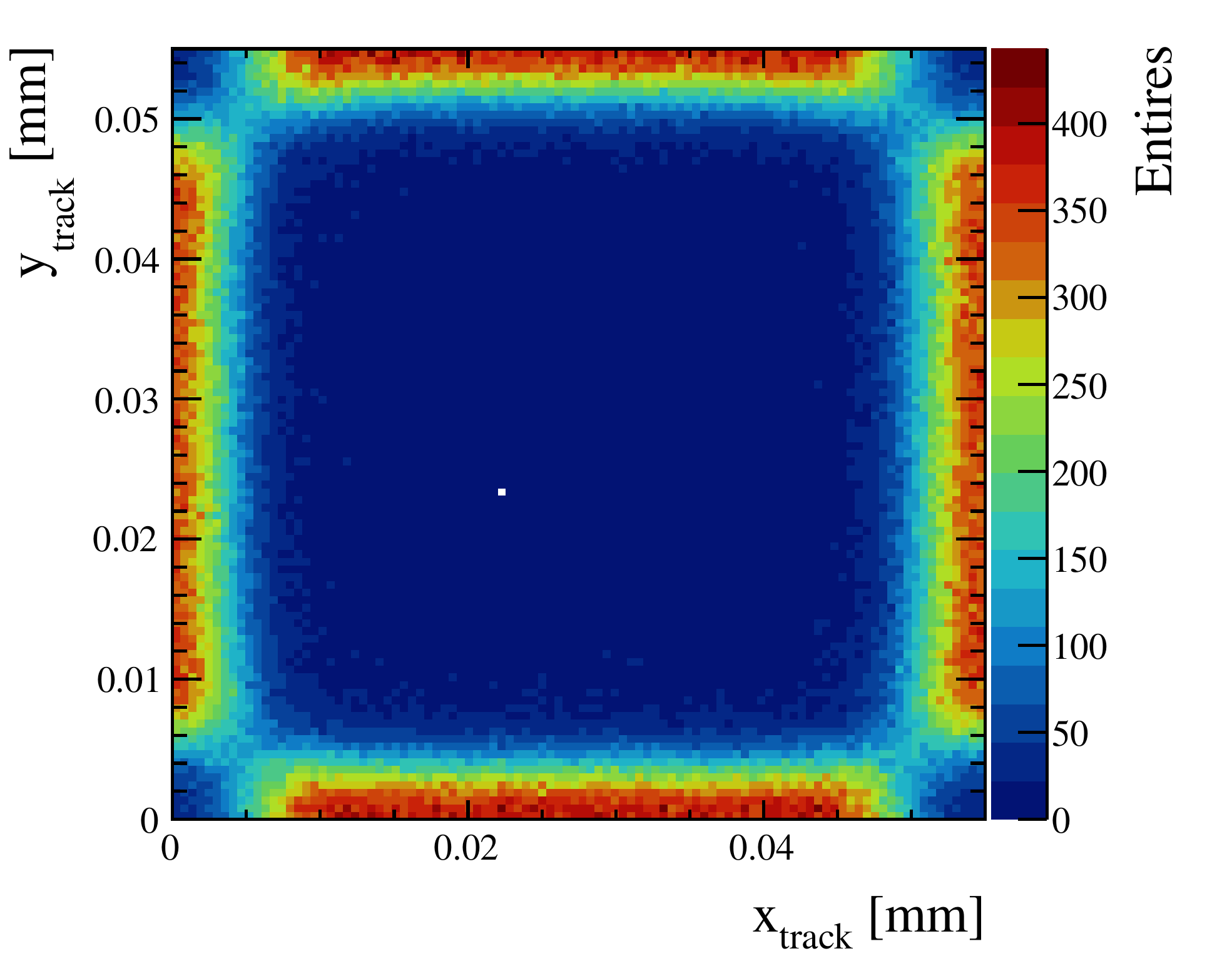}
  \includegraphics[width=0.32\textwidth]{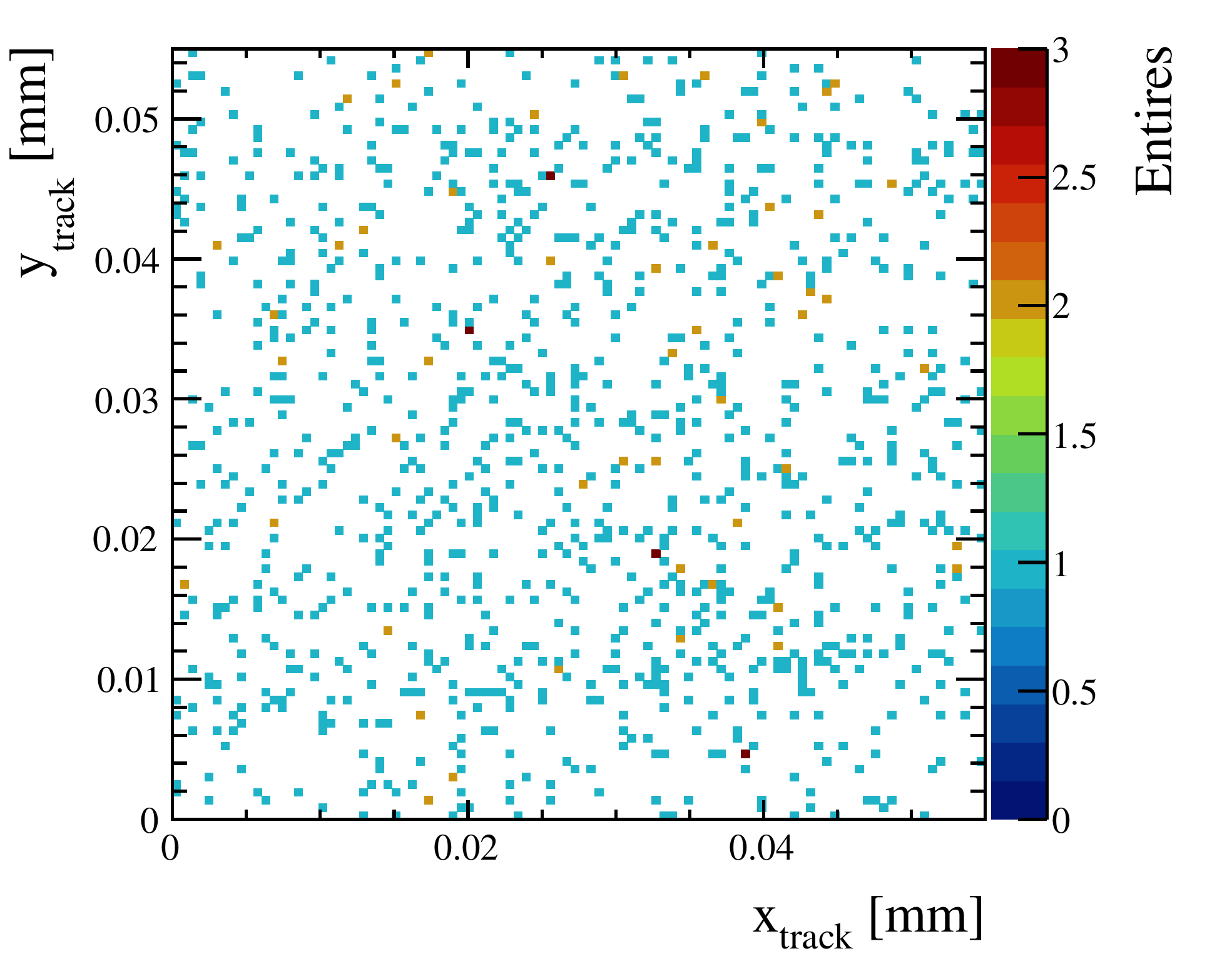}
  \includegraphics[width=0.32\textwidth]{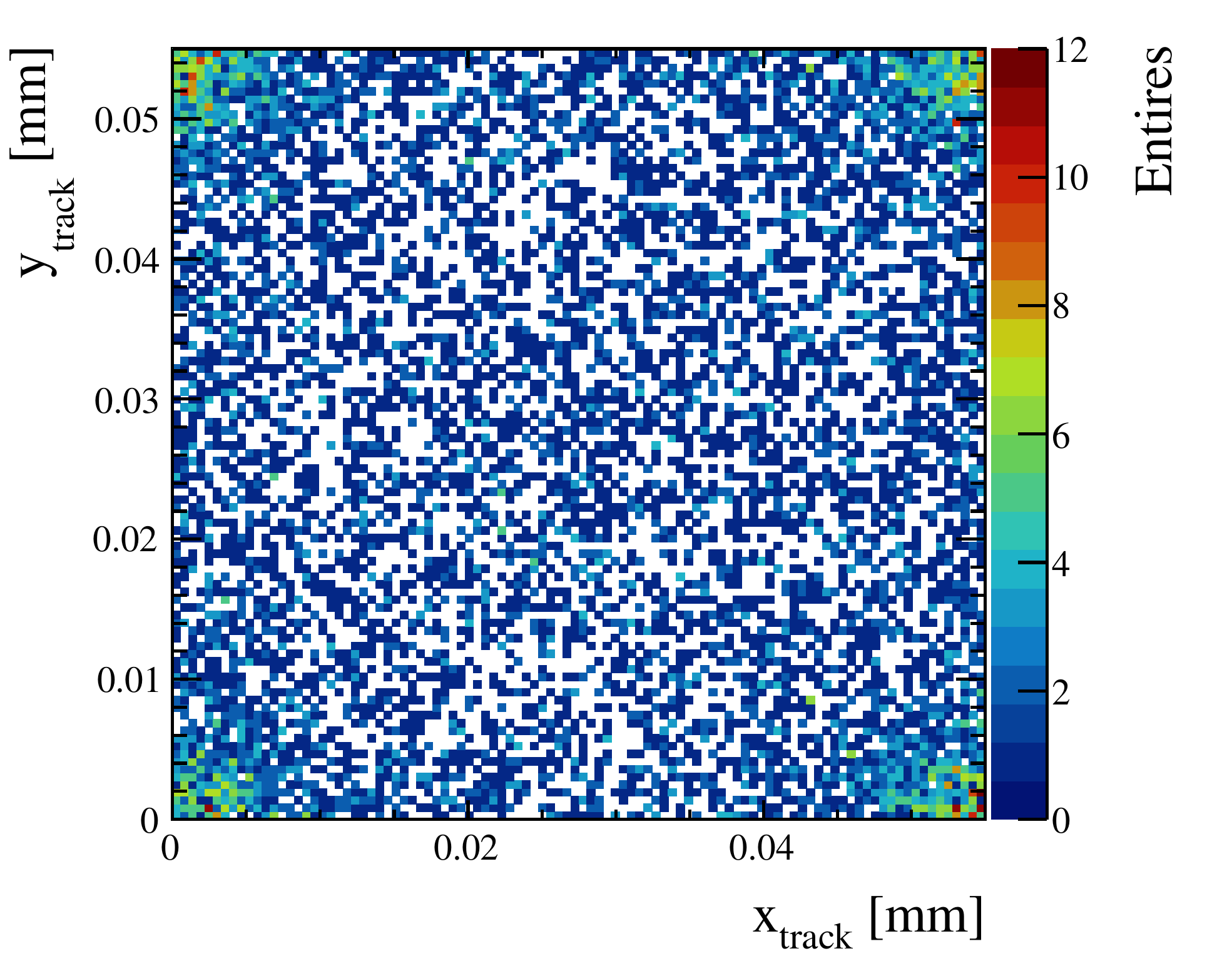}
  \includegraphics[width=0.32\textwidth]{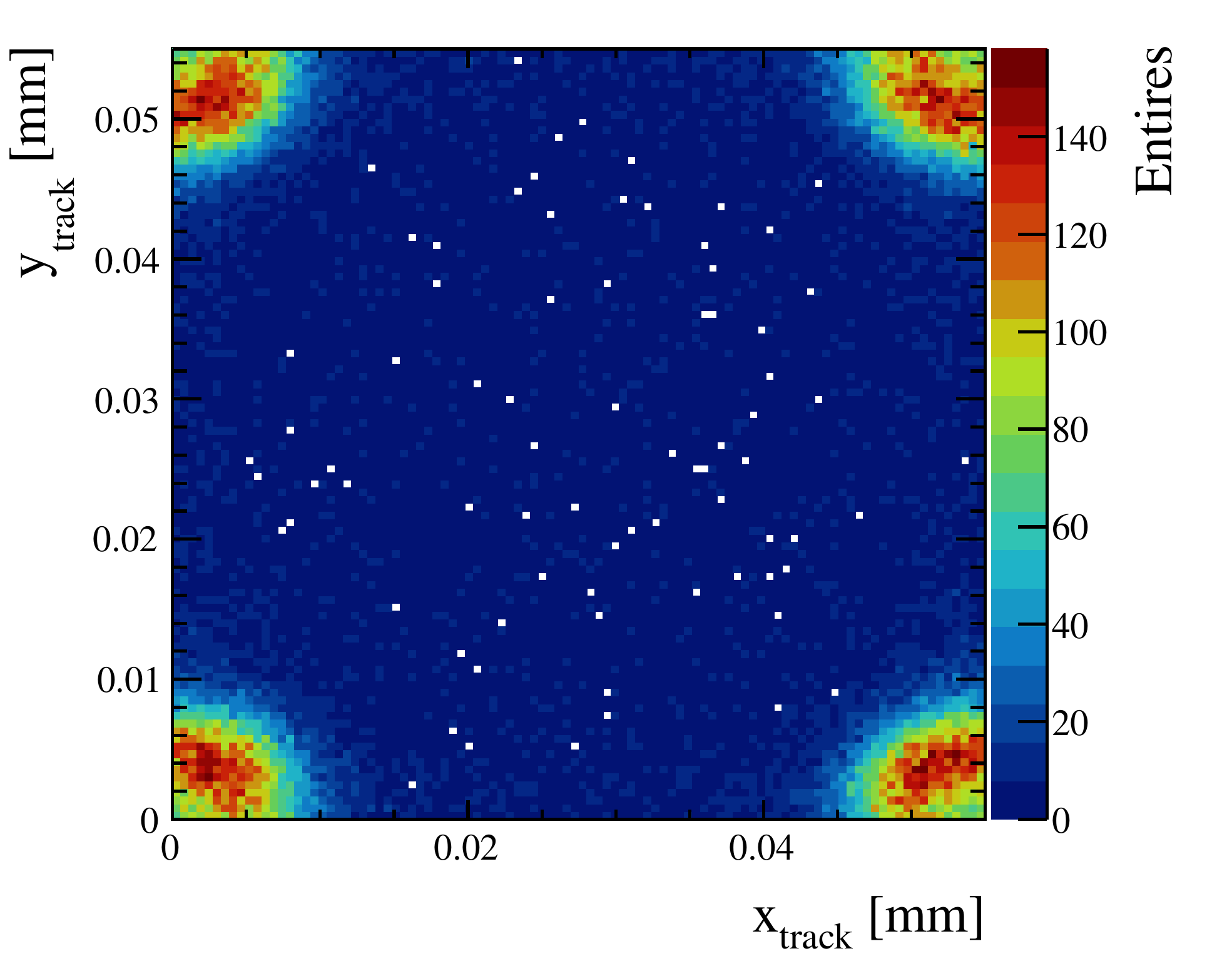}
 \includegraphics[width=0.32\textwidth]{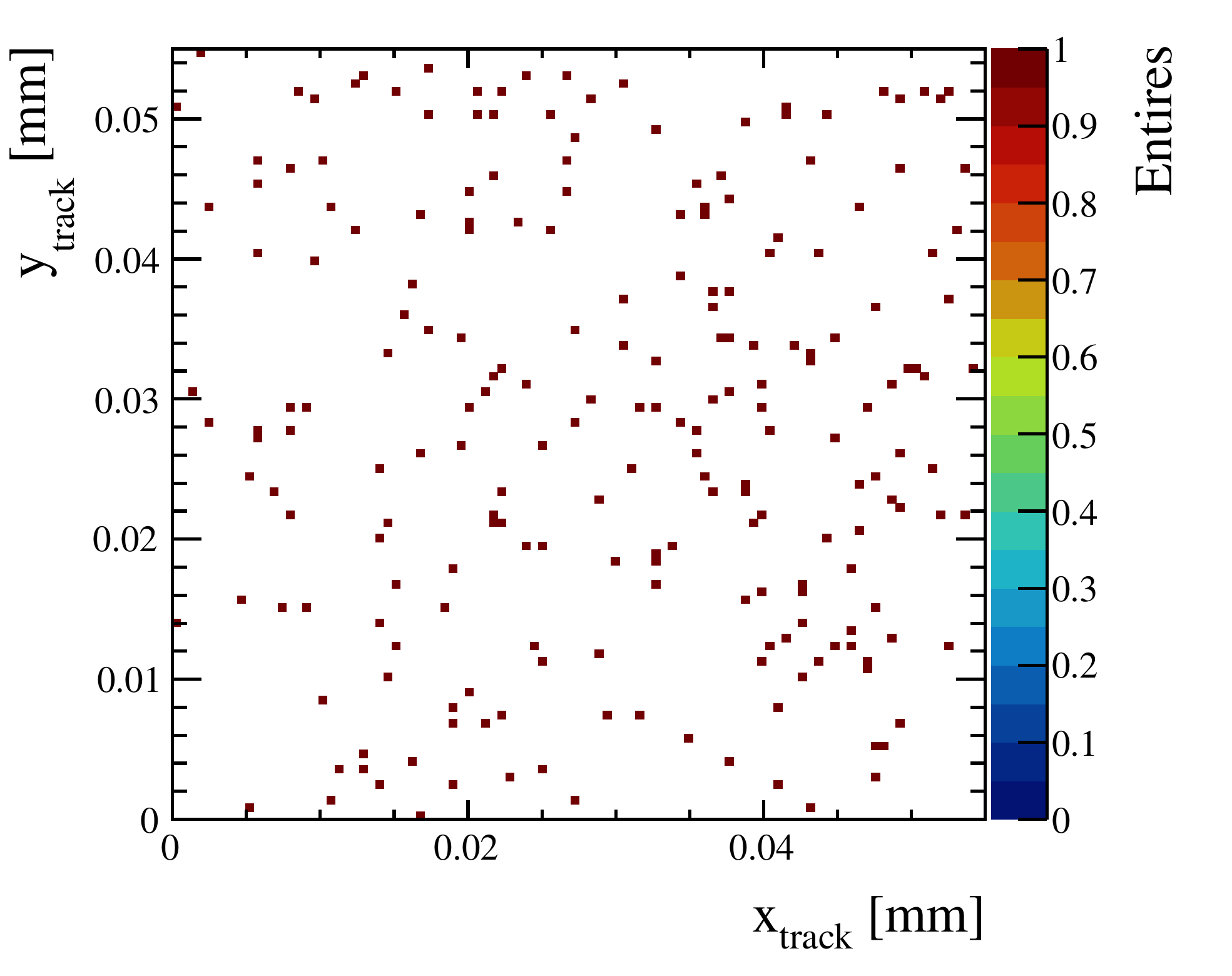}
 \includegraphics[width=0.32\textwidth]{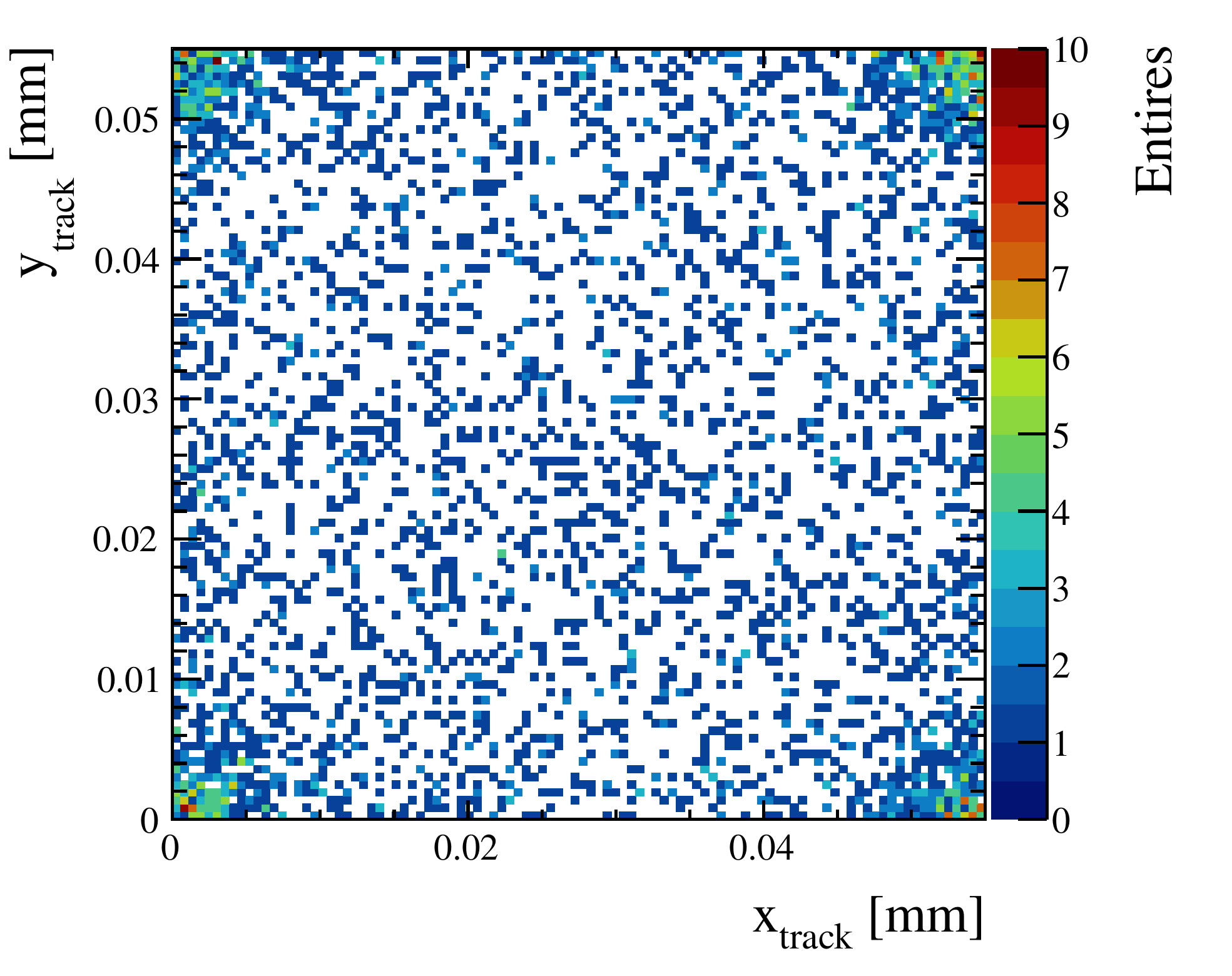}
  \includegraphics[width=0.32\textwidth]{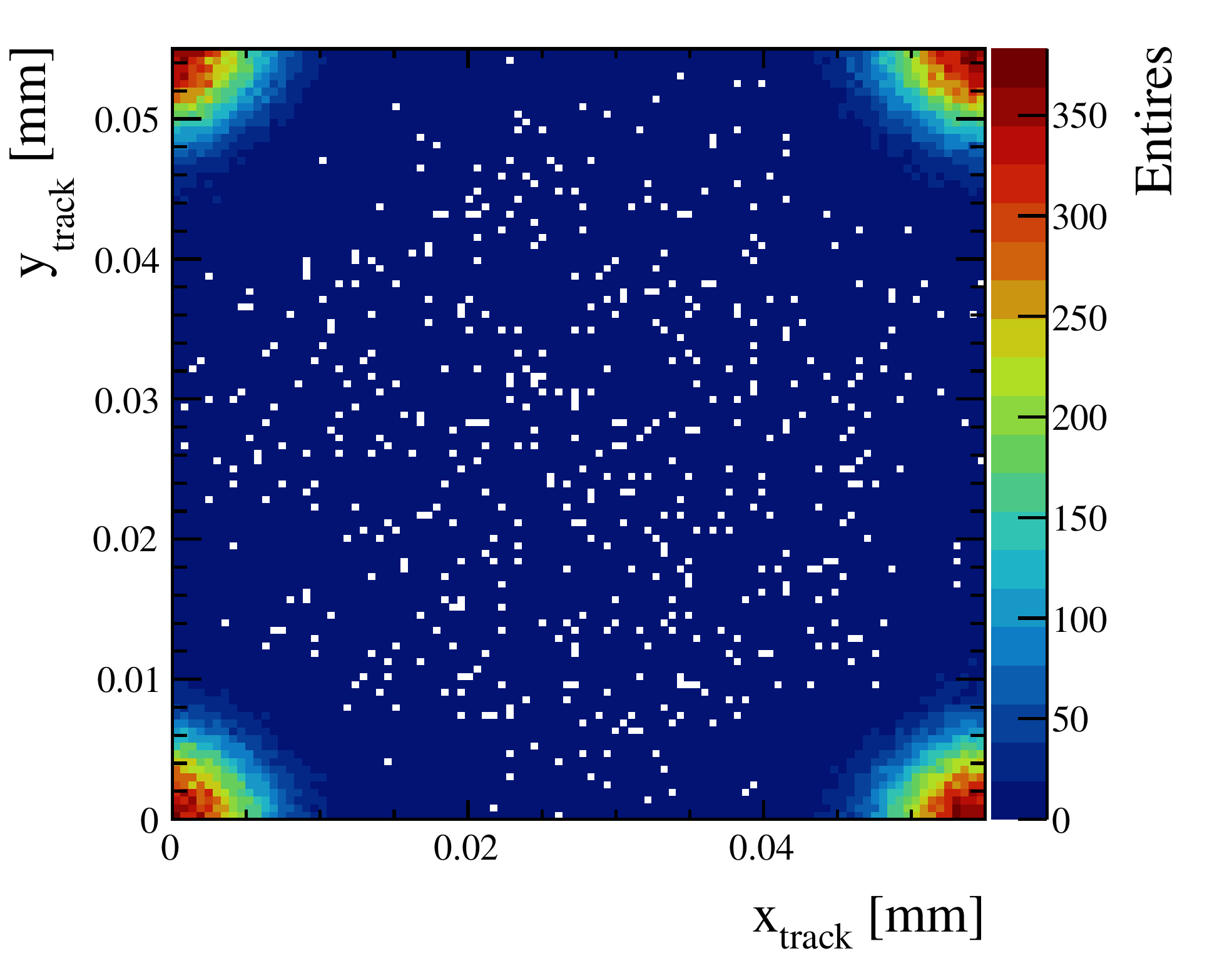}

    \caption[]{The number of tracks as a function of the intrapixel track positions depending on cluster size for three different applied voltage. The rows are size~1, size~2, size~3 and size~4 from top to bottom and the columns are 100~V, 500~V and 1000~V from left to right. The example shown is for a HPK n-on-p (S9) sensor uniformly irradiated to \maxfluence.  }
  \label{fig:post_bias_hits}
\end{figure}

The intrapixel track positions depending on cluster size 1, 2, 3 $\&$ 4 for a Micron n-on-p sensor operated around the depletion voltage and placed perpendicular to the beam was shown previously in~\Cref{fig:hits}. The same plots are shown in \Cref{fig:post_bias_hits} for a HPK sensor operated at three different applied bias voltages (100, 500, 1000~V) for a sensor uniformly irradiated to the \maxfluence. At 100~V, the majority of tracks result in a size~1 cluster due to the inefficiencies described in~\Cref{sec:efficiency}. At 500~V, the sensor is not yet fully efficient but the number of tracks resulting in size~2 clusters increases. At 1000~V, the sensor is fully efficient.

\begin{figure}[ht]
 \centering
  \includegraphics[width=0.32\textwidth]{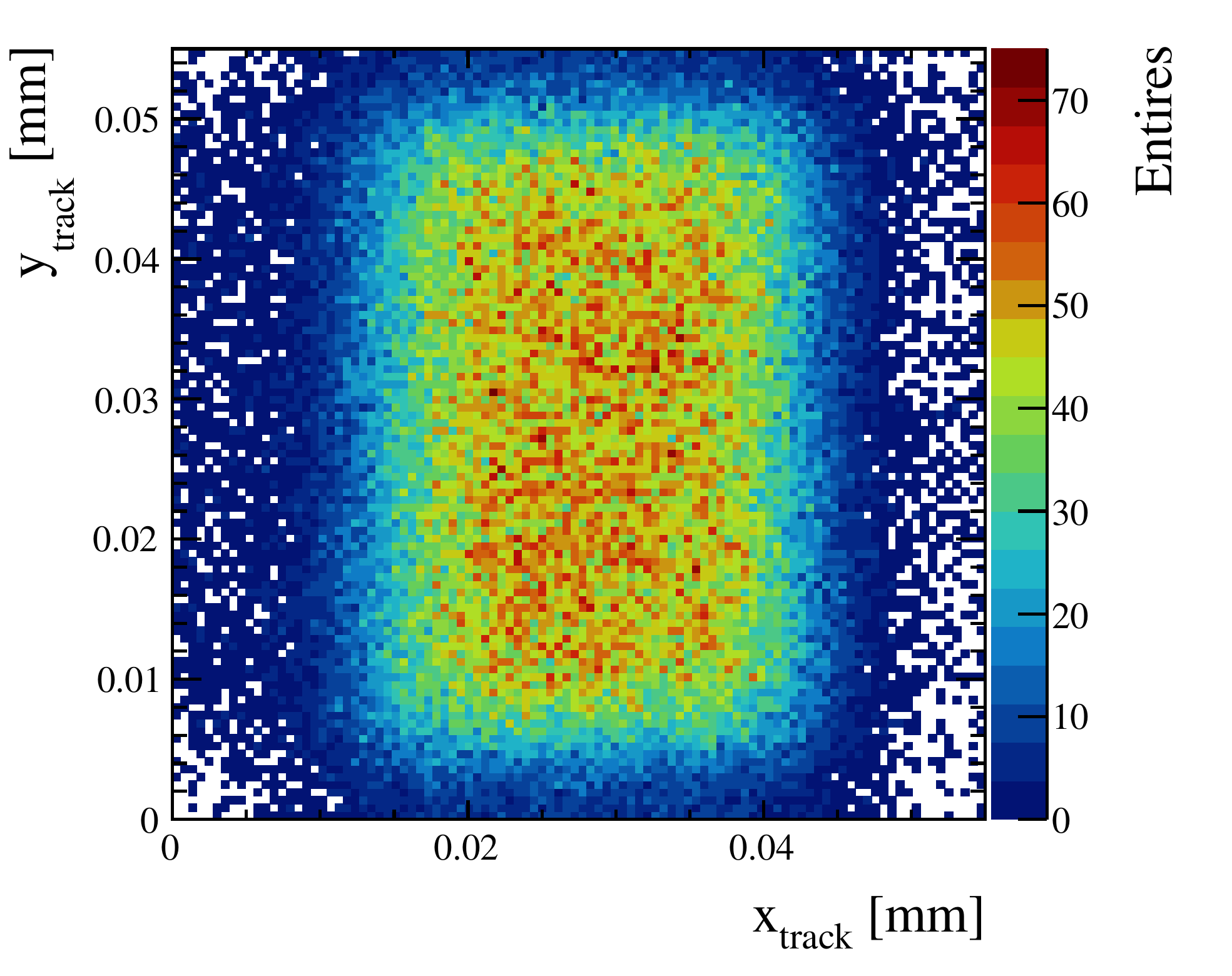}
  \includegraphics[width=0.32\textwidth]{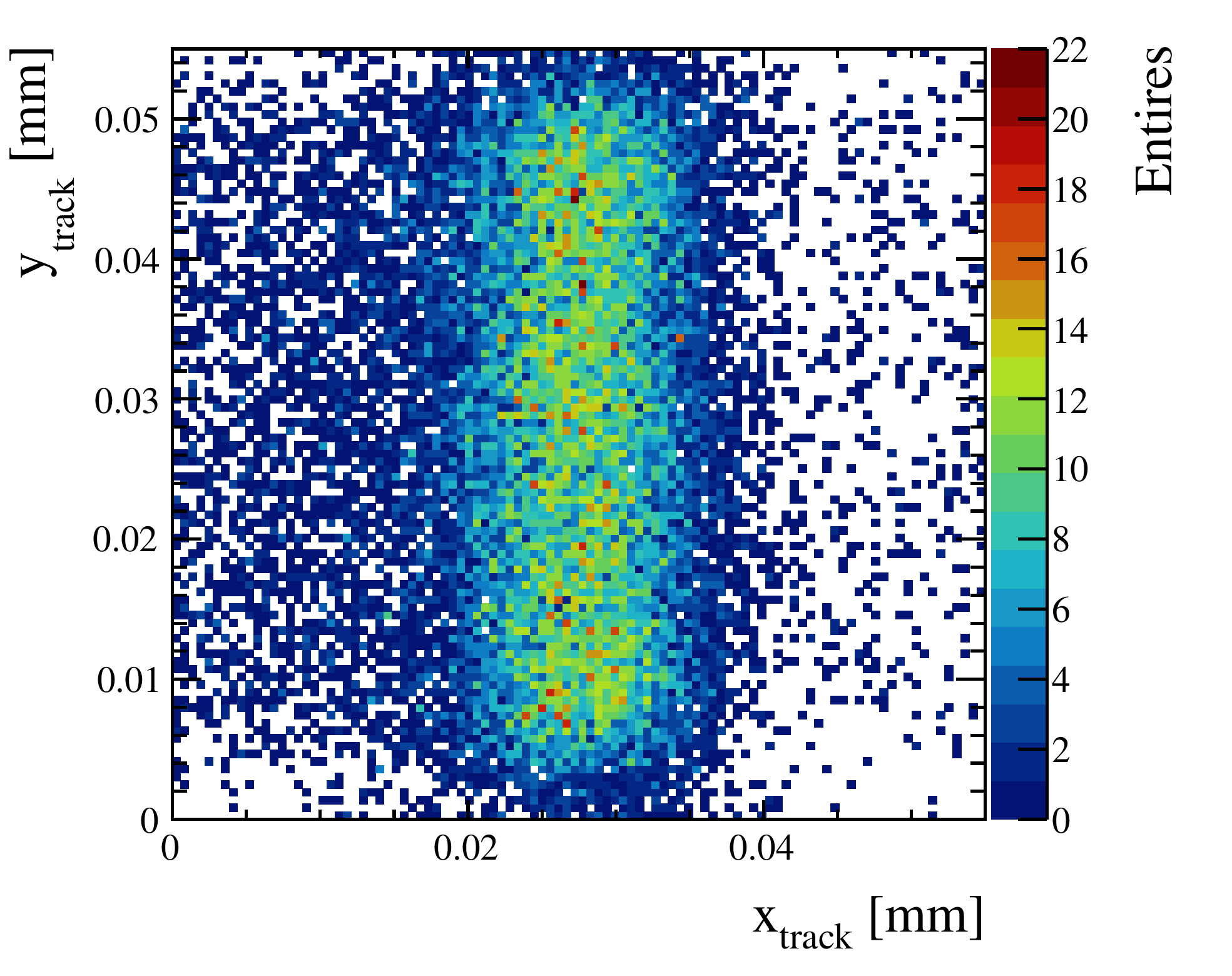}
  \includegraphics[width=0.32\textwidth]{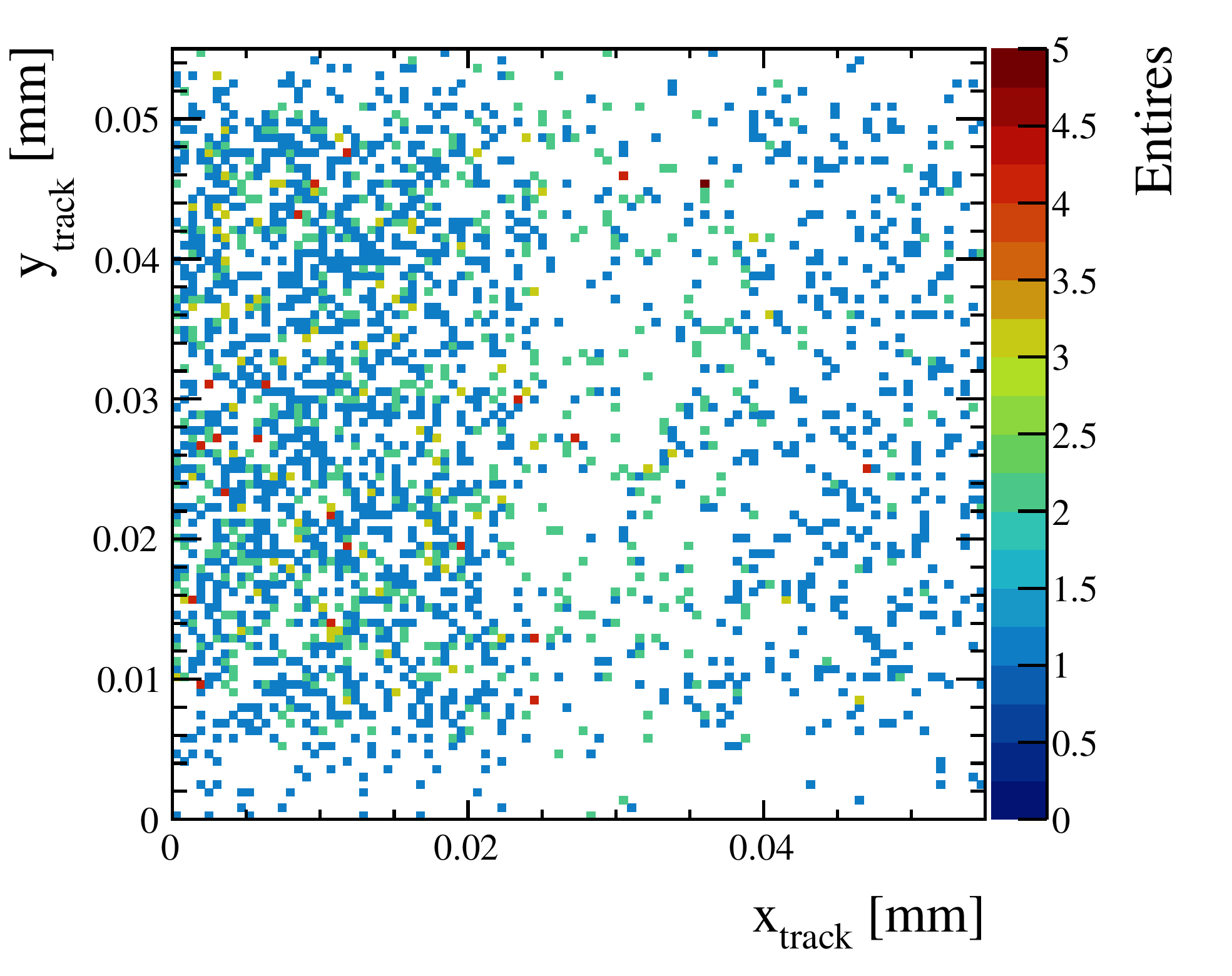}
  \includegraphics[width=0.32\textwidth]{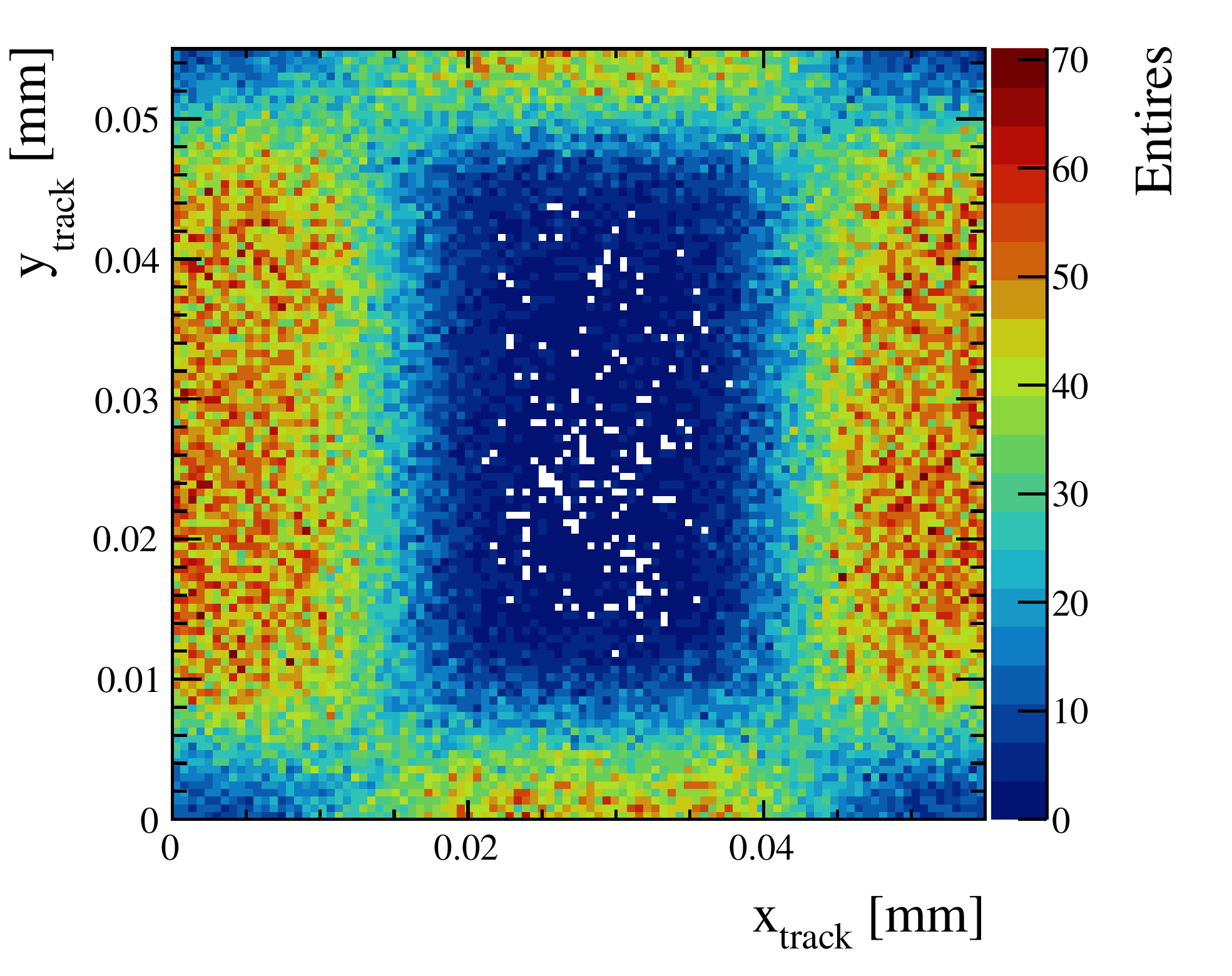}
  \includegraphics[width=0.32\textwidth]{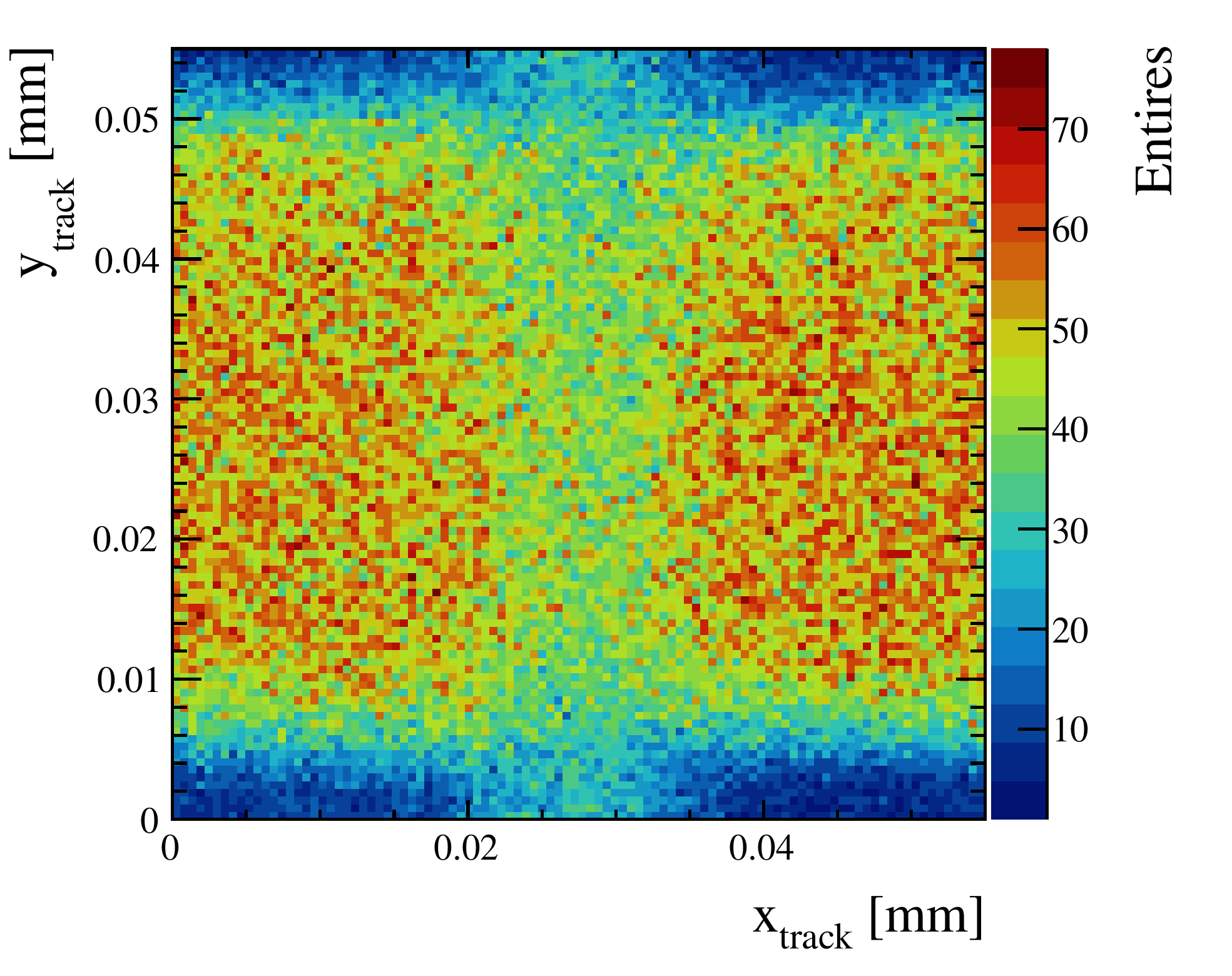}
  \includegraphics[width=0.32\textwidth]{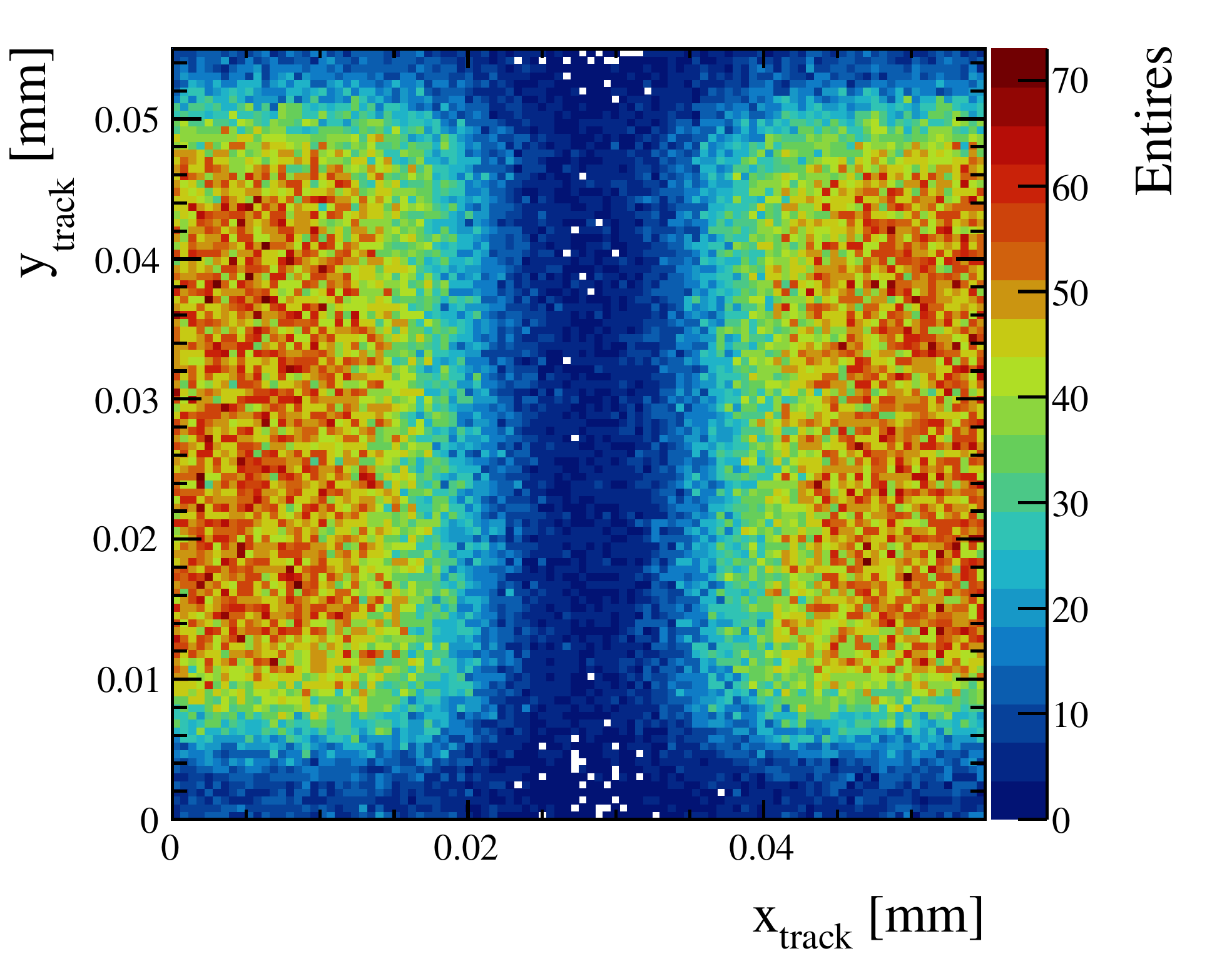}
  \includegraphics[width=0.32\textwidth]{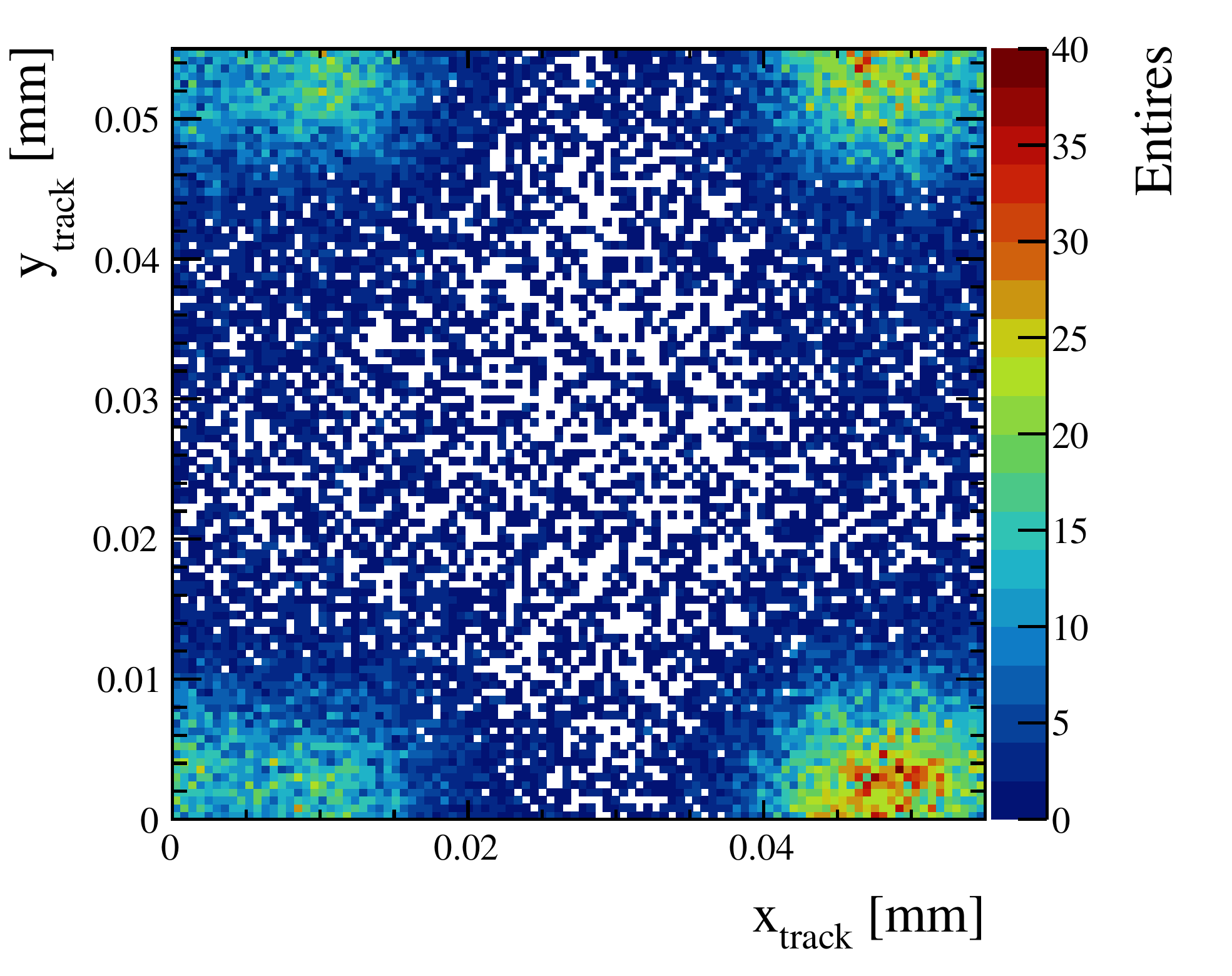}
  \includegraphics[width=0.32\textwidth]{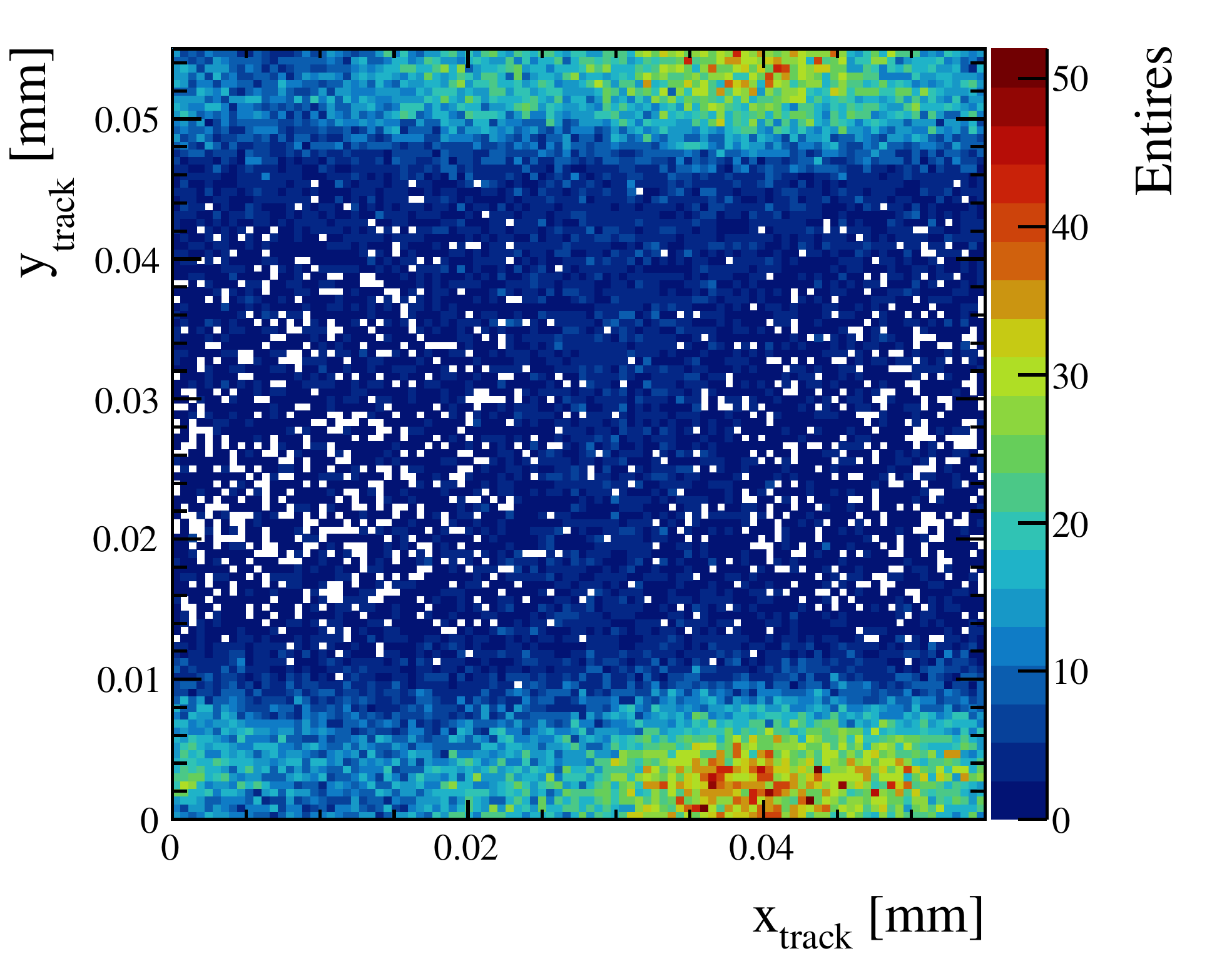}
  \includegraphics[width=0.32\textwidth]{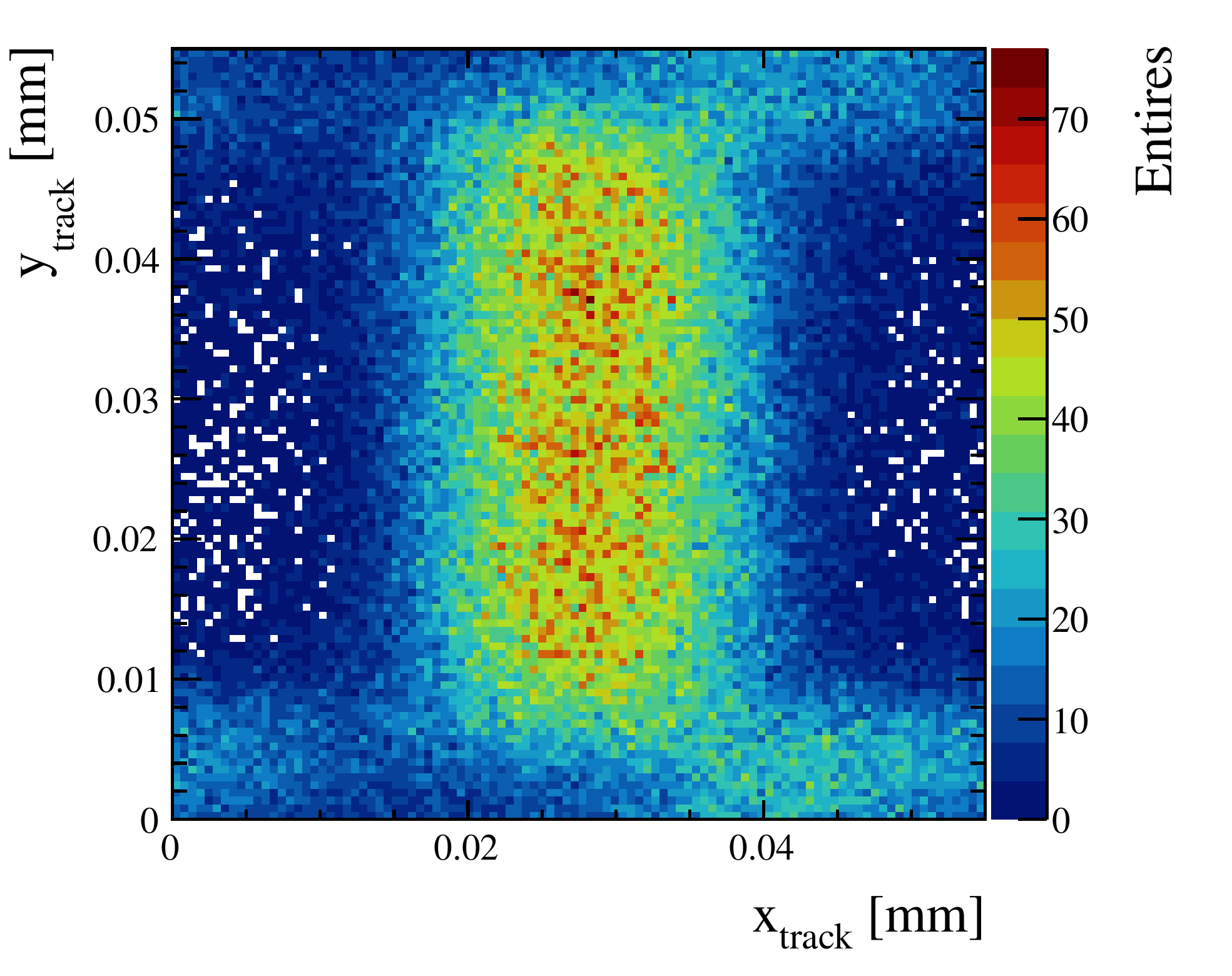}
  \includegraphics[width=0.32\textwidth]{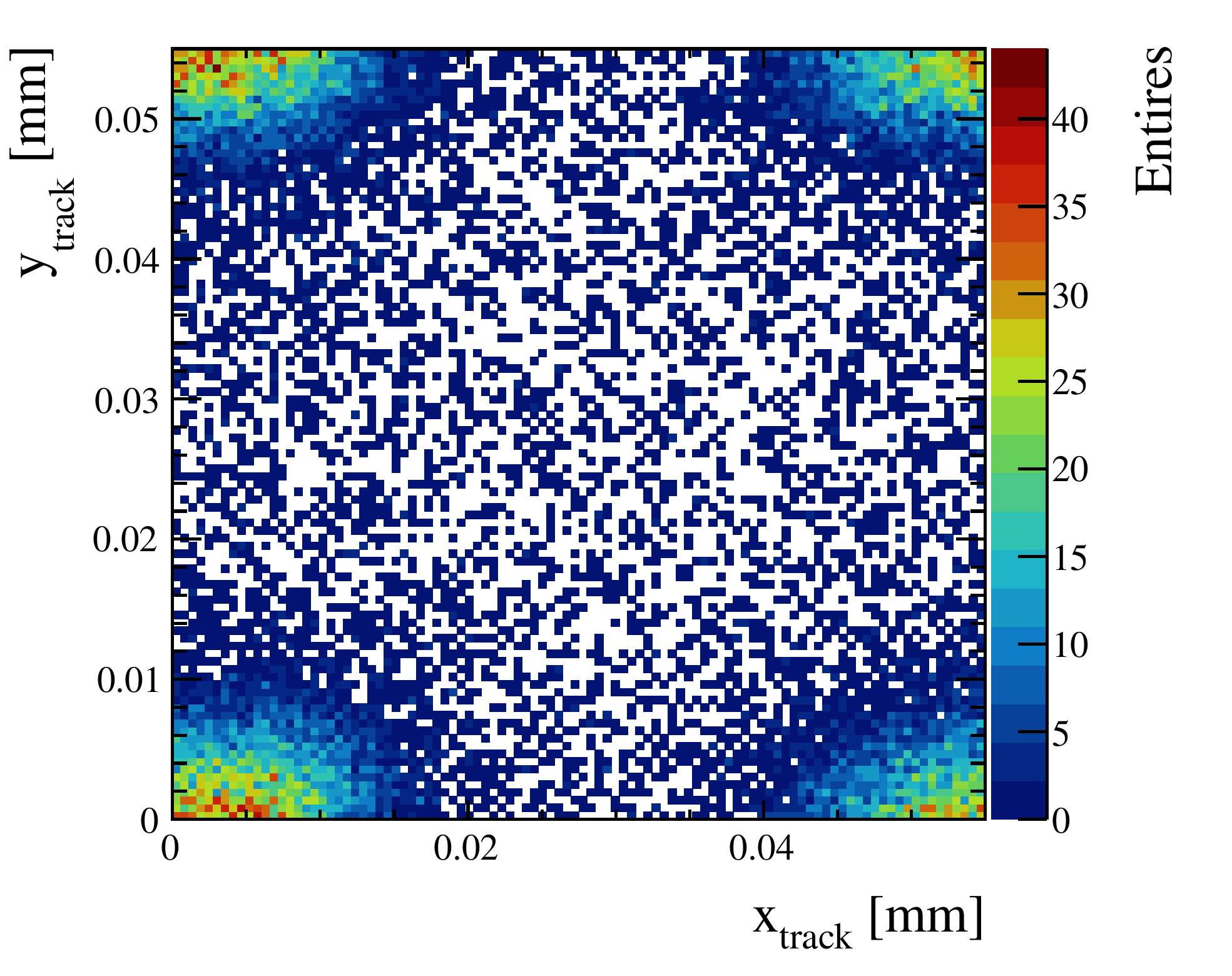}
  \includegraphics[width=0.32\textwidth]{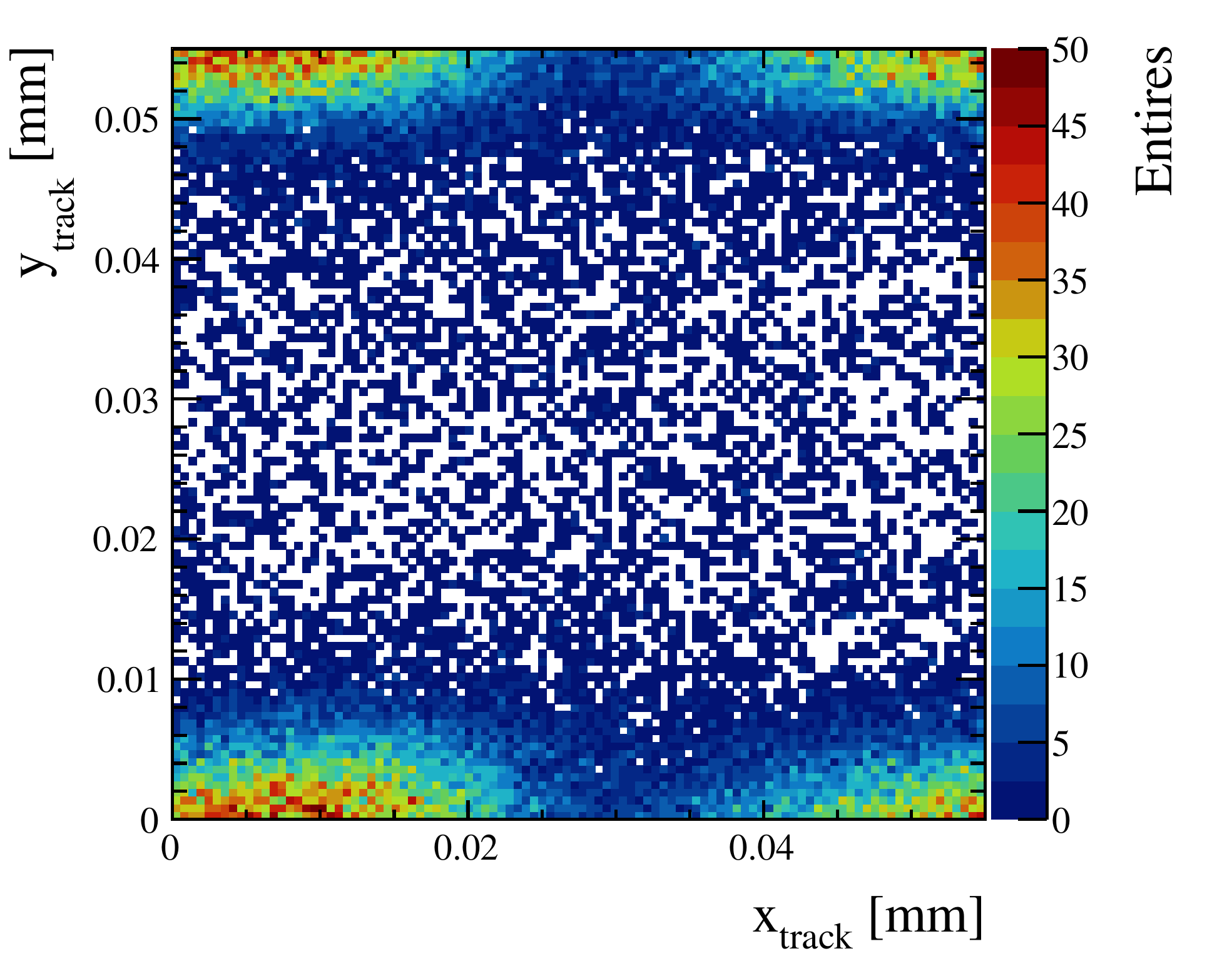}
  \includegraphics[width=0.32\textwidth]{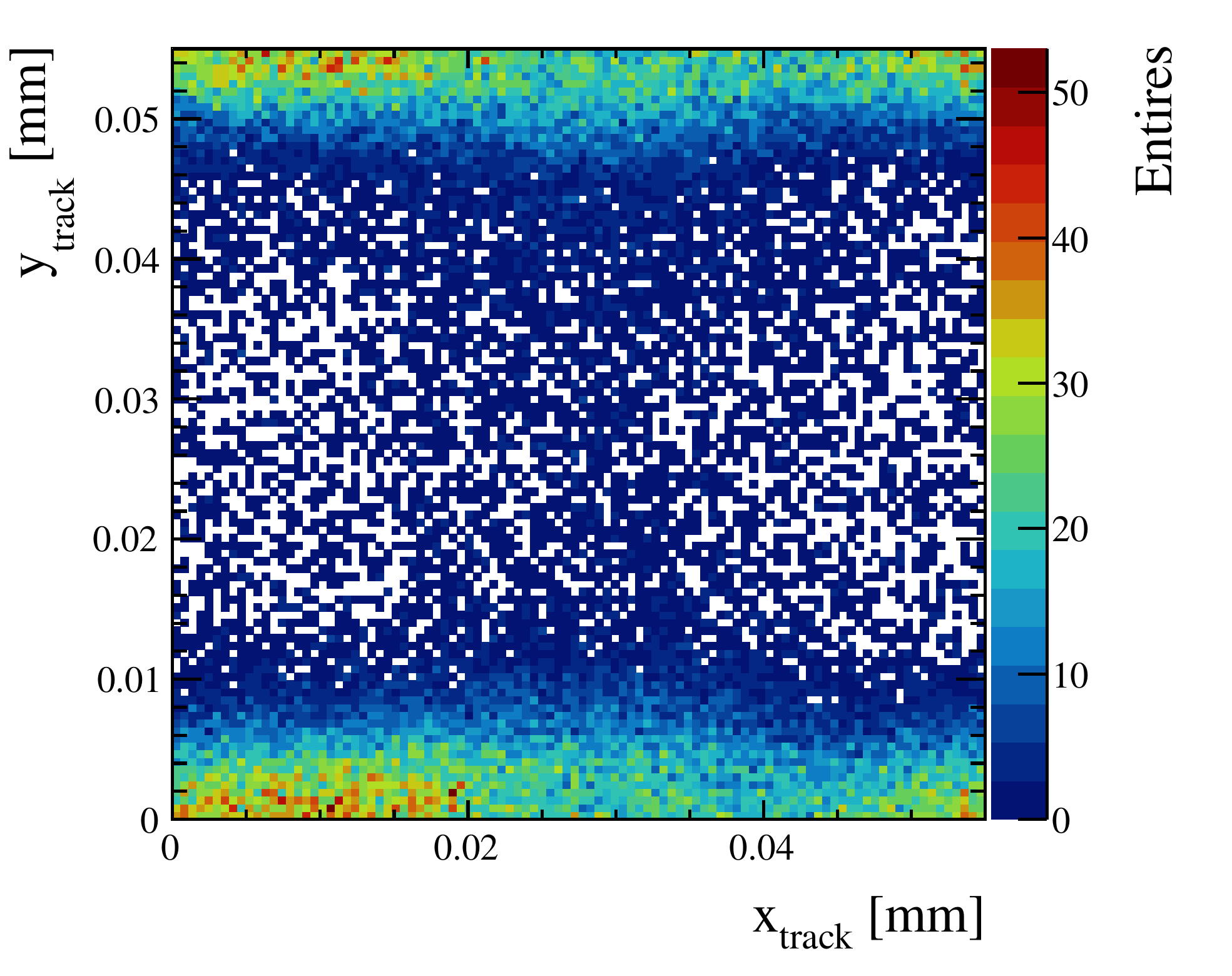}

  \caption[]{The number of tracks as a function of the intrapixel track positions depending on cluster size for a 200~$\mum$ non-irradiated Micron n-on-p sensor (S25) at angles of 8$^{\circ}$, 16$^{\circ}$ and 22$^{\circ}$ relative to the beam. The rows are size~1, size~2, size~3 and size~4 from top to bottom and the columns are 8$^{\circ}$, 16$^{\circ}$ and 22$^{\circ}$ from left to right. }
  \label{fig:hits_track}
 \end{figure}

The intrapixel track positions depending on cluster size are also shown in \Cref{fig:hits_track} for a non-irradiated Micron n-on-p sensor operated around depletion, placed at three different angles (8$^{\circ}$, 16$^{\circ}$ $\&$ 22$^{\circ}$) relative to the beam. The apparent left-right asymmetry is due to diffusion from the charges moving through longer paths in the sensor. 

\end{appendices}

\clearpage
\addcontentsline{toc}{section}{References}
\setboolean{inbibliography}{true}
\bibliographystyle{JHEP}
\bibliography{main}

\providecommand{\href}[2]{#2}\begingroup\raggedright\begin{thebibliography}{10}

\bibitem{LHCb-TDR-013}
{\scshape LHCb} collaboration, \emph{{LHCb VELO Upgrade Technical Design
  Report}},  Tech. Rep. {CERN-LHCC-2013-021}, CERN, Geneva, 2013.

\bibitem{VELOPIX}
T.~Poikela et~al., \emph{{The VeloPix ASIC}},
  \href{https://doi.org/10.1088/1748-0221/12/01/C01070}{\emph{JINST} {\bfseries
  12} (2017) C01070}.

\bibitem{Timepix3}
T.~Poikela, J.~Plosila, T.~Westerlund, M.~Campbell, M.~Gaspari, X.~Llopart
  et~al., \emph{Timepix3: A 65k channel hybrid pixel readout chip with
  simultaneous toa/tot and sparse readout},
  \href{https://doi.org/10.1088/1748-0221/9/05/C05013}{\emph{Journal of
  Instrumentation} {\bfseries 9} (2014) C05013}.

\bibitem{Akiba_2019}
K.~Akiba, M.~v. Beuzekom, H.~Boterenbrood, E.~Buchanan, J.~Buytaert,
  W.~Byczynski et~al., \emph{Lhcb velo timepix3 telescope},
  \href{https://doi.org/10.1088/1748-0221/14/05/p05026}{\emph{Journal of
  Instrumentation} {\bfseries 14} (2019) P05026–P05026}.

\bibitem{VicenteBarretoPinto:2134709}
M.~Vicente Barreto~Pinto, \emph{{Caracterização do TimePix3 e de sensores
  resistentes à radiação para upgrade do VELO}},  Master's thesis,
  Universidade Federal do Rio de Janeiro, Feb, 2016.

\bibitem{dallocco2021temporal}
E.~Dall’Occo, K.~Akiba, M.~van Beuzekom, E.~Buchanan, P.~Collins, T.~Evans
  et~al., \emph{Temporal characterisation of silicon sensors on timepix3
  asics}, \href{https://doi.org/10.1088/1748-0221/16/07/p07035}{\emph{Journal
  of Instrumentation} {\bfseries 16} (2021) P07035}.

\bibitem{Geertsema_2021}
R.~Geertsema, K.~Akiba, M.~v. Beuzekom, E.~Buchanan, C.~Burr, W.~Byczynski
  et~al., \emph{Charge collection properties of prototype sensors for the lhcb
  velo upgrade},
  \href{https://doi.org/10.1088/1748-0221/16/02/p02029}{\emph{Journal of
  Instrumentation} {\bfseries 16} (2021) P02029–P02029}.

\bibitem{JSI}
L.~Snoj et~al., \emph{{Computational analysis of irradiation facilities at the
  JSI TRIGA reactor}},
  \href{https://doi.org/https://doi.org/10.1016/j.apradiso.2011.11.042}{\emph{Applied
  Radiation and Isotopes} {\bfseries 70} (2012) 483 }.

\bibitem{Gkotse_2237333}
\emph{{IRRAD: The New 24GeV/c Proton Irradiation Facility at CERN}}, Nov, 2015.

\bibitem{DallOccoThesis}
E.~Dall'Occo, \emph{{Search for heavy neutrinos and characterisation of silicon
  sensors for the VELO upgrade}}, Ph.D. thesis, Vrije Universiteit Amsterdam,
  2020.

\bibitem{sps}
CERN, \emph{The super proton synchrotron},  tech. rep.

\bibitem{Heijhoff_2020}
K.~Heijhoff, K.~Akiba, M.~v. Beuzekom, P.~Bosch, J.~Buytaert, M.~Campbell
  et~al., \emph{Timing performance of the lhcb velo timepix3 telescope},
  \href{https://doi.org/10.1088/1748-0221/15/09/p09035}{\emph{Journal of
  Instrumentation} {\bfseries 15} (2020) P09035–P09035}.

\bibitem{Buchanan}
E.~Buchanan, \emph{{Spatial Resolution Studies for the LHCb VELO Upgrade}},
  Ph.D. thesis, University of Bristol, Oct, 2018.

\bibitem{Spieler:1010490}
H.~Spieler, \emph{{Semiconductor detector systems}}, Series on semiconductor
  science and technology. Oxford Univ. Press, Oxford, 2005,
  \href{https://doi.org/10.1093/acprof:oso/9780198527848.001.0001}{10.1093/acprof:oso/9780198527848.001.0001}.

\bibitem{Richards}
S.~E. Richards, \emph{{Characterisation of silicon detectors for the LHCb
  Vertex Locator Upgrade}}, Ph.D. thesis, University of Bristol, Nov, 2017.

\end{thebibliography}\endgroup

\end{document}